\input epsf
\magnification=\magstep1
\documentstyle{amsppt}
\pagewidth{6.4truein}\hcorrection{0.00in}
\pageheight{8.1truein}\vcorrection{0.4in}

\TagsOnRight
\NoRunningHeads
\NoBlackBoxes
\catcode`\@=11
\def\logo@{}
\footline={\ifnum\pageno>1 \hfil\folio\hfil\else\hfil\fi}
\topmatter
\title 
%
%
%
The scaling limit of the correlation of holes on the triangular lattice
with periodic boundary conditions 
\endtitle
\author Mihai Ciucu\endauthor
\thanks Research supported in part by NSF grant DMS 0100950.
\endthanks
\affil
  School of Mathematics, Georgia Institute of Technology\\
  Atlanta, Georgia 30332-0160
\endaffil
\abstract We define the correlation of holes on the triangular lattice under periodic boundary 
conditions and study its asymptotics as the distances between the holes grow to infinity. 
We prove
that
the joint correlation of an arbitrary collection of lattice-triangular holes of even sides satisfies,
for large separations between the holes, a Coulomb law and a superposition principle that perfectly 
parallel the laws of two dimensional electrostatics, with physical charges corresponding to holes, 
and their magnitude to the difference between the number of right-pointing and left-pointing unit 
triangles in each hole.

We detail this parallel by indicating that, as a consequence
of our result, the relative probabilities of finding a fixed collection of holes at given mutual
distances (when sampling uniformly at random over all unit rhombus tilings of the complement of the 
holes) approaches, for large separations between the holes, the relative probabilities of 
finding the corresponding two dimensional physical system of charges at given mutual distances.
Physical temperature corresponds to a parameter refining the background triangular lattice.

We also give an equivalent phrasing of our result in terms of covering surfaces of given holonomy. From this 
perspective, two dimensional electrostatic potential energy arises by averaging over all possible discrete 
geometries of the covering surfaces.

\endabstract 
\endtopmatter
\document

\def\mysec#1{\bigskip\centerline{\bf #1}\message{ * }\nopagebreak\par\bigskip}

\def\myref#1{\item"{[{\bf #1}]}"} 
 
\def\pf{{\it Proof.\ }} 

\def\epf{\hfill{$\square$}\smallpagebreak}
\def\epfmath{\hfill{$\square$}}
\def\cite#1{\relaxnext@
  \def\nextiii@##1,##2\end@{[{\bf##1},\,##2]}%
  \in@,{#1}\ifin@\def\next{\nextiii@#1\end@}\else
  \def\next{[{\bf#1}]}\fi\next}
\def\proclaimheadfont@{\smc}

\def\pf{{\it Proof.\ }}

\define\Z{{\Bbb Z}}
\define\Q{{\Bbb Q}}
\define\R{{\Bbb R}}
\define\C{{\Bbb C}}
\define\M{\operatorname{M}}
\define\Rep{\operatorname{Re}}

\define\ch{\operatorname{ch}}

\define\de{\operatorname{d}}

\define\twoline#1#2{\line{\hfill{\smc #1}\hfill{\smc #2}\hfill}}

\def\mypic#1{\epsffile{figs/#1}}



\define\Bout{1}
\define\ri{2}
\define\sc{3}
\define\ppone{4}
\define\ef{5}
\define\CEP{6}
\define\CKP{7}
\define\CLP{8}
\define\Fone{9}
\define\FS{10}
\define\Har{11}
\define\HKMS{12}
\define\Jor{13}
\define\Kone{14}
\define\Ktwo{15}
\define\Kthree{16}
\define\KOS{17}
\define\Muir{18}
\define\Ol{19}
\define\Perc{20}
\define\SheffRS{21}
\define\SheffGFF{22}
\define\Sl{23}
\define\Sta{24}
\define\Statwo{25}

\centerline{\bf Introduction}

\medskip
In \cite{\sc} we considered the joint correlation $\omega$ of symmetrically distributed holes on the
triangular lattice, a natural extension of the monomer-monomer correlation introduced by Fisher and 
Stephenson in \cite{\FS}. Under the assumption that one of the holes is a unit lattice triangle $u$
on the symmetry axis $\ell$ and all remaining holes are lattice triangles of side 2, oriented so that
they point away from the lattice line perpendicular to $\ell$ that supports $u$, we proved in
\cite{\sc} that, asymptotically as the distances between holes are large,
$\omega$ satisfies a multiplicative superposition principle that perfectly parallels
the superposition principle of two dimensional electrostatics (with holes corresponding to electrical
charges, and charge magnitude given by the difference between the number of up-pointing and 
down-pointing unit triangles in a hole). 
Our proof was based on explicit product formulas we
obtained in \cite{\ppone} for the number of lozenge tilings of two families of lattice regions.

The correlation we used in \cite{\sc}  was defined by including the holes inside a lattice
hexagon approaching infinite size so that the holes remain around its center. The presence of the 
boundary of the hexagon distorts the local dimer statistics. More precisely, there exists an explicit
real valued function $f$ defined on the hexagon ($f$ is the unique maximum of a certain local entropy integral;
see \cite\CLP\cite\CKP) so that in the scaling limit the local statistics of dimers at each point inside the 
hexagon is governed by $\mu_{s,t}$, where $(s,t)$ is the tilt of $f$ at that point and
$\mu_{s,t}$ is the unique invariant Gibbs measure of slope $(s,t)$ (this was conjectured by Cohn, Kenyon and
Propp in \cite{\CKP} and proved and generalized by Sheffield in \cite{\SheffRS} and Kenyon, Okounkov and 
Sheffield in \cite{\KOS}). 
It follows from the latter two papers that $\mu_{0,0}$ is the unique invariant Gibbs measure of maximal entropy. 
Since the function $f$ has tilt $(0,0)$ only at the center of the hexagon, the local dimer statistics is
distorted away from maximal entropy everywhere except at the center\footnote{This 
figure is courtesy of David Wilson.}:

\midinsert
\centerline{\mypic{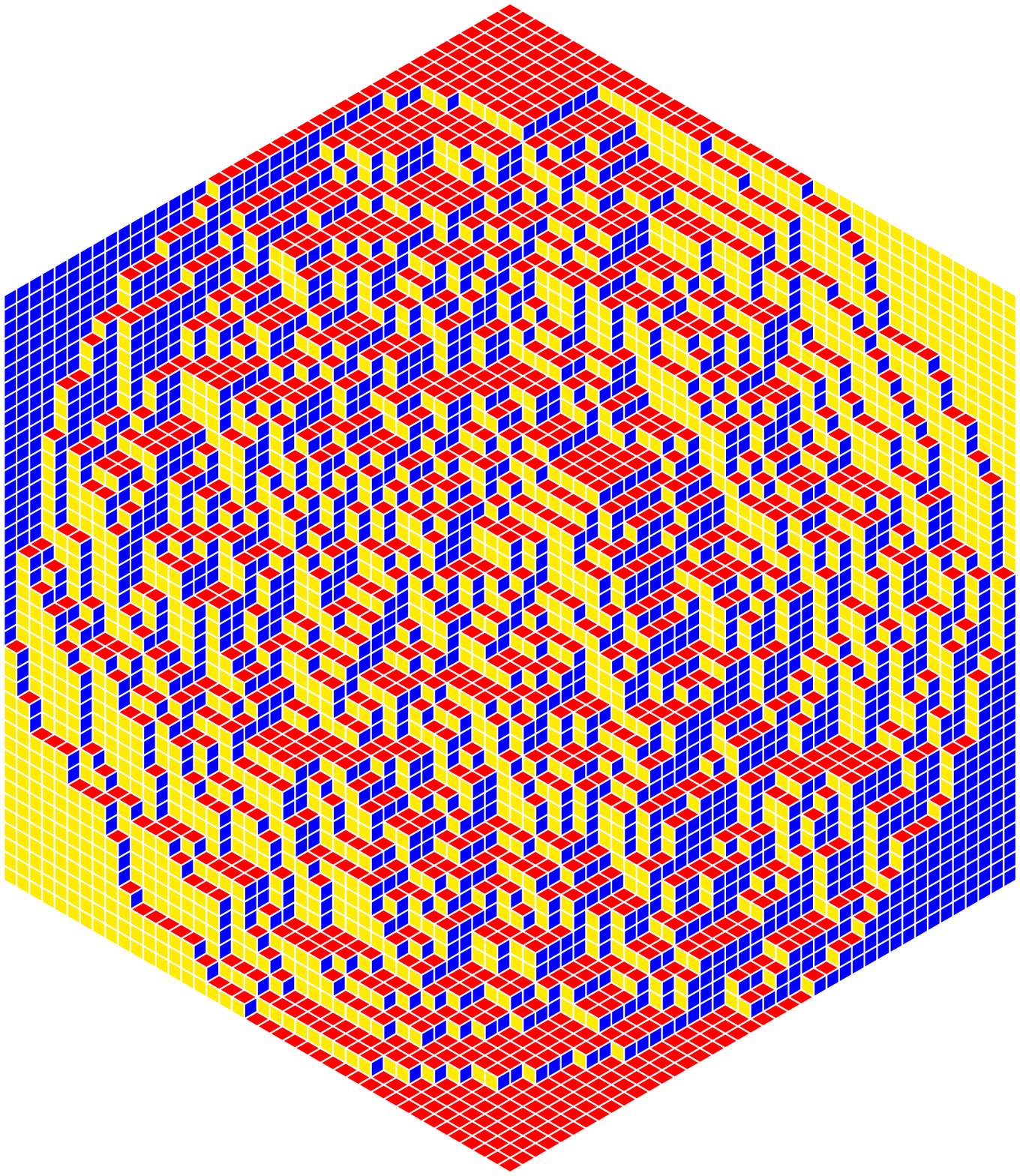}}
\centerline{{\smc } {\rm A typical lozenge tiling of a hexagon of side 40.}}
\endinsert

\flushpar
(for more details, see \cite{\CLP}; this does not 
happen in the case of a square on the square lattice considered in \cite{\FS}).
Therefore, an important goal is to 
define the correlation of holes in a different way, via regions that do not distort the dimer statistics,
and determine whether it still reduces to the superposition principle of 
electrostatics in the limit as the holes grow far apart.
Other highly desirable features are to allow general, not necessarily symmetric
distributions of the holes, as well as holes of arbitrary size.

It is 
these goals
that we 
accomplish
in this paper. We give a new definition for the correlation of 
holes by including them in a sequence
of lattice tori of size approaching infinity (see Section~1). 
We prove that this new correlation also reduces to the superposition principle of electrostatics as the 
distances between the holes grow to infinity.

We note that by \cite{\SheffRS} and \cite{\KOS} the torus measures on doubly periodic bipartite planar graphs
converge to the maximal entropy invariant Gibbs measure. Thus defining hole correlations via limits of tori
is consistent with our goal of having maximal entropy dimer statistics.

The main result of this paper, Theorem 1.1, addresses the
case of an arbitrary collection of lattice-triangular holes of even sides. It is a 
counterpart of our result \cite{\sc, Theorem 2.1}, but more general than that in three important 
ways: (1) no symmetry of the distribution of the holes is assumed here; (2) the triangular holes 
are allowed to have arbitrary even sides, and (3) the multiplicative constant has an explicit 
conceptual interpretation---it is obtained by multiplying a simple explicit constant, depending only 
on the charges of the holes, by the product of the correlations of each hole regarded on its own. 



As a consequence of Theorem 1.1, the parallel to electrostatics described in \cite{\sc}, which was 
restricted there
by assumptions of symmetry and charge, becomes now established in general---including the presence of
a ``lattice refinement'' parameter which corresponds to physical temperature. We detail this in Section
15, where we also 
give an equivalent phrasing of our result in terms of covering surfaces of given holonomy. From this 
perspective, two dimensional electrostatics arises by averaging over all possible discrete geometries 
of the covering surfaces.

Regarded from the point of view of surface models, our result raises naturally the question of describing
the limit of the average height function --- which can be suitably defined, even though as we see in Section 15
it is multivalued in general --- as the lattice spacing approaches zero. It turns out that the right
object to consider is, by contrast to other dimer surface limit shape results in the literature 
(see \cite{\CEP}\cite{\CLP}\cite{\CKP}), 
the un-normalized average height function (the normalized one approaches zero in the fine mesh limit),
and, as we show in \cite\ef, the limit surface is a sum of helicoids. This in turn is equivalent to the statement 
that a certain naturally defined discrete field (the $90^{\circ}$ rotation of the tilt of the average height
function) converges in the scaling limit to the electric field; see Section 15 for more details. 
The general surface models arising
from dimers on doubly periodic bipartite planar graphs are studied in \cite{\KOS}, where they are classified
according to the behavior of the variance of the height function. 

The defining difference between our results and the above quoted ones is the presence of holes. Indeed, the
boundaries of the holes form the only boundary present in our case. If there are no holes, 
the effects we are studying in this
paper and in \cite{\ef} are not present. Consistent with this is that for instance \cite{\KOS} focuses on the 
variance of height function differences, since the behavior of the average height function is not so interesting 
in the absence of boundary. So we could say that when specialized to the hexagonal lattice with unit weights
the results of \cite{\KOS} give the variance of the height function differences (the field mentioned in the 
previous paragraph being zero), while 
the sequel \cite{\ef} of this paper gives the value of the field created by the presence of holes.

It appears that there are very few rigorously proved results on the asymptotics of the correlation of 
non-zero charge holes in the literature (it is this case that reveals electrostatics as the governing
law; the particular instance of holes of charge zero equivalent 
to dimer-dimer correlations has 
been studied more; see e.g. \cite{\FS}). 
In fact, the only such result the author is aware of is the monomer-monomer 
correlation on the square lattice along a lattice diagonal direction obtained
by Hartwig \cite{\Har} (this was conjectured in \cite{\FS}). For a discussion of the differences between
our set-up and other discrete models for Coulomb gas in the physics literature, and for some
earlier pointers to electrostatics suggested by the study of the spin-spin correlation in the Ising
model, 
see \cite{\sc,pp.\,2--3\,and\,91}.

In order to relate he results of this paper relate to those of \cite{\ri} and \cite{\sc}, we note that
despite their very different definitions, the correlation $\hat\omega$ considered in this paper and the 
correlation $\omega$ of \cite{\sc} turn out to satisfy, up to a multiplicative constant, the same asymptotic 
superposition principle. We also note that the special case of Theorem 1.1 when one has
just two oppositely oriented lattice triangular holes of side 2 was treated in \cite{\ri} using a third,
quite different definition of correlation, and was shown there to obey exactly the same asymptotics as
the one that follows by Theorem~1.1 for $\hat\omega$ in this special case.

We conclude by mentioning that, as pointed out by the anonymous referee, in a different surface model, 
the discrete Gaussian free field model (see e.g. 
\cite{\SheffGFF}), there is a natural analog of holes.
Moreover, the question of determining the scaling limit of the average height function in the presence of these
``holes'' is much simpler than what it turns out to be in our model, and also leads to 
two dimensional electrostatics. In view of previous connections between the discrete Gaussian free field model
and the dimer model (see e.g. Boutillier \cite{\Bout} or Kenyon \cite{\Kthree}), this may offer an intuitive 
reason why the results we obtain in this paper and in \cite{\ef} should hold.

\medskip
\mysec{1. Definition of $\hat\omega$ and statement of main result}

\medskip
Draw the triangular lattice so that some of the lattice lines are vertical. For terminological brevity,
we will often refer to unit lattice triangles as {\it monomers}; a left-pointing (resp., right-pointing)
unit lattice triangle is called a {\it left-monomer} (resp., {\it right-monomer}).

For any finite union $Q$ of unit holes on the triangular lattice, define the {\it charge} $\ch(Q)$ of
$Q$ to be the number of right-monomers in $Q$ minus the number of left-monomers in $Q$. 

Let $U_1,\dotsc,U_n$ be arbitrary unit holes on the triangular lattice. Following Kenyon \cite{\Kone},
provided $\sum_{i=1}^n\ch(U_i)=0$, we define the joint correlation of $U_1,\dotsc,U_n$ by
$$
\omega_1(U_1,\dotsc,U_n):=\lim_{N\to\infty}
\frac{\M(H_{N,N}\setminus U_1\cup\cdots\cup U_n)}{\M(H_{N,N})},\tag1.1
$$
where $H_{N,N}$ is a large lattice rhombus of side $N$ whose opposite sides are identified so as to
create a torus, and
$M(R)$ denotes the number of lozenge tilings of the lattice region $R$ (a lozenge, or unit rhombus, 
is the union of any two unit lattice triangles that share an edge)
\footnote{ The total charge of the the holes needs to be zero in order for the region in the numerator
of (1.1) to contain the same number of each type of unit triangles---a necessary condition for the
existence of lozenge tilings. For any situation when the total
charge is not zero, (1.1) would assign value 0 to the joint correlation of the holes, irrespective
of their relative positions.}.

An important advantage of this definition is that, provided $U_1\cup\cdots\cup U_n$ is a union of
non-overlapping lattice triangles of side 2, \cite{\Kone} and \cite{\Ktwo} provide an 
expression for ${\omega_1}(Q_1,\dotsc,Q_n)$ as a determinant. 

To state this explicitly, we introduce 
the following system of coordinates on the mo\-no\-mers of the triangular lattice. 
Choose the origin $O$ to be at the center of a vertical unit lattice segment, and pick the 
$x$- and $y$-coordinate axes to be straight lines through $O$ of polar arguments $-\pi/3$ and 
$\pi/3$,
respectively. Coordinatize left-monomers by the (integer) coordinates of the midpoints of their 
vertical sides; coordinatize each right-monomer likewise, by the coordinates of the midpoint of its 
vertical side (thus any pair
of integers specifies a unique left-monomer and a unique right-monomer, sharing a vertical side).

Let $(l_1,l'_1),\dotsc,(l_k,l'_k)$ be the coordinates of the left-monomers contained in the union
$U_1\cup\cdots\cup U_n$ of our holes, and let the coordinates of 
the right-monomers contained in this union be $(r_1,r'_1),\dotsc,(r_k,r'_k)$ (the latter are the same
in number as the former, since we are assuming the total charge of the unit holes to be zero). Then,
provided $U_1\cup\cdots\cup U_n$ is a union of non-overlapping lattice triangles of side 2, 
it follows by \cite{\Kone} and \cite{\Ktwo} that
$$
{\omega_1}(U_1,\dotsc,U_n)=|\det\, (P(r_i-l_j,r'_i-l'_j))_{1\leq i,j\leq k}|,\tag1.2
$$
where 
$$
P(x,y):=\frac{1}{4\pi^2}\int_0^{2\pi}\int_0^{2\pi}\frac{e^{ix\theta}e^{iy\phi}d\theta\,d\phi}
{1+e^{-i\theta}+e^{-i\phi}}.\tag1.3
$$
Thus, if $Q_1,\dotsc,Q_n$ are holes that are unions of non-overlapping 
triangles of side 2, $i=1,\dotsc,n$, and $\sum_{i=1}^n\ch(Q_i)=0$, defining their correlation
by $\omega_1$ seems both natural and convenient, as the above determinant formula can be used
as a starting point for determining its asymptotics.


What if $\sum_{i=1}^n \ch(Q_i)\neq0$? We need then an appropriate extension of the correlation
${\omega_1}$ that applies also in this situation, and is manageable enough to work with so that 
its asymptotics can still be worked out. We found that the following inductively
defined correlation $\hat\omega$ does the job. This was a crucial part of our proof.

\proclaim{Definition} 
Let $T_1,\dotsc,T_n$ be lattice triangular holes of side $2$.

If $q=\sum_{i=1}^n \ch(T_i)\geq0$, define their joint correlation
$\hat{\omega}(T_1,\dotsc,T_n)$ inductively by: 

$(i)$ If $q=0$, define $\hat{\omega}$ by $(1.1)$.

$(ii)$ If $q=2s$, $s\geq1$, define
$$
\hat{\omega}(T_1,\dotsc,T_n)=\lim_{r\to\infty}(3r)^{2s}\hat{\omega}(T_1,\dotsc,T_n,W(3r,0)),
$$
where $W(3r,0)$ is the left- $($or west-$)$ pointing lattice triangle of side two whose western-most 
left-monomer has coordinates $(3r,0)$.

If $q<0$, consider the mirror images $\bar{T}_1,\dotsc,\bar{T}_n$ of our holes across a
vertical lattice line, and define 
$\hat{\omega}(T_1,\dotsc,T_n):=\hat{\omega}(\bar{T}_1,\dotsc,\bar{T}_n)$.

\endproclaim

(The fact that $\hat{\omega}$ is well-defined by the above follows from Propositions 3.2 and 4.1.)

Note that $\hat\omega$ is defined via a sequence of toroidal regions of sizes growing to infinity. 
Since, in stark contrast with
the case of the hexagonal regions used to define the correlation $\omega$ of \cite{\sc},
the periodic boundary conditions of such regions do not distort in the limit the local dimer 
statistics, the definition $\hat\omega$ for the
correlation seems much more natural from a physical point of view than the $\omega$ defined 
in \cite{\sc}. The question is: Does the asymptotics of $\hat\omega$ also lead to the 
superposition principle of electrostatics?
The object of this paper is to prove that the answer to this question is affirmative.

To state our results explicitly, pick a reference monomer $\mu$ in each hole $Q$ under 
consideration, and define $Q(a,b)$ to be the translated copy of $Q$ in which the image of $\mu$ has 
coordinates~$(a,b)$.

The main result of this paper is the following.

\proclaim{Theorem 1.1} Let $\Delta_1,\dotsc,\Delta_n$ be lattice-triangular holes of even side-lengths
$($see Figure {\rm 1.1(a)} for an illustration$)$.
Then for any family of $n$ distinct pairs of integer
multiples of~$3$, $(x_i,y_i)$, $i=1,\dotsc,n$, we have 
$$
\align
\hat\omega&(\Delta_1(Rx_1,Ry_1),\dotsc,\Delta_n(Rx_n,Ry_n))=
\left(\frac{\sqrt{3}}{2\pi}\right)^{-\frac12\{ \sum_{i=1}^n|\ch(\Delta_i)| 
                                              - |\sum_{i=1}^n\ch(\Delta_i)| \}}
\prod_{i=1}^n\hat\omega(\Delta_i)
\\
&
\times
\prod_{1\leq i<j\leq n}\de(\Delta_i(Rx_i,Ry_i),\Delta_j(Rx_j,Ry_j))^{\frac12\ch(\Delta_i)\ch(\Delta_j)}
\\
&
\ \ \ \ \ \ \ \ \ \ \ \ \ \ \ \ \ \ \ \ \ \ \ \ \ \ \ \ \ \ \ \ \ \ \ \ \ \ \ \ \ \ \ \ \ \ \ \ 
+O(R^{\frac12\sum_{1\leq i<j\leq n}\ch(\Delta_i)\ch(\Delta_j))-1}),
\endalign
$$
as $R\to\infty$, where $\de$ denotes the Euclidean distance.

\endproclaim

To prove this, it turns out to be convenient to generalize first the above statement by introducing a 
number of new
parameters. These will be essential for the evaluation of a certain determinant which represents a
key step of our calculations.

To state this generalization, let $Q_i$ be an arbitrary union of non-overlapping lattice triangles of 
side two,
for $i=1,\dotsc,n$. Since $Q_i$ is not necessarily connected, we call it
a {\it multihole}. 

Consider the case of multiholes satisfying the following three
conditions: (1) each $Q_i$ is {\it pure}, i.e., is a union of like-oriented lattice triangular holes
of side 2; (2) each $Q_i$ is {\it linear}, i.e., its constituent lattice triangular holes of side 2
have their centers along a straight (not necessarily lattice) line; and (3) the directions of these
pure and linear multiholes are all parallel among themselves (an instance of such multiholes is 
pictured in Figure 1.1(b)). 


\topinsert
\twoline{\mypic{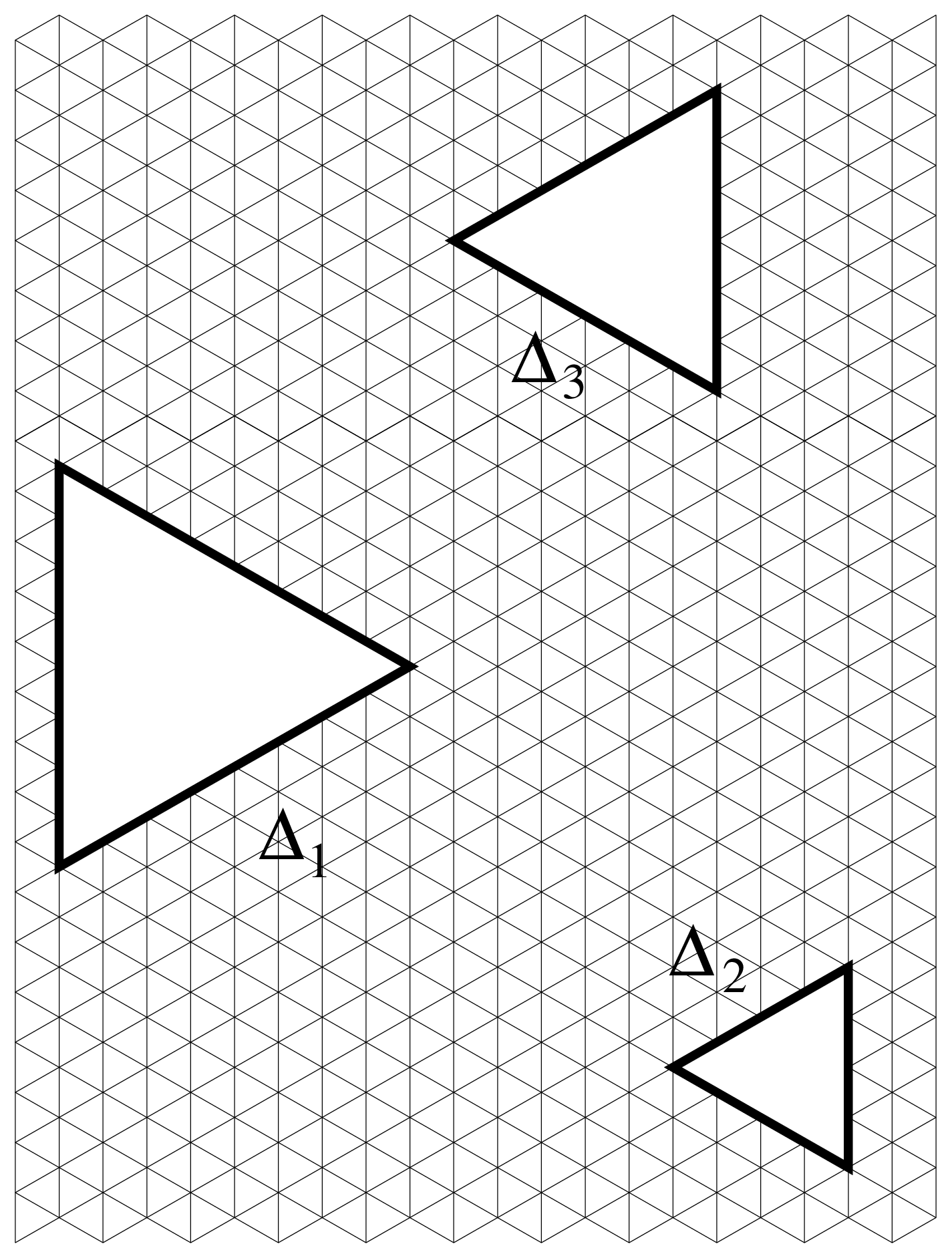}}{\mypic{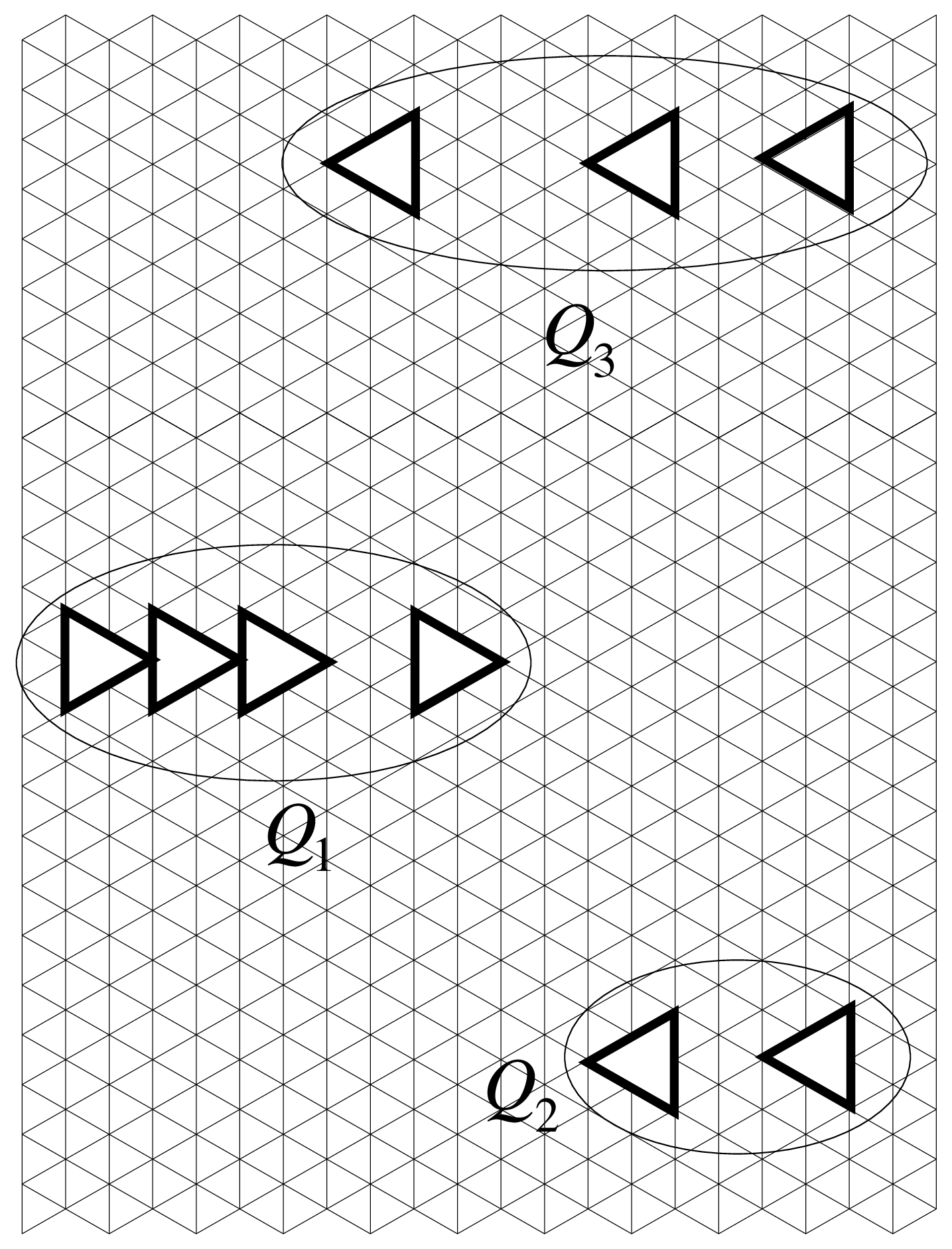}}
\twoline{{\rm (a). \ \ \ \ \  }}{\ \ \ \ \ \ \ \ \ \ {\rm (b).}}
\medskip
\centerline{ {\smc Figure~1.1.} {\rm (a) Triangular holes of even side. 
(b) An example of pure, linear, parallel} }
\centerline{multiholes (ellipses indicate that multiholes keep fixed size and shape as they}
\centerline{move away from one another).} 

\endinsert

We will say that a rational number $r$ is divisible by 3 if $r=a/b$, where $a$ and $b$ are relatively 
prime integers and $3|a$. The generalization of Theorem 1.1 we need is the following. 

\proclaim{Theorem 1.2}
If $L_1,\dotsc,L_n$ are pure, linear and parallel multiholes, and if the slope
$q\in\Q$ of their common direction satisfies $3|1-q$, then for any family of $n$ distinct pairs of integer
multiples of $3$, $(x_i,y_i)$, $i=1,\dotsc,n$, we have 
$$
\align
\hat\omega&(L_1(Rx_1,Ry_1),\dotsc,L_n(Rx_n,Ry_n))=
\left(\frac{\sqrt{3}}{2\pi}\right)^{-\frac12\{ \sum_{i=1}^n|\ch(L_i)| - |\sum_{i=1}^n\ch(L_i)| \}}
\prod_{i=1}^n\hat\omega(L_i)
\\
&
\times
\prod_{1\leq i<j\leq n}\de(L_i(Rx_i,Ry_i),L_j(Rx_j,Ry_j))^{\frac12\ch(L_i)\ch(L_j)}
\\
&
\ \ \ \ \ \ \ \ \ \ \ \ \ \ \ \ \ \ \ \ \ \ \ \ \ \ \ \ \ \ \ \ \ \ \ \ \ \ \ \ \ \ \ \ \ \ \ \ 
+O(R^{\frac12\sum_{1\leq i<j\leq n}\ch(L_i)\ch(L_j))-1}),\tag1.4
\endalign
$$
as $R\to\infty$, where $\de$ denotes the Euclidean distance.

\endproclaim

We note that the assumption $3|1-q$ 
on the slope $q$ of the common direction of the linear multiholes is
not really essential, but makes the calculations in Sections 5--7 simpler (outside this assumption, 
Newton's divided differences need to be adjusted by introducing some multiplicative constants). 
In addition, the set of rationals $q$ with $3|1-q$ is easily seen to be dense in $\Q$. 
Similarly, the divisibility by 3 of the integers $x_i$ and $y_i$ is not really essential, but simplifies
the application of the method of factor exhaustion to the evaluation of the determinant $M''$ given
by (5.11)--(5.15) (see Sections 9, 11 and 12). It clearly does
not impend on the generality of the directions of the rays through the origin that contain $(x_i,y_i)$.

Geometrically, the above result states that if the multiholes $L_1,\dotsc,L_n$ are translated away from 
each other at the same rate along fixed rays through the origin, 
with their shapes and sizes kept intact, then the asymptotics of their joint
correlation $\hat\omega$ is given by an explicit multiplicative superposition principle. This corresponds
to a refinement of the superposition principle of electrostatics: the $R$-dependent part on the right
hand side of (1.4) parallels perfectly its physical counterpart (our statistic $\ch$ corresponding to
electrical charge), but our multiplicative constant is given by the product of the correlations of the
constituents of each multihole (this would correspond to the physical charges having some ``structure'')
times an absolute constant raised to a power that depends just on the $\ch(L_i)$'s. 

Equivalently, we
could hold the location of the multiholes fixed, and refine the triangular lattice instead (more 
precisely, for each $i$, fix the geometrical position of some lattice point $p_i$ in $L_i$ and 
shrink $L_i$ around $p_i$ in step with the refinement of the lattice).
From this
perspective, Theorem~1.2 states that the scaling limit of the joint correlation of
pure, linear and parallel multiholes is given by the described refined superposition principle. As the 
lattice spacing approaches zero, the multiholes, preserving their charge, become more and more point-like.
Nevertheless, the specific way in which each multihole $L_i$ approaches this point-likeness---the 
structure of $L_i$---makes itself felt as a multiplicative contribution of $\hat\omega(L_i)$.


Theorem 1.1 follows easily from Theorem 1.2.

\topinsert
\centerline{\mypic{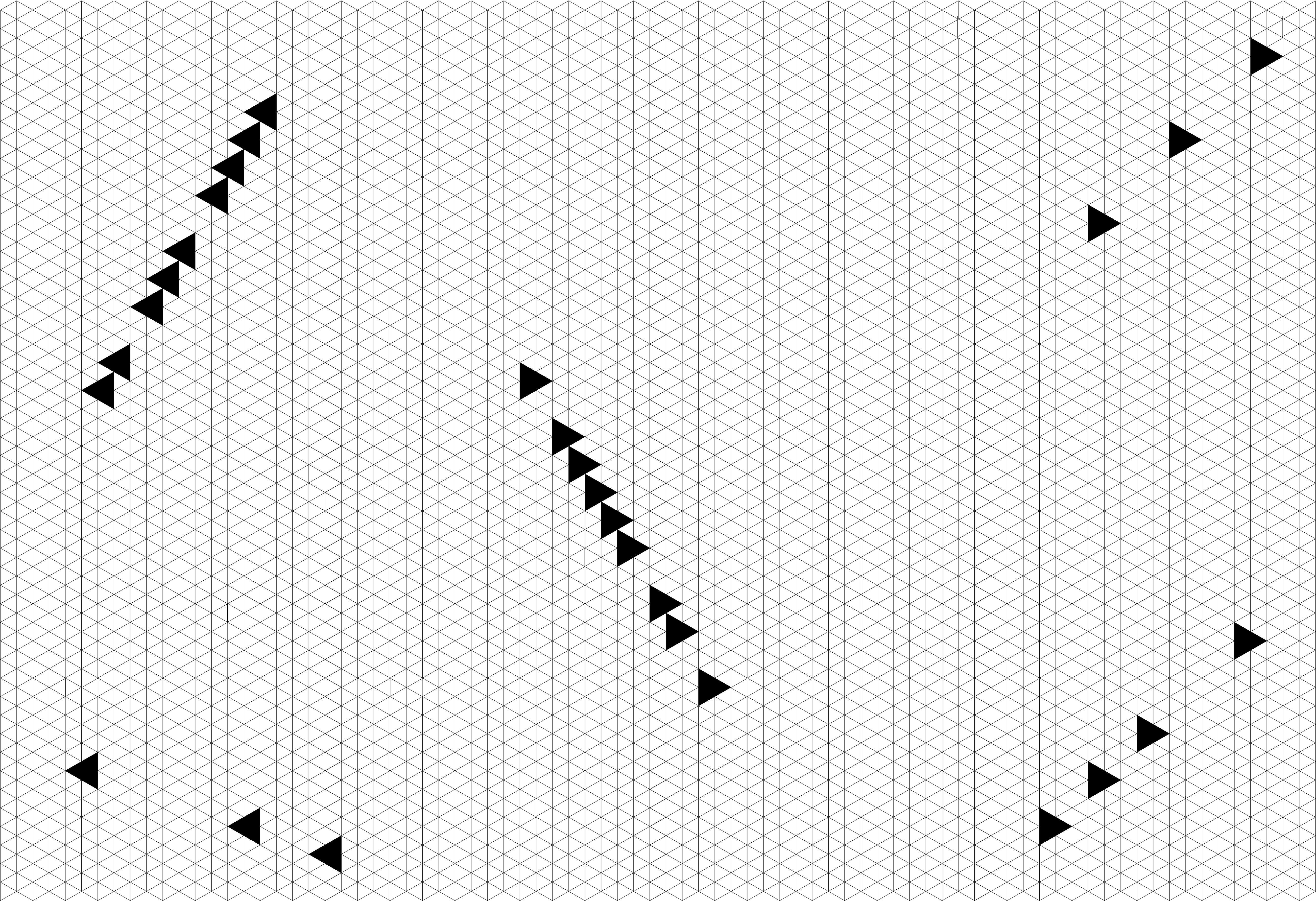}}
\bigskip
\centerline{{\smc Figure~1.2.} {\rm An instance of pure, linear multiholes of arbitrary directions. }}

\endinsert

\smallpagebreak
{\it Proof of Theorem 1.1.} By Lemma 3.1(b), due to forced unit rhombic tiles, a triangular hole of 
side $2s$ is equivalent to 
a horizontal string of $s$ contiguous triangular holes of side 2 (an instance of this is illustrated in 
Figure 1.1(b): the union of the leftmost three holes in $Q_1$ forces a total of 12 unit rhombic 
tiles; their removal creates a triangular hole of side 6). Since in our system of coordinates horizontal
lines have slope $q=1$, the statement follows by Theorem 1.2. \epf


Although the statement of Theorem 1.2 is general enough so that we can prove the determinant evaluation
it boils down to by factor exhaustion (and thus deduce Theorem 1.1), it is worth mentioning that we can
extend it further to the case when the pure, linear multiholes have arbitrary, not necessarily parallel 
directions.

\proclaim{Theorem 1.3} The statement of Theorem 1.2 holds more generally when the slopes $q_i$
of $L_i$ and $q'_j$ of $R_j$ are
arbitrary rational numbers with $3|1-q_i$, $3|1-q'_j$, $i=1,\dotsc,m$, $j=1,\dotsc,n$.

\endproclaim

Such a more general instance is pictured in Figure 1.2.

%

\medskip
\flushpar
{\smc Remark 1.4.} 
%
The 
approach employed in this paper does not allow the presence of holes shaped as 
lattice triangles of odd size. This is due to the fact that a certain advantageous employment (noticed
by Kenyon \cite{\Ktwo}) of a construction of Percus \cite{\Perc} for
obtaining a Pfaffian orientation of a subgraph from a Pfaffian orientation of the full graph, which
allows expressing the ratio between the number of perfect matchings of a subgraph and the number
of perfect matchings of the full graph as a determinant, is not applicable in this situation.
%

\mysec{2. Deducing Theorem 1.2 from Theorem 2.1 and Proposition 2.2}



Denote by $E(x,y)$ the right- (or east-) pointing lattice triangular hole of side 2 whose central 
monomer (a left-monomer) has coordinates $(x,y)$.
For any $q\in\Q$ and any list ${\bold a}=[a_1,\dotsc,a_n]$ of integers for which
$qa_i\in\Z$, and the $E(a_i,qa_i)$'s are mutually disjoint, $i=1,\dotsc,n$, define the multihole 
$E_{\bold a}^q$ by
$$
E_{\bold a}^q:=E(a_1,qa_1)\cup\cdots\cup E(a_n,qa_n).
$$
Similarly, let $W(x,y)$ be the left- (or west-) pointing lattice triangular hole of side 2 whose 
central monomer (a right-monomer) has coordinates $(x,y)$. Given $q\in\Q$ and a list of integers 
${\bold b}=[b_1,\dotsc,b_n]$ so that $qb_i\in\Z$, and the $W(b_i,qb_i)$'s are mutually disjoint, 
$i=1,\dotsc,n$, define the multihole $W_{\bold b}^q$ by
$$
W_{\bold b}^q:=W(b_1,qb_1)\cup\cdots\cup W(b_n,qb_n).
$$

By definition, any pure and linear multihole is of the form $E_{{\bold a}}^q$ or $W_{{\bold b}}^q$.
The parameter $q$ gives the slope of the line containing the centers of the constituent triangular
holes of side 2. Thus, the members of any family of pure, linear and parallel multiholes have 
one of the above two forms, and they all share the same $q$.

\proclaim{Theorem 2.1} Let $m,n\geq0$ be integers, and $q$ a rational number for which $3|1-q$. 
For $i=1,\dotsc,m$, let ${\bold a}_i=[a_{i1}<\cdots<a_{is_i}]$ be a list of $s_i$ integers so that 
$qa_{ij}\in\Z$ and the $E(a_{ij},qa_{ij})$'s are mutually disjoint, $j=1,\dotsc,s_i$. 
For $i=1,\dotsc,n$, let ${\bold b}_i=[b_{i1}<\cdots<b_{it_i}]$ be a list of $t_i$ integers so that 
$qb_{ij}\in\Z$ and the $W(b_{ij},qb_{ij})$'s are mutually disjoint, $j=1,\dotsc,t_i$. 

Then for any family of $m+n$ distinct pairs of integer multiples of $3$, $(x_i,y_i)$, $i=1,\dotsc,m$, 
$(z_j,w_j)$, $j=1,\dotsc,n$, we have, as $R\to\infty$, that
$$
\align
&\hat\omega(E_{{\bold a}_1}^q(Rx_1,Ry_1),\dotsc,E_{{\bold a}_m}^q(Rx_m,Ry_m),
W_{{\bold b}_1}^q(Rz_1,Rw_1),\dotsc,W_{{\bold b}_n}^q(Rz_n,Rw_n))=\\
&
\left(\frac{\sqrt{3}}{2\pi}\right)^{\sum_{i=1}^m s_i+\sum_{i=1}^n t_i+|\sum_{i=1}^m s_i-\sum_{i=1}^n t_i|}
(1+q+q^2)^{\sum_{i=1}^m {s_i \choose 2}+\sum_{i=1}^n {t_i \choose 2}}\\
&
\times
\prod_{i=1}^m \ \prod_{1\leq j<k\leq s_i} (a_{ij}-a_{ik})^2
\prod_{i=1}^n \ \prod_{1\leq j<k\leq t_i} (b_{ij}-b_{ik})^2\\
&
\times
\frac{
\prod_{1\leq i<j\leq m}[(x_i-x_j)^2+(x_i-x_j)(y_i-y_j)+(y_i-y_j)^2]^{s_is_j}
}
{\prod_{i=1}^m \prod_{j=1}^n [(x_i-z_j)^2+(x_i-z_j)(y_i-w_j)+(y_i-w_j)^2]^{s_it_j}}\\
&
\times
\prod_{1\leq i<j\leq n}[(z_i-z_j)^2+(z_i-z_j)(w_i-w_j)+(w_i-w_j)^2]^{t_it_j}\\
&
\times
R^{2\{\sum_{1\leq i<j\leq m}s_is_j+\sum_{1\leq i<j\leq n}t_it_j-\sum_{i=1}^m\sum_{j=1}^ns_it_j \}}\\
&
+O(R^{2\{\sum_{1\leq i<j\leq m}s_is_j+\sum_{1\leq i<j\leq n}t_it_j-\sum_{i=1}^m\sum_{j=1}^ns_it_j\} -1}).
\tag2.1
\endalign
$$

\endproclaim

\proclaim{Proposition 2.2} Let $q\in\Q$ with $3|1-q$, and let ${\bold a}=[a_1,\dotsc,a_s]$ be integers
so that $qa_i\in\Z$ and the $E(a_i,qa_i)$'s are mutually disjoint, $i=1,\dotsc,s$. Then
$$
\hat\omega(E_{\bold a}^q)=\left(\frac{\sqrt{3}}{2\pi}\right)^{2s}(1+q+q^2)^{s \choose 2}
\prod_{1\leq i<j\leq s}(a_i-a_j)^2.\tag2.2
$$

\endproclaim

Theorem 1.2 follows easily from the above two results.

\smallpagebreak
{\it Proof of Theorem 1.2.} Re-denote the integer $n$ in the statement of the theorem by $N$. 
Assume that $m$ of the $N$ given $L_i$'s point east; by assumption, they are of the form 
$E_{{\bold a}_1}^q,\dotsc,E_{{\bold a}_m}^q$, where $q\in\Q$, $3|1-q$, and 
${\bold a}_i=[a_{i1},\dotsc,a_{is_i}]$ are lists of integers. The remaining $n:=N-m$ multiholes 
are then of the form $W_{{\bold b}_1}^q,\dotsc,W_{{\bold b}_n}^q$, where 
${\bold b}_i=[b_{i1},\dotsc,b_{it_i}]$ are lists of integers.

Then the left hand side of (1.4) is clearly the same as the left hand side of (2.1). By Theorem~2.1,
to complete the proof it suffices to show that the right hand side of (2.1) is equal to the right hand
side of (1.4).

Clearly, by our definition of $\hat\omega$ in Section 1, Proposition 2.2 implies that for any $q\in\Q$, 
$3|1-q$, and any list ${\bold b}=[b_1,\dotsc,b_t]$ of integers for which $W_{\bold b}^q$ is defined,
one also has
$$
\hat\omega(W_{\bold b}^q)=\left(\frac{\sqrt{3}}{2\pi}\right)^{2t}(1+q+q^2)^{t \choose 2}
\prod_{1\leq i<j\leq t}(b_i-b_j)^2.\tag2.3
$$
By (2.2) and (2.3), the expression on the right hand side of (2.1) can be rewritten as
$$
\align
&
\left(\frac{\sqrt{3}}{2\pi}\right)^{-\sum_{i=1}^ms_i-\sum_{i=1}^n t_i+|\sum_{i=1}^m s_i-\sum_{i=1}^n t_i|}
\prod_{i=1}^m \hat\omega(E_{{\bold a}_i}^q) \prod_{i=1}^n \hat\omega(W_{{\bold b}_i}^q)\\
&
\times
\frac{
\prod_{1\leq i<j\leq m}[(Rx_i-Rx_j)^2+(Rx_i-Rx_j)(Ry_i-Ry_j)+(Ry_i-Ry_j)^2]^{s_is_j}
}
{\prod_{i=1}^m \prod_{j=1}^n [(Rx_i-Rz_j)^2+(Rx_i-Rz_j)(Ry_i-Rw_j)+(Ry_i-Rw_j)^2]^{s_it_j}}\\
&
\times
\prod_{1\leq i<j\leq n}[(Rz_i-Rz_j)^2+(Rz_i-Rz_j)(Rw_i-Rw_j)+(Rw_i-Rw_j)^2]^{t_it_j}.
\endalign
$$
However, since in our $60^\circ$-angle system of coordinates the Euclidean distance is given by
the formula 
$$
\de((a,b),(c,d))=\sqrt{(a-c)^2+(a-c)(b-d)+(b-d)^2},
$$
the quantities in the square brackets are recognized to be the squares of Euclidean distances between
the reference monomers in the pairs of multiholes 
$$(E_{{\bold a}_i}^q(Rx_i,Ry_i),E_{{\bold a}_j}^q(Rx_j,Ry_j)),$$
$$(W_{{\bold b}_i}^q(Rz_i,Rw_i),(W_{{\bold b}_j}^q(Rz_j,Rw_j)),$$
and 
$$(E_{{\bold a}_i}^q(Rx_i,Ry_i),W_{{\bold b}_j}^q(Rz_j,Rw_j)),$$
respectively.

Furthermore, it is clear
by definition that $\ch(E_{{\bold a}_i}^q)=2s_i$ and $\ch(W_{{\bold b}_j}^q)=-2t_j$, for
$i=1,\dotsc,m$, $j=1,\dotsc,n$. Therefore the expression above is just the right hand side of (1.4)
when the collection of multiholes $\{L_1,\dotsc,L_N\}$ is written out explicitly as
$\{E_{{\bold a}_1}^q,\dotsc,E_{{\bold a}_m}^q,W_{{\bold b}_1}^q,\dotsc,W_{{\bold b}_n}^q\}$.
This completes the proof of Theorem~1.2. \epf

The above arguments reduce the proof of Theorem 1.2 to proving Theorem 2.1 and Proposition~2.2.

\smallpagebreak
{\it Outline of proof of Theorem 2.1.} The arguments in the proof can be grouped in three parts.

The first part, covered by Sections 3 and 4, proceeds by providing an expression for $\hat\omega$ 
(see Proposition 3.2) as
the determinant of a certain matrix $M$ whose entries are of two types: on the one hand, values
of the function $P$ defined by (1.3), and on the other, values of certain two-variable polynomials
$U_s(a,b)$, $s\geq0$, that are the coefficients of the asymptotic series of $P(-R-1+a,-1+b)$, as 
$R\to\infty$. Proposition 4.1 provides an explicit expression for the $U_s(a,b)$'s 
in terms of powers of the finite difference operator. We obtain this expression
by Laplace's method for finding the asymptotic series of a contour integral. We use in our
derivation  Gauss' summation of 
a ${}_2 F_1$ hypergeometric function and Newton's expression for a
polynomial in terms of powers of the difference operator.

To prove the statement of Theorem 2.1 one needs then to determine the large $R$ asymptotics of $\det(M)$.
However, the matrix formed by the leading $R$-parts of the entries of $M$ turns out to be singular. 
The second part of our proof, contained in Sections 5--7, addresses this complication by finding
explicit elementary row and column operations on the matrix $M$ so that the resulting matrix 
$M'$---clearly of the same determinant as $M$---has the property that the matrix formed by the asymptotics
of its entries is non-singular. The row and column operations that achieve this turn out to be governed
by Newton's divided difference operator. It is in this part that we crucially use the pure, linear and
parallel structure of our multiholes.

Finally, in the third part (covered by Sections 8--12) we evaluate the determinant of the matrix $M''$
formed by the asymptotics
of the entries of $M'$ by the method of factor exhaustion. This well-known method of evaluating 
determinants with polynomial entries applies when the value of the determinant is guessed explicitly,
and is a polynomial with all roots and multiplicities known. The method of factor exhaustion proceeds
by showing that the determinant admits all these guessed roots with the guessed multiplicities, by 
finding a suitable number of independent row or column combinations that vanish when the variable 
equals the root in question. This, together with a comparison of the degrees,
implies that the value of the determinant is equal to the guessed
polynomial up to a multiplicative constant. An additional argument is then needed to establish this
multiplicative constant, and thus the value of the determinant.

The guessed value of $\det(M'')$ follows from the expression on the right hand side of (2.1). 
The entries of $M''$ turn out to be rational functions in the variables $q$ and $x_i$, $y_i$, 
$z_j$, $w_j$, $i=1,\dotsc,m$, $j=1,\dotsc,n$. 
In Section 9 we prove that, regarded as a polynomial in $q$, $\det(M'')$ is divisible by the factors
involving $q$ in its guessed expression. Comparing degrees in $q$ we deduce that it suffices to prove
the determinant evaluation for the matrix $M_0$ obtained from $M''$ by specializing $q=0$. 
To this end, we apply the method of factor exhaustion to $dM_0$, where $d$ is the least common multiple of
the denominators of the entries of $M_0$.

We provide vanishing row and column combinations that prove the divisibility of $\det dM_0$ by the 
factors of its guessed expression 
in 
Sections 11 and 12. The fact that the specified row and column combinations are vanishing is deduced from 
known exact evaluations of ${}_2 F_1$ and ${}_3 F_2$ hypergeometric functions. This allows us to deduce
in Section 10 the validity of the determinant evaluation up to a multiplicative constant.
Finally, in the Appendix we show that the multiplicative constant is the one dictated by 
the right hand side of (2.1).

As described above, the obtained evaluation of $\det(M'')$ determines the asymptotics of 
$\det(M)$, and thus that of $\hat\omega$. This leads to the expression on the right hand side of (2.1), 
and completes the proof of Theorem 2.1. The details are given in Section 13.\epf

{\it Outline of proof of Proposition 2.2.} We use the approach described in the first two parts of
the above outline of the proof of Theorem 2.1, applying it to the single multihole $E_{\bold a}^q$. 
The corresponding matrix $M''$ turns out to be an upper-triangular block-matrix consisting of $2\times 2$
blocks. The determinants of the diagonal blocks are readily calculated, and their product is checked to
agree with the expression (1.6). The detailed proof is presented in Section 13. \epf

The necessary changes in the proof of Theorem 1.2 that afford the more general Theorem 1.3 are presented in
Section 14.

\medskip
\mysec{3. A determinant formula for $\hat\omega$}

\medskip
Recall the toroidal regions $H_{N,N}$ considered at the beginning of Section 1. 
We say that a collection $\Cal U=\{u_1,\dotsc,u_n\}$ of holes on the triangular lattice {\it forces} 
a unit rhombic tile $t$ on the lattice if $t$ is present in all tilings of 
$H_{N,N}\setminus\{u_1,\dotsc,u_n\}$, for all $N$ large enough that $H_{N,N}$ contains $\Cal U$.
In proving our determinant formula for $\hat\omega$ we make use of the following simple observation. 
This is also needed to deduce Theorem 1.1 from Theorem 1.2.

\proclaim{Lemma 3.1} $(${\rm a}$)$. Let $\Cal U=\{u_1,\dotsc,u_n\}$ be a collection of unit holes having 
total charge zero, and let $t_1,\dotsc,t_k$ be unit rhombic tiles that are forced by $\Cal U$. Then
$$
\omega_1(u_1,\dotsc,u_n,t_1,\dotsc,t_k)=\omega_1(u_1,\dotsc,u_n).
$$

$(${\rm b}$)$. If the collection $\{T_1,\dotsc,T_n\}$ of non-overlapping lattice triangular holes of 
side $2$ forces the unit rhombic tiles $t_1,\dotsc,t_k$, then
$$
\hat\omega(T_1,\dotsc,T_n,t_1,\dotsc,t_k)=\hat\omega(T_1,\dotsc,T_n).
$$

\endproclaim

\pf Both statements follow directly from the definitions of $\omega_1$ and $\hat\omega$ in 
Section 1.~\epf


Consider the function $P$ defined by (1.3). For $a,b\in\Z$ and $s\geq0$, define $U_s(a,b)$ by
writing the asymptotic series of $P(-3r-1+a,-1+b)$, $r\to\infty$, as
$$
P(-3r-1+a,-1+b)\sim\sum_{s=0}^\infty (3r)^{-s-1}U_s(a,b).
$$
By the definition of asymptotic series (see e.g., \cite{\Ol,p.16}), this means
$$
P(-3r-1+a,-1+b)-\sum_{s=0}^{n-1} (3r)^{-s-1}U_s(a,b)=O(r^{-n}),\tag3.1
$$
for each fixed value of $n$. The fact that $P(-3r-1+a,-1+b)$ admits an asymptotic series follows from
Proposition 4.1, which also gives an explicit formula for the $U_s(a,b)$'s.

For notational brevity, it will be convenient to define the following families of $2\times2$
matrices:
$$
\spreadmatrixlines{2\jot}
A(x,y):=
\left[\matrix P(x-1,y-1)&P(x-2,y)\\
                        P(x,y-2)&P(x-1,y-1)
\endmatrix\right],\tag3.2
$$
and, for $s\geq0$,
$$
\spreadmatrixlines{2\jot}
B_s(x,y):=
\left[\matrix U_s(x,y)&U_s(x-1,y+1)\\
                        U_s(x+1,y-1)&U_s(x,y)
\endmatrix\right].\tag3.3$$

The following result extends the determinant expression (1.2) for $\omega_1$ to $\hat\omega$. 
In particular, it provides a determinant formula for the exact correlation of an arbitrary collection
of non-overlapping triangular holes of side two ((1.2) addressed this only when there were the same
number of holes of each orientation in the collection).

As in the previous section, let $E(x,y)$ and $W(x,y)$ denote the lattice triangular holes of side~2
that point east and west, respectively, and have their central monomer of coordinates $(x,y)$.

\medskip
\proclaim{Proposition 3.2} For $m\geq n$ we have
$$
\hat\omega(E(a_1,b_1),\dotsc,E(a_m,b_m),W(c_1,d_1),\dotsc,W(c_n,d_n))=
\left|\det M \right|,
$$
where $M$ is the $m\times m$ matrix
$$
\spreadmatrixlines{4\jot}
\align
\left[\matrix 
{\scriptstyle A(a_1-c_1,b_1-d_1)}&\!\!\!\!{\scriptstyle A(a_1-c_2,b_1-d_2)}&\!\!\!\!
{\scriptstyle \cdots}&\!\!\!\!{\scriptstyle A(a_1-c_n,b_1-d_n)}&\!\!\!\!
                  {\scriptstyle B_0(a_1,b_1)}&\!\!\!\!{\scriptstyle B_1(a_1,b_1)}&\!\!\!\!
{\scriptstyle \cdots}&\!\!\!\!{\scriptstyle B_{m-n-1}(a_1,b_1)}
\\
{\scriptstyle A(a_2-c_1,b_2-d_1)}&\!\!\!\!{\scriptstyle A(a_2-c_2,b_2-d_2)}&\!\!\!\!
{\scriptstyle \cdots}&\!\!\!\!{\scriptstyle A(a_2-c_n,b_2-d_n)}&\!\!\!\!
                  {\scriptstyle B_0(a_2,b_2)}&\!\!\!\!{\scriptstyle B_1(a_2,b_2)}&\!\!\!\!
{\scriptstyle \cdots}&\!\!\!\!{\scriptstyle B_{m-n-1}(a_2,b_2)}
\\
{\scriptstyle \cdot}&\!\!\!\!{\scriptstyle \cdot}&\!\!\!\!
{\scriptstyle \cdots}&\!\!\!\!{\scriptstyle \cdot}&\!\!\!\!
{\scriptstyle \cdot}&\!\!\!\!{\scriptstyle \cdot}&\!\!\!\!
{\scriptstyle \cdots}&\!\!\!\!{\scriptstyle \cdot}
\\
{\scriptstyle \cdot}&\!\!\!\!{\scriptstyle \cdot}&\!\!\!\!
{\scriptstyle \cdots}&\!\!\!\!{\scriptstyle \cdot}&\!\!\!\!
{\scriptstyle \cdot}&\!\!\!\!{\scriptstyle \cdot}&\!\!\!\!
{\scriptstyle \cdots}&\!\!\!\!{\scriptstyle \cdot}
\\
{\scriptstyle \cdot}&\!\!\!\!{\scriptstyle \cdot}&\!\!\!\!
{\scriptstyle \cdots}&\!\!\!\!{\scriptstyle \cdot}&\!\!\!\!
{\scriptstyle \cdot}&\!\!\!\!{\scriptstyle \cdot}&\!\!\!\!
{\scriptstyle \cdots}&\!\!\!\!{\scriptstyle \cdot}
\\
{\scriptstyle A(a_m-c_1,b_m-d_1)}&\!\!\!\!{\scriptstyle A(a_m-c_2,b_m-d_2)}&
\!\!\!\!{\scriptstyle \cdots}&\!\!\!\!{\scriptstyle A(a_m-c_n,b_m-d_n)}&\!\!\!\!
                  {\scriptstyle B_0(a_m,b_m)}&\!\!\!\!{\scriptstyle B_1(a_m,b_m)}&
\!\!\!\!{\scriptstyle \cdots}&\!\!\!\!{\scriptstyle B_{m-n-1}(a_m,b_m)}
\endmatrix\right]
\\
\tag3.4
\endalign
$$

\endproclaim

\pf We proceed by induction on $m-n$. Let $m-n=0$. By definition, when the total charge of the holes
is zero, $\hat\omega$ is the same as $\omega_1$. Therefore, by
Lemma 3.1(a), the value of $\hat\omega$ on the left hand side above 
remains unchanged if we replace each $E(a_i,b_i)$ by the pair of right-monomers of coordinates
$(a_i-1,b_i)$ and $(a_i,b_i-1)$, and each $W(c_j,d_j)$ by the pair of left-monomers of coordinates
$(c_j,b_j+1)$ and $(c_j+1,b_j)$. Since in this collection the $4m$ monomers come in adjacent
pairs, Theorem 2.3 of \cite{\Ktwo} is applicable, and, combined with (1.2),
provides an expression for their joint correlation $\hat\omega$ as 
$$
\left|\det\,\left(P(r_i-l_j,r'_i-l'_j)\right)_{1\leq i<j\leq 2m}\right|,
$$
where $(r_i,r'_i)$ and $(l_i,l'_i)$, $i=1,\dotsc,2m$, are the coordinates of the right-, respectively
left-monomers in the collection. It is immediate to check that when listing the coordinates of the
right-monomers in the order $[(a_i-1,b_i),\,(a_i,b_i-1):\,i=1,\dotsc,m]$, and those
of the left-monomers in the order $[(c_i,b_i+1),\,(c_i+1,b_i):\,i=1,\dotsc,m]$, the above matrix
becomes precisely the $m=n$ specialization of matrix (3.4). 

Suppose now the statement holds for $m-n=k$. By the definition of $\hat\omega$ and by the induction
hypothesis we have
$$
\spreadlines{2\jot}
\spreadmatrixlines{4\jot}
\align
&
\hat\omega(E(a_1,b_1),\dotsc,E(a_{n+k+1},b_{n+k+1}),W(c_1,d_1),\dotsc,W(c_n,d_n))\\
&
=\lim_{r\to\infty}(3r)^{2k+2}
\hat\omega(E(a_1,b_1),\dotsc,E(a_{n+k+1},b_{n+k+1}),W(c_1,d_1),\dotsc,W(c_n,d_n),W(3r,0))\\
&
=\lim_{r\to\infty}(3r)^{2k+2}\\
&\!\!\!\!\!\!\!\!
\times
\left|
\det\left[\matrix
{\scriptstyle A(a_1-c_1,b_1-d_1)}&{\scriptstyle \cdots}&{\scriptstyle A(a_1-c_n,b_1-d_n)}
&{\scriptstyle A(a_1-3r,b_1)}&{\scriptstyle B_0(a_1,b_1)}
&{\scriptstyle \cdots}&{\scriptstyle B_{k-1}(a_1,b_1)}
\\
{\scriptstyle A(a_2-c_1,b_2-d_1)}&{\scriptstyle \cdots}&{\scriptstyle A(a_2-c_n,b_2-d_n)}
&{\scriptstyle A(a_2-3r,b_2)}&{\scriptstyle B_0(a_2,b_2)}
&{\scriptstyle \cdots}&{\scriptstyle B_{k-1}(a_2,b_2)}
\\
\cdot&\cdot&\cdot&\cdot&\cdot&\cdot
\\
\cdot&\cdot&\cdot&\cdot&\cdot&\cdot
\\
\cdot&\cdot&\cdot&\cdot&\cdot&\cdot
\\
{\scriptstyle A(a_{N}-c_1,b_{N}-d_1)}&{\scriptstyle \cdots}&{\scriptstyle A(a_{N}-c_n,b_{N}-d_n)}
&{\scriptstyle A(a_{N}-3r,b_{N})}&{\scriptstyle B_0(a_{N},b_{N})}
&{\scriptstyle \cdots}&{\scriptstyle B_{k-1}(a_{N},b_{N})}
\endmatrix\right]
\right|,
\\
\tag3.5
\endalign
$$
where $N=n+k+1$.
By (3.1)--(3.3), one sees that adding suitable multiples of the last $2k$ columns to columns $2n+1$
and $2n+2$, the $2\times2$ block at the intersection of rows $2i-1$, $2i$ and columns $2n+1$, $2n+2$
in the above matrix becomes:
$$
\spreadmatrixlines{2\jot}
\left[\matrix
(3r)^{-k-1}U_k(a_i,b_i)+O(r^{-k-2})&(3r)^{-k-1}U_k(a_i-1,b_i+1)+O(r^{-k-2})\\
(3r)^{-k-1}U_k(a_i+1,b_i-1)+O(r^{-k-2})&(3r)^{-k-1}U_k(a_i,b_i)+O(r^{-k-2})
\endmatrix\right],\tag3.6
$$
for $i=1,\dotsc,N$. The contributions of the terms in the expansion of the determinant in which at 
least one of the $O(r^{-k-2})$ parts is chosen is clearly $O(r^{-2k-3})$. Therefore, omitting the 
$O(r^{-k-2})$ terms in these two columns does not change the limit (3.5). After omitting these terms,
the factors $(3r)^{-k-1}$ can be factored out along these two columns, canceling
the factor $(3r)^{2k+2}$ in the limit (3.5). By (3.3), the $2\times 2$ block resulting this way 
from (3.6) is just $B_k(a_i,b_i)$. Interchange the pairs of columns $(2n+1,2n+2k+1)$ and 
$(2n+2,2n+2k+2)$
of the matrix resulting this way from the matrix in (3.5) to complete the induction step. \epf

%

\bigskip
\mysec{4. An exact formula for $U_s(a,b)$}

\medskip
The goal of this section is to prove the following result.

\proclaim{Proposition 4.1} Let $P$ be given by $(1.3)$, and let $a,b\in\Z$. Then the coefficients
of the asymptotic series 
$$
P(-3r-1+a,-1+b)\sim\sum_{s=0}^\infty (3r)^{-s-1}U_s(a,b),\ \ \ r\to\infty
$$
are given by
$$
U_s(a,b)=-\frac{i}{2\pi}\left.\left[\zeta^{a-b-1}(1-D\zeta^{-1})^{-b}-
         \zeta^{-a+b+1}(1-D\zeta)^{-b}\right](x^s)\right|_{x=a+b-1},\tag4.1
$$
where $D$ is the difference operator $($for a function $f$ defined on $\Z$, $Df$
is defined by $Df(x)=f(x+1)-f(x)$$)$, and $\zeta=e^{2\pi i/3}$.

\endproclaim

Consider the functions
$$
\align
p(t)&=-\ln(-1-t)\tag4.2\\
Q(t)&=t^{-b}(-1-t)^{-a}.\tag4.3
\endalign
$$

In proving Proposition 4.1 we will use some properties of these functions, which are stated 
in the following three lemmas.

\proclaim{Lemma 4.2} The power series expansion of $Q(t)$ around $t=z$ is
$$
Q(t)=z^{-b}(-1-z)^{-a}\sum_{k\geq0}\left[\frac{(-1)^k}{k!\,z^k(1+z)^k}
\sum_{l=0}^k{k \choose l}(a+b+l)_{k-l}(b)_l z^{k-l}\right](t-z)^k.\tag4.4
$$

\endproclaim

\pf By Taylor's formula, it suffices to show that
$$
\left.\frac{d^k}{dt^k} t^{-b}(-1-t)^{-a}\right|_{t=z}=z^{-b}(-1-z)^{-a}\left[\frac{(-1)^k}
{z^k(1+z)^k} \sum_{l=0}^k{k \choose l}(a+b+l)_{k-l}(b)_l z^{k-l}\right].\tag4.5
$$
One readily sees that
$$
\frac{d^l}{dt^l} t^{-b}=(-1)^l(b)_l t^{-b-l}
$$
and
$$
\frac{d^l}{dt^l} (-1-t)^{-a}=(a)_l (-1-t)^{-a-l},
$$
for $l\geq0$. Therefore, by Leibniz's formula for the derivative of a product, we obtain
$$
\align
\left.\frac{d^k}{dt^k} t^{-b}(-1-t)^{-a}\right|_{t=z}
&=\sum_{l=0}^k {k \choose l}(-1)^l (b)_l(a)_{k-l}z^{-b-l}(-1-z)^{-a-(k-l)}\\
&= z^{-b}(-1-z)^{-a}(-1)^k \sum_{l=0}^k {k \choose l}(a)_{k-l}(b)_l(1+z)^{-(k-l)}z^{-l}.
\endalign
$$
Thus, to prove (4.5) it suffices to show that
$$
\sum_{l=0}^k {k \choose l} (a)_{k-l}\,(b)_l\,z^{-l}(1+z)^{-(k-l)}=
(1+z)^{-k}\sum_{l=0}^k {k \choose l} (a+b+l)_{k-l}\,(b)_l\,z^{-l}.\tag4.6
$$
Multiplying by $(1+z)^k$, both sides above become polynomials in $z^{-1}$ of degree
at most $k$. Therefore (4.6) will follow provided we check that the coefficient of $z^{-n}$ is 
the same on both sides, for all $0\leq n\leq k$. This is readily seen to amount to showing that
$$
\sum_{l=n}^k
{k \choose l}{l \choose l-n}(a)_{k-l}\,(b)_l={k \choose n}(a+b+n)_{k-n}\,(b)_n,\tag4.7
$$
for all $0\leq n\leq k$.

The left hand side of (4.7) can be expressed in terms of a hypergeometric function
\footnote{The hypergeometric function of parameters
$a_1,\dotsc,a_p$ and $b_1,\dotsc,b_q$ is defined by
$${}_p F_q\!\left[\matrix a_1,\dotsc,a_p\\ b_1,\dotsc,b_q\endmatrix;
z\right]=\sum _{k=0} ^{\infty}\frac {(a_1)_k\cdots(a_p)_k}
{k!\,(b_1)_k\cdots(b_q)_k} z^k\ ,$$
where $(a)_0:=1$ and $(a)_k:=a(a+1)\cdots (a+k-1)$ for $k\geq1$.}
by writing
$$
\spreadlines{2\jot}
\align
{k \choose l}&=\frac{(-1)^l(-k)_l}{l!}=\frac{(-1)^l(-k)_n}{n!}\frac{(-k+n)_{l-n}}{(n+1)_{l-n}}\\
{l \choose l-n}&=\frac{l!}{n!\,(l-n)!}=\frac{n!\,(n+1)_{l-n}}{n!\,(l-n)!}=\frac{(n+1)_{l-n}}{(l-n)!}
\endalign
$$
$$
\spreadlines{2\jot}
\align
(a)_{k-l}&=(a)_{k-n}\frac{(a)_{k-l}}{(a)_{k-n}}=\frac{(-1)^{l-n}(a)_{k-n}}{(-a-k+n+1)_{l-n}}\\
(b)_l&=(b)_n\,(b+n)_{l-n}.
\endalign
$$
Indeed, the left hand side of (4.7) becomes
$$
\spreadlines{2\jot}
\align
\sum_{l=n}^k {k \choose l}{l \choose l-n}&(a)_{k-l}\,(b)_l\\
&=\frac{ (-1)^n(-k)_n\,(b)_n\,(a)_{k-n} }{ n! }
\sum_{l=n}^k \frac{(-k+n)_{l-n}\,(n+1)_{l-n}\,(b+n)_{l-n}}{(n+1)_{l-n}\,(l-n)!\,(-a-k+n+1)_{l-n}}\\
&=\frac{(-1)^n (-k)_n\,(a)_{k-n}\,(b)_n}{n!}\,
{}_2 F_1\!\!\left[\matrix -k+n,b+n\\ -a-k+n+1\endmatrix;1\right].\tag4.8
\endalign
$$

By Gauss' formula (see e.g. \cite{\Sl, (1.7.6),\,Appendix (III.3)}), for $a,b,c\in\C$ with $\Rep\,(c-a-b)>0$ and $c\neq0,-1,-2,\dotsc,$ 
one has
$$
{}_2 F_1\left[\matrix a,b\\ c\endmatrix;1\right]=
\frac{\Gamma(c)\Gamma(c-a-b)}{\Gamma(c-a)\Gamma(c-b)}.\tag4.9
$$

Let $b$, $k\geq n$ be fixed, and choose $a$ to be a large enough negative integer so that 
$-a-b-n+1>0$ and $-a-k+n+1>0$.
Then by (4.9) we get
$$
\spreadlines{2\jot}
\align
{}_2 F_1\!\!\left[\matrix -k+n,b+n\\ -a-k+n+1\endmatrix;1\right]
&=\frac{\Gamma(-a+1-(k-n))}{\Gamma(-a+1)}\frac{\Gamma(-a-b-k+1+(k-n))}{\Gamma(-a-b-k+1)}\\
&=\frac{1}{(-a+1-k+n)_{k-n}} \,(-a-b-k+1)_{k-n},
\endalign
$$
where at the last equality we used repeatedly that $\Gamma(x+1)=x\Gamma(x)$. Therefore, by (4.8)
equality (4.7) amounts to
$$
\frac{ (-1)^n(-k)_n }{ n! }(b)_n\,(a)_{k-n} \frac{(-a-b-k+1)_{k-n}}{(-a+1-k+n)_{k-n}} 
=
{k \choose n}(a+b+n)_{k-n}\,(b)_n.
$$
Dividing by ${k \choose n}=\frac{ (-1)^n(-k)_n }{ n! }$ and then multiplying by the denominator
leads to having the same factors on both sides. This proves (4.7) for large enough negative integers
$a$. Since both sides of (4.7) are polynomials in $a$, it follows that (4.7) holds for general $a$.
This completes the proof of the Lemma.\epf

\proclaim{Lemma 4.3} For $z\neq-1$ the function $p(t)$ given by $(4.2)$ is analytic in a neighborhood
of $z$, and the reversion of its power series expansion
$$
v:=p(t)-p(z)=p_0(t-z)+p_1(t-z)^2+\cdots
$$
is given by
$$
t-z=(1+z)(e^{-v}-1).\tag4.10
$$
\endproclaim

\pf The first part of the statement is clear. The expression for the reversion of the series
also follows easily by exponentiating $-v=-p(t)+p(z)=\ln(-1-t)-\ln(-1-z)$:
$$
\frac{-1-t}{-1-z}=e^{-v} \Leftrightarrow 
t-z+z+1=(1+z)e^{-v} \Leftrightarrow 
t-z=(1+z)(e^{-v}-1).
$$
\epf

Clearly, for the function $p(t)$ given by (4.2), $p'(t)=-1/(1+t)$, so
$$
\frac{1}{p'(t)}=-(1+z)-(t-z).\tag4.11
$$

\proclaim{Lemma 4.4} Replacing $t-z$ in $(4.4)$ and $(4.11)$ by the expression given by $(4.10)$ 
leads to
$$
\frac{Q(t)}{p'(t)}=z^{-b}(-1-z)^{-a}\sum_{k\geq0}\left[-\frac{1+z}{k!}
\sum_{l=0}^k \left. \frac{(b)_l}{l!}D^l(x^k)\right|_{x=a+b-1} z^{-l}\right] v^k, \tag4.12
$$
where $D$ is the difference operator.

\endproclaim

\pf Denoting by $[v^k] f(v)$ the coefficient of $v^k$ in the series $f(v)$, we need to show that
for all $k\geq0$ one has
$$
\align
&
[v^k]\frac{Q(t)}{p'(t)}=
[v^k]\left\{z^{-b}(-1-z)^{-a}\sum_{s\geq0}\frac{1}{s!}
\left[\sum_{l=0}^s {s \choose l}(a+b+l)_{s-l}\,(b)_{l}\,z^{-l}\right](1-e^{-v})^s\right\}
\\
&
\times
\left\{-(1+z)-(1+z)(e^{-v}-1)\right\}=
-z^{-b}(-1-z)^{-a}\frac{1+z}{k!}\sum_{l=0}^k\left.\frac{(b)_l}{l!}D^l(x^k)\right|_{x=a+b-1} z^{-l}.
\\
\tag4.13
\endalign
$$

Since
$$
[v^k]\,e^{-(j+1)v}=[v^k]\sum_{t\geq0}\frac{(-j-1)^t v^t}{t!}=\frac{(-1)^k}{k!}(j+1)^k,
$$
and thus
$$
[v^k]\,(1-e^{-v})^s e^{-v}=[v^k]\sum_{j=0}^s(-1)^j{s \choose j}e^{-(j+1)v}
=\sum_{j=0}^s(-1)^j{s \choose j}\frac{(-1)^k}{k!}(j+1)^k,
$$
we obtain that the left hand side of (4.13) equals
$$
\spreadlines{2\jot}
\align
&
-z^{-b}(-1-z)^{-a}(1+z)
\sum_{s\geq0}\frac{1}{s!}
\left[\sum_{j=0}^s \frac{(-1)^{j+k}}{k!}{s \choose j}(j+1)^k\right]
\\
&
\ \ \ \ \ \ \ \ \ \ \ \ \ \ \ \ \ \ \ \ \ \ \ \ \ \ \ \ \ \ \ \ \ \ \ \ \ \ \ \ 
\times
\left[\sum_{l=0}^s {s \choose l}(a+b+l)_{s-l}\,(b)_l\,z^{-l}\right].\tag4.14
\endalign
$$

Thus, equality (4.13) is equivalent to
$$
\align
(-1)^k
\sum_{s\geq0}\frac{1}{s!}
\left[\sum_{j=0}^s (-1)^{j} {s \choose j} (j+1)^k\right]
&\left[\sum_{l=0}^s {s \choose l}(a+b+l)_{s-l}\,(b)_l\,z^{-l}\right]
\\
&
=
\sum_{l=0}^k\left.\frac{(b)_l}{l!}D^l(x^k)\right|_{x=a+b-1} z^{-l},
\endalign
$$
which in turn, by extracting the coefficients of $z^{-n}$ and dividing by $(-1)^k(b)_n$, amounts to
$$
\sum_{s\geq0}\frac{1}{s!}
\left[\sum_{j=0}^s (-1)^{j} {s \choose j} (j+1)^k\right]
{s \choose n}(a+b+n)_{s-n}
=
\frac{(-1)^k}{n!}\left.D^n(x^k)\right|_{x=a+b-1},\tag4.15
$$
for all $n\geq0$. Using the formula for the powers of the difference operator (see e.g. 
\cite{\Sta,p.36})
$$
D^s f(u)=\sum_{j=0}^s (-1)^{s-j} {s \choose j} f(u+j),\tag4.16
$$
the sum in the brackets in (4.15) is readily seen to be just $(-1)^sD^s x^k|_{x=1}$. Therefore
(4.15) becomes
$$
\sum_{s\geq n}\frac{(-1)^s}{s!} {s \choose n} (a+b+n)_{s-n}\,D^s x^k|_{x=1}
=
\frac{(-1)^k}{n!}D^n x^k|_{x=a+b-1}.
$$
We show more generally that for any polynomial $f(x)$ one has
$$
\sum_{s\geq n}\frac{(-1)^s}{s!} {s \choose n} (c+n)_{s-n}\,D^s\!f(x)|_{x=1}
=
\frac{1}{n!}D^n\!f(-x)|_{x=c-1}.\tag4.17
$$

By the definition of $D$ it follows that $\left.D[f(-x)]\right|_{x=c}=-D[f(x)]|_{x=-c-1}$.
Repeated application of this implies
$$
D^n[f(-x)]|_{x=c}=(-1)^nD^n[f(x)]|_{x=-c-n}.
$$
Using this, the right hand side of (4.17) becomes
$$
\frac{(-1)^n}{n!}D^n\!f(x)|_{x=1-c-n}.
$$
On the other hand, the left hand side of (4.17) equals
$$
\sum_{s\geq n}(-1)^s\frac{(c+n)_{s-n}}{n!\,(s-n)!}D^s\!f(x)|_{x=1}
=
\frac{(-1)^n}{n!}\sum_{s\geq n} {-c-n \choose s-n}D^s\!f(x)|_{x=1}.
$$
Therefore, (4.17) is equivalent to
$$
\sum_{j\geq0}{-c-n \choose j}D^{n+j}\!f(x)|_{x=1}=D^n\!f(x)|_{x=1-c-n}.\tag4.18
$$
However, this is readily deduced from Newton's formula for a polynomial $f(x)$ in terms of
powers of the difference operator (see e.g. \cite{\Jor,(4),\,p.75}):
$$
f(x)=f(d)+{x-d \choose 1}Df(x)|_{x=d}+{x-d \choose 2}D^2\!f(x)|_{x=d}+\cdots
$$
Indeed, taking $x=1-c-n$, $d=1$ in the above equality, and applying then $D^n$ to both sides,  
we obtain equality (4.18). This completes the proof of (4.12). \epf

We are now ready to present the proof of the main result of this section.

{\it Proof of Proposition 4.1.} By \cite{\Kone,(4.11)}, for $x\leq-1$ the function $P$ 
given by (1.3) can be written as
$$
P(x,y)=-\frac{i}{2\pi}\int_{e^{2\pi i/3}}^{e^{4\pi i/3}} t^{-y-1}(-1-t)^{-x-1}dt,\tag4.19
$$
where the integral is a contour integral along the the counterclockwise oriented arc of the 
unit circle connecting $e^{2\pi i/3}$ to $e^{4\pi i/3}$. 
Therefore, for all $R\geq a$ we have 
$$
\spreadlines{4\jot}
\align
\!\!\!\!\!\!\!\!\!\!
P(-R-1+a,-1+b)&=-\frac{i}{2\pi}\int_{e^{2\pi i/3}}^{e^{4\pi i/3}} t^{-b}(-1-t)^{R-a}dt\\
&=-\frac{i}{2\pi}\int_{e^{2\pi i/3}}^{e^{4\pi i/3}} e^{-R[-\ln(-1-t)]}t^{-b}(-1-t)^{-a}dt\\
&=-\frac{i}{2\pi}\left\{\int_{e^{2\pi i/3}}^{e^{\pi i/3}} e^{-Rp(t)}Q(t)dt-
\int_{e^{4\pi i/3}}^{e^{\pi i/3}} e^{-Rp(t)}Q(t)dt\right\},\tag4.20
\endalign
$$
where $p(t)$ and $Q(t)$ are given by (4.2)--(4.3).
We find the asymptotic series for large $R$ of the two integrals in the curly braces
using Laplace's method for contour
integrals as presented in \cite{\Ol,Theorem~6.1,\,p.125}.

Denote the two integrals by
$$
\spreadlines{3\jot}
\align
I_1(R)&=\int_{e^{2\pi i/3}}^{-1} e^{-Rp(t)}Q(t)dt\\
I_2(R)&=\int_{e^{4\pi i/3}}^{-1} e^{-Rp(t)}Q(t)dt.
\endalign
$$

One readily verifies that $I_1(R)$ satisfies conditions $(i)$--$(v)$ of \cite{\Ol,pp.121--122}. 
Indeed,
$p(t)$ and $Q(t)$ are clearly independent of $R$ and holomorphic in the open disk
$D(e^{2\pi i/3},1)$, which contains the interior of the integration path $\Cal P$; this checks
conditions $(i)$ and $(ii)$. Condition $(iii)$ requires 
$p(t)$ and $Q(t)$ to admit power series expansions around $t=e^{2\pi i/3}$; this is clear from their
definitions. Condition $(iv)$ requires $I_1(R)$ to converge at $-1$ absolutely and uniformly for $R$
large enough; this is readily seen to be the case. Finally, to check condition $(v)$ we need to
show that $\Rep\{p(t)-p(e^{2\pi i/3})\}>0$ for $t\in(e^{2\pi i/3},-1)$. Since 
$\Rep p(t)=\Rep\{-\ln(-1-t)\}=-\ln|-1-t|$, and $|-1-t|\leq1$ throughout $\Cal P$, with equality
only for $t=e^{2\pi i/3}$, the latter requirement is verified as well.

Therefore, \cite{\Ol,Theorem~6.1,\,p.125} applies and the asymptotics series of $I_1(R)$ is given by 
$$
\frac{I_1(R)}{e^{-Rp(\zeta)}}\sim\sum_{s=0}^\infty \Gamma\left(\frac{s+\lambda}{\mu}\right)
\frac{a_s}{R^{(s+\lambda)/\mu}},\tag4.21
$$
where $\zeta=e^{2\pi i/3}$, $\mu$ is the smallest power of $t-z$ with a non-zero coefficient in 
the series expansion of $p(t)-p(\zeta)$, $\lambda-1$ is the smallest power of $t-\zeta$ with 
a non-zero coefficient in the 
expansion of $Q(t)$, and $a_s$ is the coefficient of $v^s$ in the power series expansion (4.12)
of $Q(t)/p'(t)$. Since the value of $p_0$ in Lemma 4.3 is $p'(\zeta)=-1/(1+\zeta)\neq0$, we have 
$\mu=1$. Lemma 4.4 shows that $\lambda=1$, and also provides the explicit form of the $a_s$'s. 
Therefore (4.21) becomes
$$
\frac{I_1(R)}{ e^{-Rp(\zeta)}  }\sim \zeta^{-b}(-1-\zeta)^{-a+1}
\sum_{s=0}^\infty \left[\sum_{l=0}^s \frac{(b)_l}{l!}D^l(x^k)|_{x=a+b-1}\zeta^{-l}\right]
\frac{1}{R^{s+1}}.
\tag4.22
$$
The sum in the brackets is, by the binomial theorem, just $(1-D\zeta^{-1})^{-b}(x^k)|_{x=a+b-1}$.
Since $-1-\zeta=\zeta^{-1}$ and $e^{-Rp(\zeta)}=(-1-\zeta)^R=\zeta^{-R}$, we obtain from (4.22) the
explicit asymptotic series of $I_1(R)$ to be
$$
\frac{I_1(R)}{ \zeta^{-R}  }\sim \zeta^{a-b-1}
\sum_{s=0}^\infty \frac{(1-D\zeta^{-1})^{-b}(x^k)|_{x=a+b-1}}{R^{s+1}}.
\tag4.23
$$

The value of the integral $I_2(R)$ is readily seen to be the complex conjugate of $I_1(R)$. It follows
then from (4.23) that
$$
\frac{I_2(R)}{ \zeta^{R}  }\sim \zeta^{-a+b+1}
\sum_{s=0}^\infty \frac{(1-D\zeta)^{-b}(x^k)|_{x=a+b-1}}{R^{s+1}}.
\tag4.24
$$
Substituting (4.23)--(4.24) with $R$ replaced by $3r$ in (4.20) and using $\zeta^3=1$ we obtain (4.1).
\epf

\bigskip
\medskip
\mysec{5. Asymptotic singularity and Newton's divided difference operator}

\bigskip
\bigskip
\medskip

Direct application of Proposition 3.2 to the collection of $\sum_{i=1}^m s_i + \sum_{j=1}^n t_j$ triangular
holes of side~2 that the multiholes 
$E^q_{{\bold a}_1},\dotsc,E^q_{{\bold a}_m},W^q_{{\bold b}_1},\dotsc,W^q_{{\bold b}_n}$ consist of
leads to the following result.

Let $S=\sum_{i=1}^m s_i$ and $T=\sum_{j=1}^n t_j$.

\bigskip
\bigskip
\flushpar
{\smc Lemma 5.1.} {\it For $S\geq T$,
the correlation on the left hand side of $(2.1)$ can be written as}
$$
\spreadlines{3\jot}
\align
&
\hat\omega(E_{{\bold a}_1}^q(Rx_1,Ry_1),\dotsc,E_{{\bold a}_m}^q(Rx_m,Ry_m),
W_{{\bold b}_1}^q(Rz_1,Rw_1),\dotsc,W_{{\bold b}_n}^q(Rz_n,Rw_n))
\\
&
\ \ \ \ \ \ \ \ \ \ \ \ \ \ \ \ \ \ \ \ \ \ \ \ \ \ \ \ \ \ \ \ 
=|\det(M)|
,
\endalign
$$
{\it where} 
$$
M=\left[\matrix C_1&\cdot&\cdot&\cdot&C_n&D \endmatrix\right]\tag5.1 
$$
{\it is a $2S\times 2S$ matrix whose blocks are given by}

\medskip

\midinsert
\centerline{\mypic{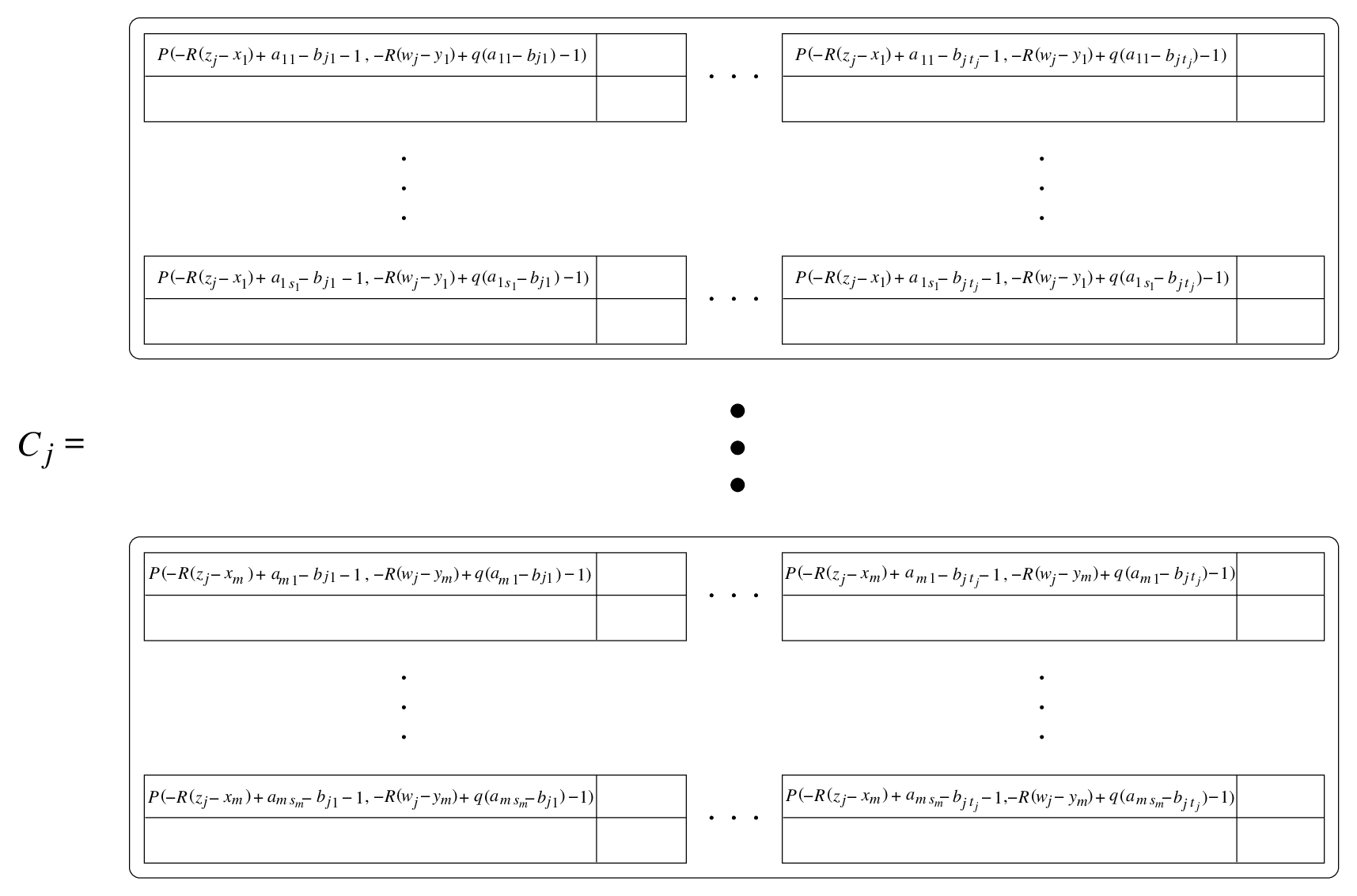}}
{\rm \line{\hfill{(5.2)}}}
\endinsert

\flushpar
{\it for $j=1,\dotsc,n$, and}

\newpage
\midinsert
\centerline{\mypic{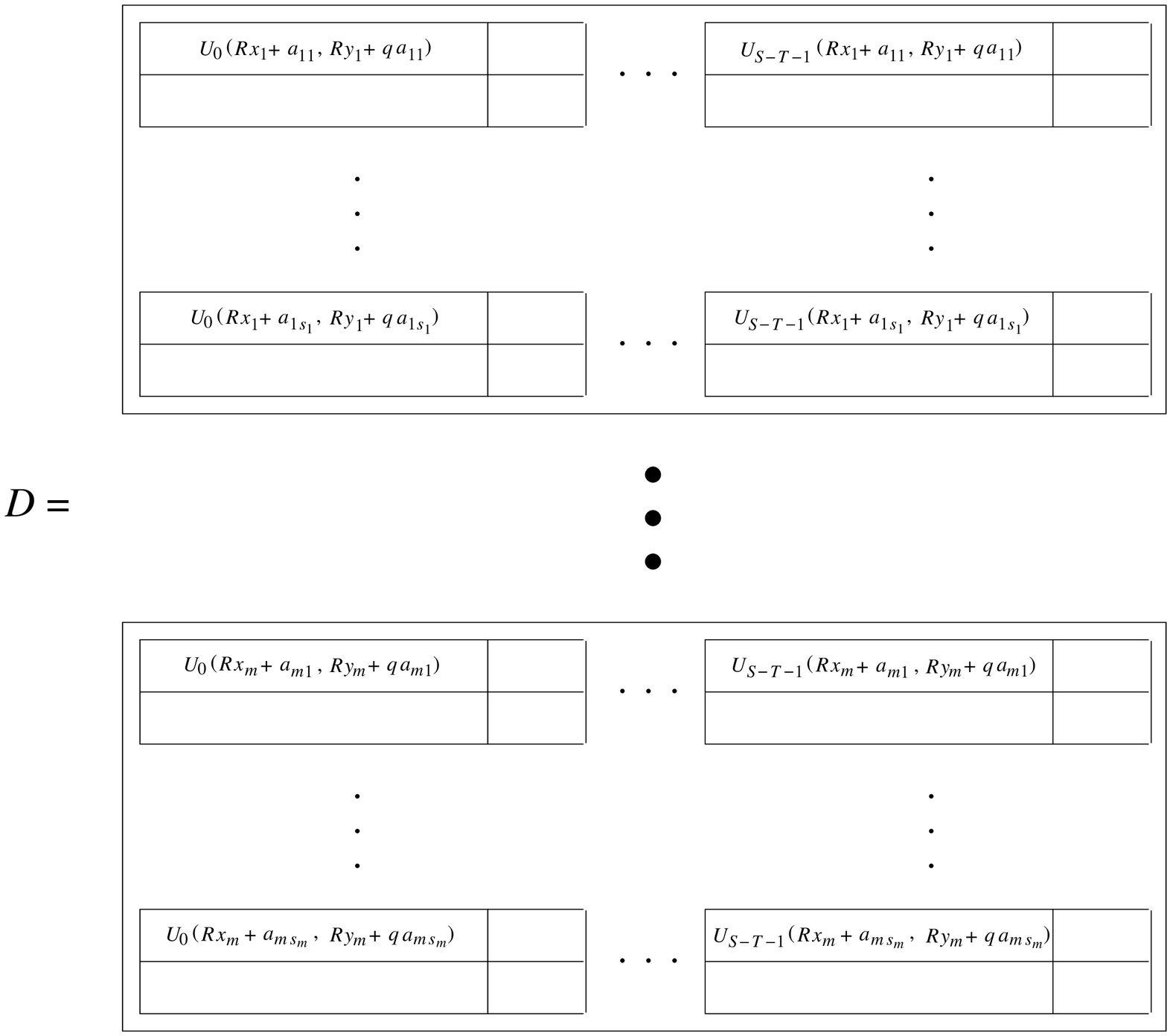}} 
{\rm \line{\hfill{(5.3)}}}
\endinsert

\flushpar
$($$C_j$ {\it has $2S$ rows and $2t_j$ columns, $D$ has $2S$ rows and $2S-2T$ columns}$)$;
{\it here for a function $f$ of two arguments the box}

\medskip

\midinsert
\centerline{\mypic{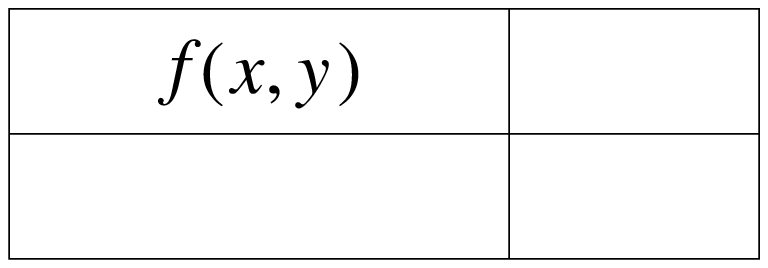}}
\endinsert

\flushpar
{\it denotes the $2\times2$ matrix}
$$
\left[\matrix f(x,y)&f(x-1,y+1)\\
                        f(x+1,y-1)&f(x,y)
\endmatrix\right],
$$
{\it and the large boxes with rounded corners outline $2s_i\times 2t_j$ blocks of $(3.4)$ that 
correspond to a given pair $(E_{{\bold a}_j},W_{{\bold b}_i})$ of multiholes}.


\medskip
Therefore, in order to determine the asymptotics of the correlation of the multiholes, it suffices
to find the asymptotics of $\det(M)$ as $R\to\infty$. 

In working out the asymptotics of the determinant of a matrix whose entries are functions
of a parameter two situations can occur. 
Consider the matrix formed by the asymptotic approximations of the
entries of the original matrix. If this matrix is non-singular, then its determinant gives the
asymptotics of the determinant of the original matrix. On the other hand, the matrix of approximants
may be singular. Then to find the asymptotics of the determinant of the original matrix, one needs
in general to work with higher order terms of the asymptotic series of the entries, and both finding
these higher terms explicitly and evaluating the resulting determinant gets usually very complicated.

It turns out that our matrix $M$ falls into the second category: the matrix of the asymptotic 
approximants (as $R\to\infty$) of its entries is singular unless $s_1=\cdots=s_m=t_1=\cdots=t_n=1$.

Fortunately, however, it also turns out that we can find suitable determinant-preserving 
row and column 
operations on $M$ that transform it into a matrix $M'$ which falls into the first category. The 
row and column operations that achieve this are patterned after the operator involved in the
following result.

\proclaim{Lemma 5.2} Let $X$ be an $n\times n$ matrix, and suppose rows $i_1,\dotsc,i_k$ in it are
the vectors $f(c_1),\dotsc,f(c_k)$, respectively, where $f$ is some vector function. Consider the
operator $\mu$ that acts on rows $i_1,\dotsc,i_k$ of $X$ by transforming them as
$$
\left[\matrix
f(c_1)\\
f(c_2)\\
.\\
.\\
.\\
f(c_k)
\endmatrix\right]
\mapsto
\left[\matrix
{\Cal D}^0f(c_1)\\
(c_2-c_1){\Cal D}^1 f(c_1)\\
.\\
.\\
.\\
(c_k-c_1)(c_k-c_2)\dotsc(c_k-c_{k-1}){\Cal D}^{k-1}f(c_1)
\endmatrix\right],\tag5.4
$$
where $\Cal D$ is Newton's divided difference operator, whose powers are defined inductively
by ${\Cal D}^0 f=f$ and 
${\Cal D}^r f(c_i)=({\Cal D}^{r-1} f(c_{i+1})- {\Cal D}^{r-1} f(c_{i}))/(c_{i+r}-c_i)$.
Then $\det(\mu(X))=\det(X)$.

\endproclaim

\pf By \cite{\Jor,(2),\,p.20}, for any function $f$ and any sequence $c_1,\dotsc,c_l$ one has
$$
\align
f(c_l)=f(c_1)+(c_l-c_1){\Cal D}f(c_1)&+(c_l-c_1)(c_l-c_2){\Cal D}^2 f(c_1)+\cdots\\
&+(c_l-c_1)(c_l-c_2)\cdots(c_l-c_{l-1}){\Cal D}^{l-1} f(c_1).\tag5.5
\endalign
$$

Define a sequence of matrices $X_j$ as follows. Let $X_0=\mu(X)$; let $X_1$ be the matrix obtained 
from $X_0$ by replacing row $i_2$ by the sum of rows $i_2$ and $i_1$; let $X_2$ be the matrix obtained
from $X_1$ by replacing row $i_3$ by the sum of rows $i_3$ and $i_2$; and so on. 
Clearly, rows $i_1,\dotsc,i_k$ of $X_{k-1}$ are then
$R_{i_1},R_{i_1}+R_{i_2},\dotsc,R_{i_1}+\cdots+R_{i_k}$,
respectively, where $R_i$ stands for row $i$ of $\mu(X)$. By (5.5), these are exactly rows
$i_1,\dotsc,i_k$ of $X$. Thus, $X_{k-1}=X$. Since $X_{k-1}$ was obtained by a sequence of elementary
row operations from $\mu(X)$, it follows that $\det(\mu(X))=\det(X_{k-1})=\det(X)$.\epf

We now define the matrix $M'$ that resolves the asymptotic singularity of $M$ described before the 
statement of Lemma 5.2.

By its definition in the statement of Lemma 5.1, it is apparent that for any $i=1,\dotsc,m$, rows
$2\sum_{k=1}^{i-1}s_k+1,2\sum_{k=1}^{i-1}s_k+3,\dotsc,2\sum_{k=1}^{i-1}s_k+2s_i-1$ of $M$ are of 
the form $f(a_{i1}),f(a_{i2}),\dotsc,f(a_{is_i})$, for a suitable vector function $f$. Apply the 
operator (5.4) on these $s_i$ rows of $M$, for 
each $i=1,\dotsc,m$. 

By the same token, rows
$2\sum_{k=1}^{i-1}s_k+2,2\sum_{k=1}^{i-1}s_k+4,\dotsc,2\sum_{k=1}^{i-1}s_k+2s_i$ of $M$ are also of 
the form $f(a_{i1}),f(a_{i2}),\dotsc,f(a_{is_i})$, for a suitable vector function $f$. Apply the 
operator (5.4) on these $s_i$ rows of $M$ as well, for each $i=1,\dotsc,m$. Let $M_1$ be the 
resulting matrix.

In the same fashion, for any fixed $j=1,\dotsc,n$, it is clear from (5.1) that
the columns of $M$ with indices 
$2\sum_{l=1}^{j-1}t_l+1,2\sum_{l=1}^{j-1}t_l+3,\dotsc,2\sum_{l=1}^{j-1}t_l+2t_j-1$, on the one 
hand, and those with indices 
$2\sum_{l=1}^{j-1}t_l+2,2\sum_{l=1}^{j-1}t_l+4,\dotsc,2\sum_{l=1}^{j-1}t_l+2t_j$, on the other,
are each of the form $f(-b_{j1})$, $f(-b_{j2}),\dotsc,f(-b_{jt_j})$, for a suitable vector function 
$f$.
Furthermore, the same is true for the corresponding two $t_j$-tuples of columns in matrix $M_1$. 
We define $M'$ to be the matrix obtained from $M_1$ by applying the operator (5.4)---which clearly
works equally well for columns---along each of these
$t_j$-tuples of columns, for $j=1,\dotsc,n$. Thus $M'$ is obtained from $M$ by applying
operator (5.4) for a total of $2m+2n$ times.

To describe the entries of $M'$ the following terminology will be convenient. We refer to the first $2T$
columns of $M$ as the {\it $P$-part of $M$}; the last $2S-2T$ columns form the {\it $U$-part of $M$}
(see~(5.1)--(5.3)).
The $P$-part consists of an $m\times n$ array of blocks, the block $P_{ij}$ in position $(i,j)$ being 
in turn
a $s_i\times t_j$ array of $2\times 2$ matrices\footnote{The block $P_{ij}$ can be thought of as 
corresponding to the pair $(E^q_{{\bold a}_i},W^q_{{\bold b}_j})$ of multiholes; then its $2\times 2$
block in position $(k,l)$ corresponds to the pair $(E(a_{ik},qa_{ik}),W(b_{jl},qb_{jl}))$ of triangular
holes of side 2.}. 
The $U$-part of $M$ is a column of $m$ blocks,
the block $U_{i}$ in position $i$ being an $s_i \times (S-T)$ array of $2\times 2$ matrices. 
{\it Block-row $i$} consists of $P_{i1},P_{i2},\dotsc,P_{im},U_i$; {\it block-column $j$} consists of
$P_{1j},P_{2j},\dotsc,P_{nj}$ (it is precisely this block-column that is displayed in (5.2)). 
The $2\times 2$ block-matrices naturally group the $2t_j$ columns of block-column $j$ into $t_j$ pairs
of consecutive columns; we call them {\it bi-columns}; bi-column $k$ of a block-column consists of 
columns $2k-1$ and $2k$ of that block-column. {\it Bi-rows} are defined analogously.

Since the block structure of $M_1$ and $M'$ is clearly the same 
as that of $M$, we use the above terminology for them as well.

With this terminology, a concise description of our constructions is that we first obtain $M_1$ 
from $M$
by applying operator (5.4) along the top halves of the bi-rows in block-row $i$, and then along
the bottom halves of the same bi-rows, for all $i=1,\dotsc,m$. $M'$ is obtained from $M_1$
by applying operator (5.4) along the left halves of its bi-columns in block-column $j$, and then along
the right halves of the same bi-columns, for $j=1,\dotsc,n$.

The $2\times 2$ matrix in position $(k,l)$ in the
$(i,j)$-block of the $P$-part of $M$ ($1\leq k\leq s_i$, $1\leq l\leq t_j$) is
$$
\spreadmatrixlines{2\jot}
\left[\matrix
{\scriptscriptstyle{P(-R(z_j-x_i)+a_{ik}-b_{jl}-1,-R(w_j-y_i)+q(a_{ik}-b_{jl})-1)}}&
{\scriptscriptstyle{P(-R(z_j-x_i)+a_{ik}-b_{jl}-2,-R(w_j-y_i)+q(a_{ik}-b_{jl}))}}\\
{\scriptscriptstyle{P(-R(z_j-x_i)+a_{ik}-b_{jl},-R(w_j-y_i)+q(a_{ik}-b_{jl})-2)}}&
{\scriptscriptstyle{P(-R(z_j-x_i)+a_{ik}-b_{jl}-1,-R(w_j-y_i)+q(a_{ik}-b_{jl})-1)}}
\endmatrix\right].
$$

By our construction, the corresponding $2\times 2$ matrix in $M'$ is 
$$
\spreadmatrixlines{2\jot}
\align
&
\alpha_k^{(i)}\beta_l^{(j)}{\Cal D}^{l-1}_b{\Cal D}^{k-1}_a
\\
&
\left[\matrix
{\scriptscriptstyle{P(-R(z_j-x_i)+a_{i1}-b_{j1}-1,-R(w_j-y_i)+q(a_{i1}-b_{j1})-1)}}&
{\scriptscriptstyle{P(-R(z_j-x_i)+a_{i1}-b_{j1}-2,-R(w_j-y_i)+q(a_{i1}-b_{j1}))}}\\
{\scriptscriptstyle{P(-R(z_j-x_i)+a_{i1}-b_{j1},-R(w_j-y_i)+q(a_{i1}-b_{j1})-2)}}&
{\scriptscriptstyle{P(-R(z_j-x_i)+a_{i1}-b_{j1}-1,-R(w_j-y_i)+q(a_{i1}-b_{j1})-1)}}
\endmatrix\right]\!\!,\\
\tag5.6
\endalign
$$
where the powers of ${\Cal D}$ act entry-wise, 
${\Cal D}^{k-1}_a$ acting on the sequence $a_{i1},a_{i2},\dotsc,a_{is_i}$, ${\Cal D}^{l-1}_b$ on 
the sequence $-b_{j1},-b_{j2},\dotsc,-b_{jt_j}$, and 
$$
\align
\alpha_k^{(i)}&=(a_{ik}-a_{i1})(a_{ik}-a_{i2})\cdots(a_{ik}-a_{i,k-1}),\tag5.7\\
\beta_l^{(j)}&=(-b_{jl}+b_{j1})(-b_{jl}+b_{j2})\cdots(-b_{jl}+b_{j,l-1}).\tag5.8
\endalign
$$

On the other hand, the $2\times 2$ submatrix of $M$ at the intersection of rows $2k-1$ and $2k$
of block-row $i$
with columns $2T+2l-1$ and $2T+2l$ ($1\leq k\leq s_i$, $1\leq l\leq S-T$)
is by (5.1)--(5.3)
$$
\spreadmatrixlines{2\jot}
\left[\matrix
U_{l-1}(Rx_i+a_{ik},Ry_i+qa_{ik})&U_{l-1}(Rx_i+a_{ik}-1,Ry_i+qa_{ik}+1)\\
U_{l-1}(Rx_i+a_{ik}+1,Ry_i+qa_{ik}-1)&U_{l-1}(Rx_i+a_{ik},Ry_i+qa_{ik})
\endmatrix\right].
$$
By our construction, the corresponding $2\times 2$ submatrix of $M_1$---and thus of $M'$, as our
applications of the operator (5.4) to $M_1$ do not affect the $U$-part of $M_1$---is
$$
\spreadmatrixlines{2\jot}
\alpha_k^{(i)}{\Cal D}^{k-1}_a
\left[\matrix
U_{l-1}(Rx_i+a_{i1},Ry_i+qa_{i1})&U_{l-1}(Rx_i+a_{i1}-1,Ry_i+qa_{i1}+1)\\
U_{l-1}(Rx_i+a_{i1}+1,Ry_i+qa_{i1}-1)&U_{l-1}(Rx_i+a_{i1},Ry_i+qa_{i1})
\endmatrix\right],\tag5.9
$$
with $\alpha_k^{(i)}$ given by (5.7).

Thus in short, the $2\times 2$ blocks that form $M'$---the matrix of the same determinant as $M$, 
but whose entries' approximants will turn out to form a non-singular matrix---are given by (5.6)--(5.9).

The asymptotics of the entries of (5.6) and (5.9) is worked out in the next two sections. 
The formulas we obtain there lead us to the following result.

\proclaim{Proposition 5.3} The asymptotics of the determinant of the matrix $M$ in the statement
of Lemma 5.1 is given by
$$
\align
\det(M)=(-1)^S\left(\frac{1}{2\pi}\right)^{2S}\prod_{i=1}^m \ &\prod_{1\leq j<k\leq s_i} (a_{ij}-a_{ik})^2
\prod_{i=1}^n \ \prod_{1\leq j<k\leq t_i} (b_{ij}-b_{ik})^2 
\\
&
\times
\det(M'')\,\,
R^{2\{\sum_{1\leq i<j\leq m}s_is_j+\sum_{1\leq i<j\leq n}t_it_j-\sum_{i=1}^m\sum_{j=1}^ns_it_j \}}
\\
&
+O(R^{2\{\sum_{1\leq i<j\leq m}s_is_j+\sum_{1\leq i<j\leq n}t_it_j-\sum_{i=1}^m\sum_{j=1}^ns_it_j\} 
-1}),\tag5.10
\endalign
$$
where 
$$
M''=\left[\matrix C''&D'' \endmatrix\right]\tag5.11 
$$
is the $2S\times 2S$ matrix whose blocks are given by

\midinsert
\centerline{\mypic{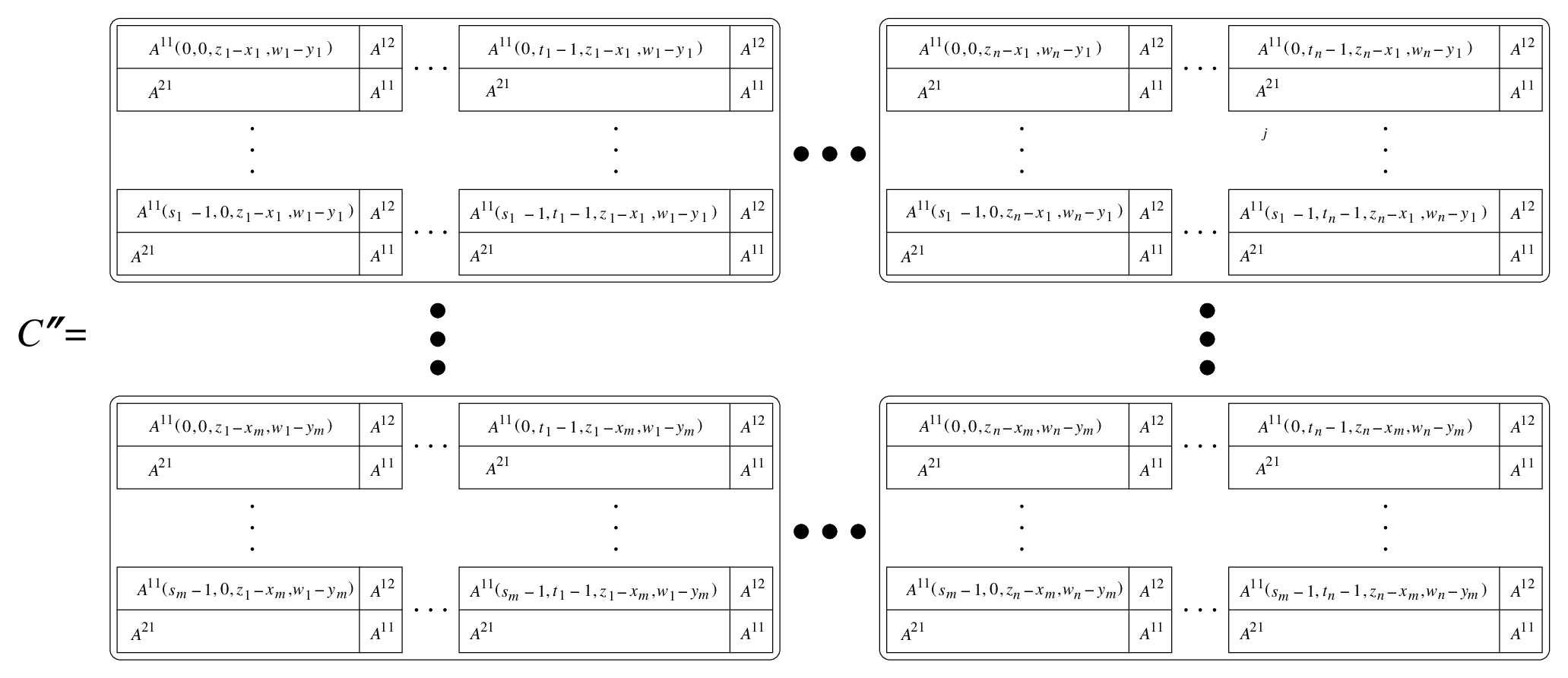}}
\line{\hfill{{\rm (5.12)}}}
\flushpar
{\rm {\it and}}
\endinsert
\midinsert
\centerline{\mypic{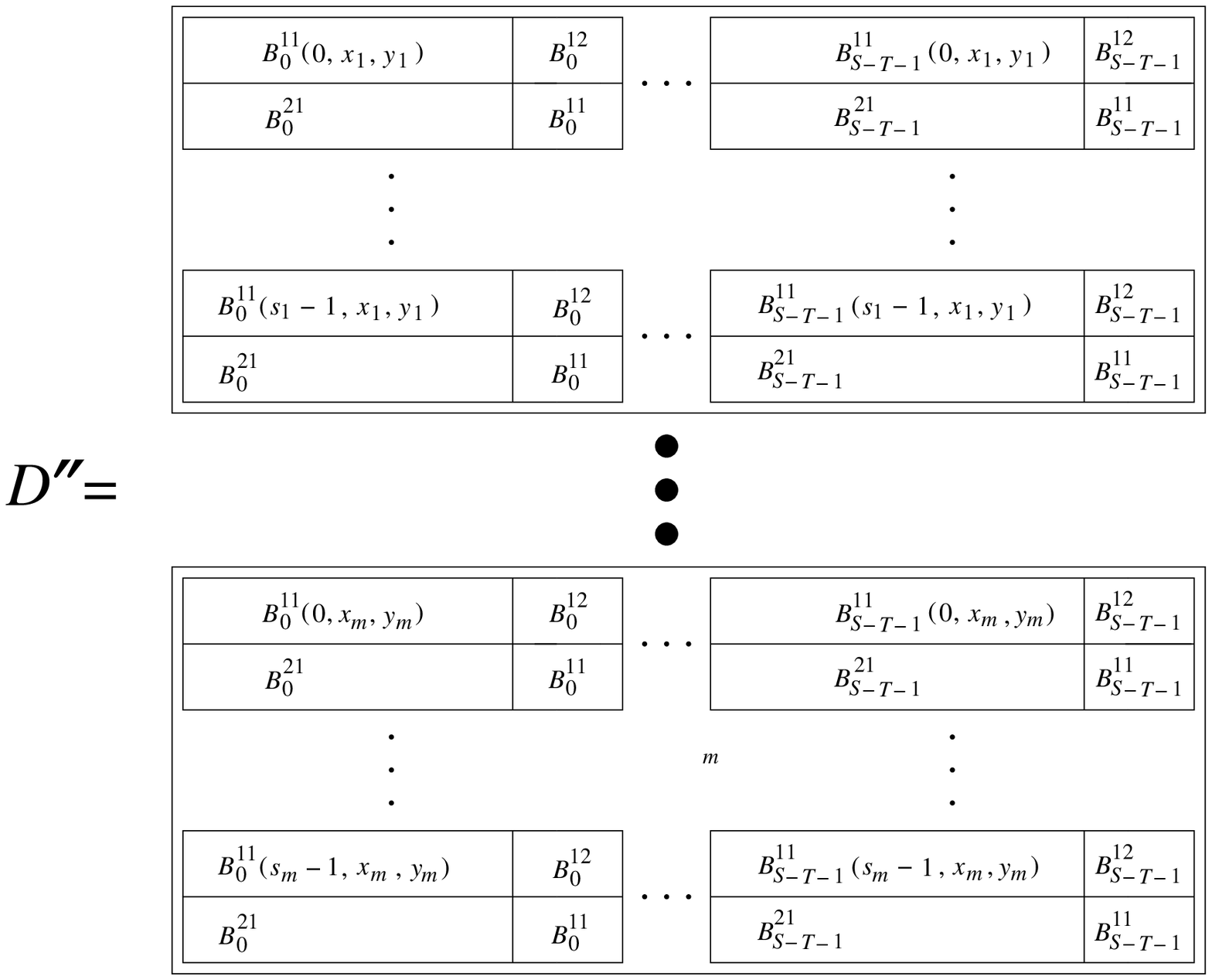}}
\line{\hfill{{\rm (5.13)}}}
\endinsert
\flushpar
with\footnote{
Equality (5.14) defines simultaneously three functions, $A^{11}$, $A^{12}$ and $A^{21}$; for 
instance, to obtain
the expression for $A^{12}$, one chooses the middle quantities in the curly braces on the right
hand side. Similarly in (5.15).}
$$
\align
A^{\left\{\matrix {\scriptstyle{11}}\\{\scriptstyle{12}}\\{\scriptstyle{21}}
\endmatrix\right\}}(k,l,u,v)=
{k+l \choose k}
\left[
\left\{\matrix{\scriptstyle{1}}\\{\scriptstyle{\zeta^{-2}}}\\{\scriptstyle{\zeta^2}}\endmatrix\right\}
\zeta^{-1}\frac{(1-q\zeta)^{k+l}}{(u-v\zeta)^{k+l+1}}
-
\left\{\matrix{\scriptstyle{1}}\\{\scriptstyle{\zeta^2}}\\{\scriptstyle{\zeta^{-2}}}\endmatrix\right\}
\zeta\frac{(1-q\zeta^{-1})^{k+l}}{(u-v\zeta^{-1})^{k+l+1}}
\right],
\\
\tag5.14
\endalign
$$
and
$$
\align
&
B_l^{\left\{\matrix {\scriptstyle{11}}\\{\scriptstyle{12}}\\{\scriptstyle{21}}
\endmatrix\right\}}
(k,u,v)=
{l \choose k}\!\!
\left[
\left\{\matrix{\scriptstyle{1}}\\{\scriptstyle{\zeta^{-2}}}\\{\scriptstyle{\zeta^{2}}}\endmatrix\right\}
\zeta^{-1} (1-q\zeta)^{k} (u-v\zeta)^{l-k}
\right.
\\
&\ \ \ \ \ \ \ \ \ \ \ \ \ \ \ \ \ \ \ \ \ \ \ \ \ \ \ \ \ \ \ \ \ \ \ \ \ \ \ \ \ \ \ \ \ \ \ \ 
\left.
-
\left\{\matrix{\scriptstyle{1}}\\{\scriptstyle{\zeta^{2}}}\\{\scriptstyle{\zeta^{-2}}}\endmatrix\right\}
\zeta (1-q\zeta^{-1})^{k} (u-v\zeta^{-1})^{l-k}
\right]
\\
\tag5.15
\endalign
$$
\flushpar
$($here $\zeta=e^{2\pi i/3}$; in $(5.12)$ and $(5.13)$ the arguments within each 
$2\times 2$ box are the same; for brevity, they are indicated only for the top left entry$)$.

\endproclaim

\pf By Proposition 7.1, as $R\to\infty$, the entries of the $2\times 2$ matrix (5.6) have their 
asymptotics given by
$$
\spreadmatrixlines{1\jot}
-\frac{i}{2\pi}
\frac{\alpha_k^{(i)}\beta_l^{(j)}}{R^{k+l-1}}
\left[\matrix
{\scriptstyle A^{11}(k-1,l-1,z_j-x_i,w_j-y_i)}&\!{\scriptstyle A^{12}(k-1,l-1,z_j-x_i,w_j-y_i)}\\
{\scriptstyle A^{21}(k-1,l-1,z_j-x_i,w_j-y_i)}&\!{\scriptstyle A^{11}(k-1,l-1,z_j-x_i,w_j-y_i)}
\endmatrix\right]
+O(R^{-k-l}).
$$

Therefore, the asymptotic approximations (as $R\to\infty$) of the entries in the $(i,j)$ block of 
the $P$-part of $M'$ form the matrix\footnote{To distinguish it from a variable index, we denote
in this section the complex number $i$ by $\sqrt{-1}$.}

$$
\spreadlines{0\jot}
\spreadmatrixlines{1\jot}
\align
&-\frac{\sqrt{-1}}{2\pi}\times\\
&\!\!\!\!\!\!\!\!\!\!\!\!
\left[\matrix
\frac{\alpha_1^{(i)}\beta_1^{(j)}A^{11}(0,0,z_j-x_i,w_j-y_i)}{R}&\!\!\!\!
\frac{\alpha_1^{(i)}\beta_1^{(j)}A^{12}}{R}&
.&.&.&
\frac{\alpha_{1}^{(i)}\beta_{t_j}^{(j)}A^{11}(0,t_j-1,z_i-x_j,w_i-y_j)}{R^{t_j}}&\!\!\!\!
\frac{\alpha_{1}^{(i)}\beta_{t_j}^{(j)}A^{12}}{R^{t_j}}
\\
\frac{\alpha_1^{(i)}\beta_{1}^{(j)}A^{11}}{R}&\!\!\!\!
\frac{\alpha_1^{(i)}\beta_{1}^{(j)}A^{12}}{R}&
.&.&.&
\frac{\alpha_{1}^{(i)}\beta_{t_j}^{(j)}A^{11}}{R^{t_j}}&\!\!\!\!
\frac{\alpha_{1}^{(i)}\beta_{t_j}^{(j)}A^{12}}{R^{t_j}}
\\
\\
\frac{\alpha_2^{(i)}\beta_1^{(j)}A^{11}(1,0,z_i-x_j,w_i-y_j)}{R^2}&\!\!\!\!
\frac{\alpha_2^{(i)}\beta_1^{(j)}A^{12}}{R^2}&
.&.&.&
\frac{\alpha_2^{(i)}\beta_{t_j}^{(j)}A^{11}(1,t_j-1,z_i-x_j,w_i-y_j)}{R^{t_j+1}}&\!\!\!\!
\frac{\alpha_2^{(i)}\beta_{t_j}^{(j)}A^{12}}{R^{t_j+1}}
\\
\frac{\alpha_2^{(i)}\beta_1^{(j)}A^{11}}{R^2}&\!\!\!\!
\frac{\alpha_2^{(i)}\beta_1^{(j)}A^{12}}{R^2}&
.&.&.&
\frac{\alpha_2^{(i)}\beta_{t_j}^{(j)}A^{11}}{R^{t_j+1}}&\!\!\!\!
\frac{\alpha_2^{(i)}\beta_{t_j}^{(j)}A^{12}}{R^{t_j+1}}
\\
\\
.&.&.&.&.&.&.\\
.&.&.&.&.&.&.\\
.&.&.&.&.&.&.
\\
\\
\frac{\alpha_{s_i}^{(i)}\beta_{1}^{(j)}A^{11}(s_i-1,0,z_i-x_j,w_i-y_j)}{R^{s_i}}&\!\!\!\!
\frac{\alpha_{s_i}^{(i)}\beta_{1}^{(j)}A^{12}}{R^{s_i}}&
.&.&.&
\frac{\alpha_{s_i}^{(i)}\beta_{t_j}^{(j)}A^{11}(s_i-1,t_j-1,z_i-x_j,w_i-y_j)}{R^{s_i+t_j-1}}&\!\!\!\!
\frac{\alpha_{s_i}^{(i)}\beta_{t_j}^{(j)}A^{12}}{R^{s_i+t_j-1}}
\\
\frac{\alpha_{s_i}^{(i)}\beta_{1}^{(j)}A^{11}}{R^{s_i}}&\!\!\!\!
\frac{\alpha_{s_i}^{(i)}\beta_{1}^{(j)}A^{12}}{R^{s_i}}&
.&.&.&
\frac{\alpha_{s_i}^{(i)}\beta_{t_j}^{(j)}A^{11}}{R^{s_i+t_j-1}}&\!\!\!\!
\frac{\alpha_{s_i}^{(i)}\beta_{t_j}^{(j)}A^{12}}{R^{s_i+t_j-1}}
\endmatrix\right]\!,
\\
\endalign
$$
{\rm \line{\hfill{(5.16)}}}

\medskip
\flushpar
and the difference between each entry of $M'$ and its approximant shown in (5.16) is $O(R^{h-1})$,
if the order of the approximant is $R^h$ (as in (5.12), the arguments of $A^{11}$, $A^{12}$ and
$A^{21}$ are the same within each $2\times 2$ block in (5.16); to economize space, they are
indicated only for the top left entry in each such block). 

On the other hand, by Proposition 6.1, matrix (5.9) can be written as
$$
\spreadmatrixlines{1\jot}
-\frac{\sqrt{-1}}{2\pi}
\frac{\alpha_k^{(i)}}{R^{k-l}}
\left[\matrix
B^{11}_{l-1}(k-1,x_i,y_i)&B^{12}_{l-1}(k-1,x_i,y_i)\\
B^{21}_{l-1}(k-1,x_i,y_i)&B^{11}_{l-1}(k-1,x_i,y_i)
\endmatrix\right]
+O(R^{k-l-1}).
$$

Thus, the asymptotic approximations of the entries in the $i$-th block of  
the $U$-part of $M'$ form the matrix

$$
\!\!
\spreadmatrixlines{1\jot}
-\frac{\sqrt{-1}}{2\pi}
\left[\matrix
\frac{\alpha_1^{(i)}B_0^{11}(0,x_i,y_i)}{R^0}&
\frac{\alpha_1^{(i)}B_0^{12}}{R^0}&
.&.&.&
\frac{\alpha_{1}^{(i)}B_{S-T-1}^{11}(0,x_i,y_i)}{R^{1-(S-T)}}&
\frac{\alpha_{1}^{(i)}B_{S-T-1}^{12}}{R^{1-(S-T)}}
\\
\frac{\alpha_1^{(i)}B_0^{21}}{R^0}&
\frac{\alpha_1^{(i)}B_0^{11}}{R^0}&
.&.&.&
\frac{\alpha_{1}^{(i)}B_{S-T-1}^{21}}{R^{1-(S-T)}}&
\frac{\alpha_{1}^{(i)}B_{S-T-1}^{11}}{R^{1-(S-T)}}
\\
\\
\frac{\alpha_2^{(i)}B_0^{11}(1,x_i,y_i)}{R^{1}}&
\frac{\alpha_2^{(i)}B_0^{12}}{R^{1}}&
.&.&.&
\frac{\alpha_{2}^{(i)}B_{S-T-1}^{11}(1,x_i,y_i)}{R^{2-(S-T)}}&
\frac{\alpha_{2}^{(i)}B_{S-T-1}^{12}}{R^{2-(S-T)}}
\\
\frac{\alpha_2^{(i)}B_0^{21}}{R^{1}}&
\frac{\alpha_2^{(i)}B_0^{11}}{R^{1}}&
.&.&.&
\frac{\alpha_{2}^{(i)}B_{S-T-1}^{21}}{R^{2-(S-T)}}&
\frac{\alpha_{2}^{(i)}B_{S-T-1}^{11}}{R^{2-(S-T)}}
\\
\\
.&.&.&.&.&.&.\\
.&.&.&.&.&.&.\\
.&.&.&.&.&.&.
\\
\\
\frac{\alpha_{s_i}^{(i)}B_{0}^{11}(s_i-1,x_i,y_i)}{R^{s_i-1}}&
\frac{\alpha_{s_i}^{(i)}B_{0}^{12}}{R^{s_i-1}}&
.&.&.&
\frac{\alpha_{s_i}^{(i)}B_{S-T-1}^{11}(s_i-1,x_i,y_i)}{R^{s_i-(S-T)}}&
\frac{\alpha_{s_i}^{(i)}B_{S-T-1}^{12}}{R^{s_i-(S-T)}}
\\
\frac{\alpha_{s_i}^{(i)}B_{0}^{21}}{R^{s_i-1}}&
\frac{\alpha_{s_i}^{(i)}B_{0}^{11}}{R^{s_i-1}}&
.&.&.&
\frac{\alpha_{s_i}^{(i)}B_{S-T-1}^{21}}{R^{s_i-(S-T)}}&
\frac{\alpha_{s_i}^{(i)}B_{S-T-1}^{11}}{R^{s_i-(S-T)}}
\endmatrix\right],\tag5.17
$$
with the same conventions about arguments and the same statement about 
the difference between exact values of entries
of $M'$ and their above approximations as in (5.16).

The forms (5.16) and (5.17) of the blocks of $M'$ clearly show that all the factors $-\sqrt{-1}/(2\pi)$
factor out along the rows, for a total contribution of $(1/2\pi)^{2S}$ to $\det(M')$.
It is also apparent from (5.16) and (5.17) that all the $\alpha$'s factor out along the rows of $M'$,
giving, by (5.7), a combined contribution of 
$$
\prod_{i=1}^m \prod_{k=1}^{s_i} \left(\alpha_k^{(i)}\right)^2=
\prod_{i=1}^m \ \prod_{1\leq j<k\leq s_i} (a_{ij}-a_{ik})^2.
$$

In a similar fashion, all the $\beta$'s factor out along the columns of $M'$, and by (5.8) they 
combine to a multiplicative factor of 
$$
\prod_{j=1}^n \prod_{l=1}^{t_j} \left(\beta_l^{(j)}\right)^2=
\prod_{i=1}^n \ \prod_{1\leq j<k\leq t_i} (b_{ij}-b_{ik})^2.
$$

Furthermore, the factors in $R$ can also be factored out along the rows and columns of $M'$.
Indeed, multiply the rows in block-row $i$ by $R^1,R^1,R^2,R^2,\dotsc,R^{s_i},R^{s_i}$,
respectively (from top to bottom), for $i=1,\dotsc,m$. By (5.16)--(5.17), 
this will make the degrees in $R$
of the entries constant along the columns. More precisely, in block-column $j$, the degree in $R$ 
becomes 0 along 
columns 1 and 2, $-1$ along columns 3 and 4, and so on, reaching degree $-(t_i-1)$ along columns 
$2t_j-1$ and $2t_j$ of block-column $j$; and in the $U$-part the degrees in $R$ become $1$ for the 
first two columns, $2$ for the next two, and so on, reaching degree $S-T$ for the last two columns of 
the $U$-part. Factoring out these powers of $R$ along the columns, and taking into account the factors
we multiplied the rows by, one obtains that the overall exponent of $R$ that gets factored out of
the determinant of $M'$ is
$$
-2\sum_{i=1}^m\sum_{k=1}^{s_i}k
-2\sum_{j=1}^n\sum_{l=1}^{t_j}(l-1)
+2\sum_{j=1}^{S-T}j
=
-\sum_{i=1}^m s_i(s_i+1)
-\sum_{j=1}^n t_j(t_j-1)
+(S-T)(S-T-1),
$$
which is readily seen to equal 
$2\sum_{1\leq i<j\leq m}s_is_j+2\sum_{1\leq i<j\leq n}t_it_j-2\sum_{i=1}^m\sum_{j=1}^ns_it_j$.
However, after factoring out this way the $-\sqrt{-1}/(2\pi)$'s, $\alpha$'s, $\beta$'s and the powers of 
$R$ from the blocks (5.16) and (5.17) of $M'$, the resulting matrix is precisely the matrix $M''$ 
given by (5.11)--(5.15). This completes the proof. \epf

\medskip
\mysec{6. The asymptotics of the entries in the $U$-part of $M'$}

\medskip
In this section we determine the asymptotics of the entries of (5.9) as $R\to\infty$. These 
asymptotics follow from the following result. When Newton's divided difference
operator $\Cal D$ acts on a function $f$ of more than one variable,
we write ${\Cal D}_x f$ to indicate that $\Cal D$ acts on $f$ regarded as a function of the variable
$x$.  

\proclaim{Proposition 6.1} Let $q\in\Q$ satisfying $3|1-q$.
Then for any fixed $c,d\in\Z$, $u,v\in3\Z$, and $0\leq k,l\in\Z$, we have
$$
\spreadlines{1\jot}
\align
&
{\Cal D}_x^k U_l(Ru+x+c,Rv+qx+d)|_{x=a_1}=
\\
&\! 
-\frac{i}{2\pi}{l \choose k}\left[\zeta^{c-d-1}(1-q\zeta)^k(u-v\zeta)^{l-k}
-\zeta^{-c+d+1}(1-q\zeta^{-1})^k(u-v\zeta^{-1})^{l-k}\right]R^{l-k}
\\
&
\ \ \ \ \ \ \ \ \ \ \ \ \ \ \ \ \ \ \ \ \ \ \ \ \ \ \ \ \ \ \ \ \ \ \ \ 
+O(R^{l-k-1}),\tag6.1
\endalign
$$
where $\zeta=e^{2\pi i/3}$ and ${\Cal D}_x^k$ acts 
with respect to a fixed integer sequence $a_1,a_2,\dotsc$ with the property 
$qa_j\in\Z$, $j\geq1$. 

\endproclaim

In our proof of the above result we use the following three preliminary lemmas.

\proclaim{Lemma 6.2} For any $a,b\in\Z$ and $0\leq k\in\Z$, we have
$$
(1-D\zeta^{-1})^{-b}(x^k)|_{x=a+b-1}=(a-b\zeta)^k+\ 
{\text{\rm monomials in $a$ and $b$ of joint degree $<k$}},\tag6.2
$$
and 
$$
(1-D\zeta)^{-b}(x^k)|_{x=a+b-1}=(a-b\zeta^{-1})^k+\ 
{\text{\rm monomials in $a$ and $b$ of joint degree $<k$}}.\tag6.3
$$

\endproclaim

\pf By repeated application of the fact that 
$(x+1)^{k-n}-x^{k-n}=(k-n)x^{k-n-1}+O(x^{k-n-2})$ as $x\to\infty$, one
obtains that
$$
D^n x^k=k(k-1)\cdots(k-n+1)x^{k-n}+O(x^{k-n-1}),\ \ \ x\to\infty.\tag6.4
$$
Since
$$
(1-D\zeta^{-1})^{-b}(x^k)=x^k-{-b \choose 1}\zeta^{-1}Dx^k+{-b \choose 2}\zeta^{-2}D^2x^k-\cdots,
$$
we get by (6.4) that
$$
\align
(1-D\zeta)^{-b}(x^k)|_{x=a+b-1}=
&\,(a+b)^k+b\zeta^{-1}k(a+b)^{k-1}+\frac{b^2}{2!}\zeta^{-2}k(k-1)(a+b)^{k-2}\\
+&\frac{b^3}{3!}\zeta^{-3}k(k-1)(k-2)(a+b)^{k-3}+\cdots\\
+&\ {\text{\rm monomials in $a$ and $b$ of joint degree $<k$}}\tag6.5
\endalign
$$
(we used here that the portion of ${-b \choose l}$ of maximal degree is $\frac{b^l}{l!}$). However,
the leading part on the right hand side in (6.5) can be written as
$$
\spreadlines{1\jot}
\align
&(a+b)^k+{k \choose 1}(a+b)^{k-1}b\zeta^{-1}+{k \choose 2}(a+b)^{k-2}(b\zeta^{-1})^2+\cdots\\
=[&(a+b)+b\zeta^{-1}]^k\\
&\!\!\!\!\!\!\!\!\!
=(a-b\zeta)^k,
\endalign
$$
since $1+\zeta^{-1}=-\zeta$. Thus, (6.2) follows from (6.5). Replacing $\zeta^{-1}$ by $\zeta$ in the
above argument proves (6.3). \epf

\proclaim{Lemma 6.3} Let $h_n(x_1,\dotsc,x_s)$ be the complete homogeneous symmetric function of
degree $n$ in the variables $x_1,\dotsc,x_s$ $($$h_n:=0$ for $n<0$, $h_0=1$, 
$h_1=\sum_{k} x_k$, $h_2= \sum_{k} x_k^2 + \sum_{k<l} x_kx_l$, and so on; see e.g. 
{\rm \cite{\Statwo,p.294}}$)$.
Then for any 
$0\leq n,k\in\Z$ we have
$$
{\Cal D}^k x^n|_{x=a_1}=h_{n-k}(a_1,\dotsc,a_{k+1}).\tag6.6
$$

\endproclaim

\pf We proceed by induction on $k$. For $k=0$, ${\Cal D}^0 x^n|_{x=a_1}=a_1^n$, which is clearly
the same as $h_n(a_1)$. The induction step follows by the calculation
$$
\align
{\Cal D}^k x^n|_{x=a_1}&=\frac{{\Cal D}^{k-1} x^n|_{x=a_2}-{\Cal D}^{k-1} x^n|_{x=a_1}}{a_{k+1}-a_1}\\
&=\frac{h_{n-k+1}(a_2,\dotsc,a_{k+1})-h_{n-k+1}(a_1,\dotsc,a_{k})}{a_{k+1}-a_1}\\
&=\sum_{r=0}^{n-k+1}
\frac{a_{k+1}^r h_{n-k+1-r}(a_2,\dotsc,a_k)-a_{1}^r h_{n-k+1-r}(a_2,\dotsc,a_k)}{a_{k+1}-a_1}\\
&=\sum_{r=0}^{n-k} h_{n-k-r}(a_2,\dotsc,a_k)(a_1^r+a_1^{r-1}a_{k+1}+\dotsc+a_{k+1}^r)\\
&=h_{n-k}(a_1,\dotsc,a_{k+1}),
\endalign
$$
where at the second equality we used the induction hypothesis. \epf

\proclaim{Lemma 6.4} For any $0\leq k,l\in\Z$ and $A,B,C\in\C$ we have
$$
{\Cal D}_x^k(Ax+RB+C)^l|_{x=a_1} = 
{l \choose k}A^kB^{l-k}R^{l-k}+O(R^{l-k-1}),\ \ \ R\to\infty.\tag6.7
$$

\endproclaim

\pf Expand the power on the left hand side by the binomial theorem and use (6.6) to obtain
$$
\align
{\Cal D}_x^k(Ax+RB+C)^l|_{x=a_1}&=\sum_{n=0}^l {l \choose n}A^n(RB+C)^{l-n}h_{n-k}(a_1,\dotsc,a_{k+1})\\
&={l \choose k} A^kB^{l-k}R^{l-k}h_0(a_1,\dotsc,a_{k+1})+O(R^{l-k-1}).
\endalign
$$
Since $h_0=1$, this completes the proof. \epf

{\it Proof of Proposition 6.1.} By (4.1) and Lemma 6.2 we obtain
$$
\spreadlines{1\jot}
\align
U_l(a,b)=&-\frac{i}{2\pi}[\zeta^{a-b-1}(a-b\zeta)^l-\zeta^{-a+b+1}(a-b\zeta^{-1})^l]\\
&+\ {\text{\rm monomials in $a$ and $b$ of joint degree $<l$}}.\tag6.8
\endalign
$$
In particular,
$$
\spreadlines{1\jot}
\align
&
U_l(Ru+x+c,Rv+qx+d)\\
&\ \ \ \ \ \ \ \ \ \ \ \ \ \ \ \ \ \ \ \ \ \ \ 
=-\frac{i}{2\pi}\{\zeta^{R(u-v)+x(1-q)+c-d-1}[(1-q\zeta)x+R(u-v\zeta)+c-d\zeta]^l
\\
&\ \ \ \ \ \ \ \ \ \ \ \ \ \ \ \ \ \ \ \ \ \ \ \ \ \  
-\zeta^{-R(u-v)-x(1-q)-c+d+1}[(1-q\zeta^{-1})x+R(u-v\zeta^{-1})+c-d\zeta^{-1}]^l\}
\\
&\ \ \ \ \ \ \ \ \ \ \ \ \ \ \ \ \ \ \ \ \ \ \ \ \ \ 
+\sum_{\alpha,\beta\geq0 \atop \alpha+\beta<l}c_{\alpha,\beta}(Ru+x+c)^{\alpha}(Rv+qx+d)^{\beta},\tag6.9
\endalign
$$
where the $c_{\alpha,\beta}$'s are independent of $R$ and $x$. 

By hypothesis, $R(u-v)+x(1-q)$ is a multiple of 3. Thus, since $\zeta=e^{2\pi i/3}$, the exponents
of $\zeta$ in front of the square brackets in (6.9) simplify to $c-d-1$ and $-c+d+1$, respectively.

By Lemma 6.4, as $R\to\infty$ we have
$$
{\Cal D}_x^k [(1-q\zeta)x+R(u-v\zeta)+c-d\zeta]^l|_{x=a_1}={l \choose k}(1-q\zeta)^k(u-v\zeta)^{l-k}
R^{l-k}+O(R^{l-k-1})
$$
and 
$$
\align
&
{\Cal D}_x^k [(1-q\zeta^{-1})x+R(u-v\zeta^{-1})+c-d\zeta^{-1}]^l|_{x=a_1}=
{l \choose k}(1-q\zeta^{-1})^k(u-v\zeta^{-1})^{l-k}
R^{l-k}
\\
&
\ \ \ \ \ \ \ \ \ \ \ \ \ \ \ \ \ \ \ \ \ \ \ \ \ \ \ \ \ \ \ \ \ \ \ \ \ \ \ \ \ \ \ \ \ \ \ \ \ \ \ \ \ \ \ \ \ \ \ \ \ \ \ \ \ \ \ \ \ \ \ \ \ \ \ \ \ \  
+O(R^{l-k-1}).
\endalign
$$
On the other hand, the result of applying ${\Cal D}_x^k$ to the sum in (6.9) is clearly $O(R^{l-k-1})$.
Applying ${\Cal D}_x^k$ to both sides of (6.9) and using the above observations we obtain (6.1).~\epf

\mysec{7. The asymptotics of the entries in the $P$-part of $M'$}

In this section we determine the asymptotics of the entries of (5.6) as $R\to\infty$. The resulting
formulas are contained in the following result. As in the previous section, a variable at the index 
of Newton's divided difference operator $\Cal D$ indicates that the function upon which it acts is
regarded as a function of that variable.

\proclaim{Proposition 7.1} 
Let $q\in\Q$ satisfying $3|1-q$.
Then for any fixed $c,d\in\Z$, $u,v\in3\Z$, $(u,v)\neq(0,0)$, and $0\leq k,l\in\Z$, we have
$$
\spreadlines{1\jot}
\align
&
{\Cal D}_y^l\left.\left\{{\Cal D}_x^k P(-Ru+x+y+c,-Rv+q(x+y)+d)|_{x=a_1}\right\}\right|_{y=b_1}=
\\
&\! 
-\frac{i}{2\pi}{k+l \choose l}\!\!\left[\!\zeta^{c-d-1}\frac{(1-q\zeta)^{k+l}}{(u-v\zeta)^{k+l+1}}
-\zeta^{-c+d+1}\frac{(1-q\zeta^{-1})^{k+l}}{(u-v\zeta^{-1})^{k+l+1}}\right]\frac{1}{R^{k+l+1}}
+O(R^{-k-l-2}),\tag7.1
\endalign
$$
where $\zeta=e^{2\pi i/3}$, ${\Cal D}_x^k$ acts with respect to a fixed integer sequence 
$a_1,a_2,\dotsc$, ${\Cal D}_y^l$ acts with respect to another integer sequence $b_1,b_2,\dotsc$,
and  $qa_j,qb_j\in\Z$ for all $j\geq1$. 

\endproclaim

In our proof we will employ the following preliminary results.

\proclaim{Lemma 7.2} For $0\leq n,k\in\Z$ and indeterminates $q$, $z$, $c_0,c_1,\dotsc,$ one has
$$
\spreadlines{2\jot}
\spreadmatrixlines{3\jot}
\align
&
\frac{\sum_{j=0}^k\frac{\displaystyle{(-1)^j(qc_0)_j}}{\displaystyle{j!}}
\frac{\displaystyle{(c_0)_{k-j}}}{\displaystyle{(k-j)!}}z^{k-2j}}
{(c_0-c_1)(c_0-c_2)\cdots(c_0-c_n)}+
\frac{\sum_{j=0}^k\frac{\displaystyle{(-1)^j(qc_1)_j}}{\displaystyle{j!}}
\frac{\displaystyle{(c_1)_{k-j}}}{\displaystyle{(k-j)!}}z^{k-2j}}
{(c_1-c_0)(c_1-c_2)\cdots(c_1-c_n)}
+\cdots
\\
&\ \ \ \ \ \ \ \ \ \ \ \ \ \ \ \ 
+\frac{\sum_{j=0}^k\frac{\displaystyle{(-1)^j(qc_n)_j}}{\displaystyle{j!}}
\frac{\displaystyle{(c_n)_{k-j}}}{\displaystyle{(k-j)!}}z^{k-2j}}
{(c_n-c_0)(c_n-c_1)\cdots(c_n-c_{n-1})}=
\left\{\matrix 
0,&{\text {\rm if $k<n$,}}\\
\frac{1}{n!}(z-qz^{-1})^n,&{\text {\rm if $k=n$.}}
\endmatrix\right.
\endalign
$$

\endproclaim

\pf Multiplying by $\prod_{0\leq j<l\leq n}(c_l-c_j)$, the statement becomes equivalent to the 
polynomial identities
$$
\spreadlines{2\jot}
\spreadmatrixlines{3\jot}
\align
&
(-1)^n\prod_{0\leq j<l\leq n \atop j,l\neq 0}(c_l-c_j)\sum_{j=0}^k \frac{(-1)^j(qc_0)_j}{j!}
\frac{(c_0)_{k-j}}{(k-j)!}z^{k-2j}+\cdots
\\
&\ \ \ \ \ \ \ \ \ \ \ \ \ \ \ \ \ \ \ \ 
+(-1)^{0}\prod_{0\leq j<l\leq n \atop j,l\neq n}(c_l-c_j)\sum_{j=0}^k \frac{(-1)^j(qc_n)_j}{j!}
\frac{(c_n)_{k-j}}{(k-j)!}z^{k-2j}
\\
&\ \ \ \ \ \ \ \ \ \ \ \ \ \ \ \ \ \ \ \ 
=
\left\{\matrix 
0,&{\text {\rm if $k<n$,}}
&\ \ \ \ \ \ \ \ \ \ \ \ \ (7.2)\\
\frac{\displaystyle{\prod_{0\leq j<l\leq n}(c_l-c_j)}}{\displaystyle{n!}}
{\displaystyle{\sum_{j=0}^n{n \choose j}(-1)^jq^jz^{n-2j}}},
&{\text {\rm if $k=n$.}}
&\ \ \ \ \ \ \ \ \ \ \ \ \ (7.3)
\endmatrix\right.
\endalign
$$

We prove (7.3) first, so let $k=n$. Agreement of the coefficients of $z^{n-2s}$ on the two sides 
of~(7.3) amounts to
$$
\spreadlines{2\jot}
\align
&
(-1)^n\prod_{0\leq j<l\leq n \atop j,l\neq 0}(c_l-c_j)\frac{(-1)^s(qc_0)_s}{s!}
\frac{(c_0)_{n-s}}{(n-s)!}+\cdots
\\
&
\ \ \ \ \ \ \ \ \ \ \ \ \ \ \ \ \ \ \ \ \ \ \ \ \ \ \ \ \ \ \ \ \ \ \ \ \ \ \ \ \ \ \ \ \ \ \ \ \ \ \ \ 
\ \ \ \ 
+(-1)^{0}\prod_{0\leq j<l\leq n \atop j,l\neq n}(c_l-c_j)\frac{(-1)^s(qc_n)_s}{s!}
\frac{(c_n)_{n-s}}{(n-s)!}
\\
&\ \ \ \ \ \ \ \ \ \ \ \ \ \ \ \ \ \ \ \ 
=
\frac{1}{n!}{n \choose s}(-1)^sq^s\prod_{0\leq j<l\leq n}(c_l-c_j).
\endalign
$$
Simplifying this identity we obtain that (7.3) is equivalent to the set of identities
$$
\align
&
(-1)^n(qc_0)_s (c_0)_{n-s}\prod_{0\leq j<l\leq n \atop j,l\neq 0}(c_l-c_j)+\cdots
+(-1)^{0}(qc_n)_s (c_n)_{n-s}\prod_{0\leq j<l\leq n \atop j,l\neq n}(c_l-c_j)
\\
&
\ \ \ \ \ \ \ \ \ \ \ \ \ \ \ \ \ \ \ \ \ \ \ \ \ \ \ \ \ \ \ \ \ \ \ \ \ \ \ \ 
=
q^s \prod_{0\leq j<l\leq n}(c_l-c_j),\tag7.4
\endalign
$$
for $0\leq s\leq n$.

Let $0\leq m<p\leq n$, and assume $c_m=c_p$. We claim that this causes the left hand side of (7.4)
to be 0. Indeed, let $\Delta(x_0,\dotsc,x_t)=\prod_{0\leq j<l\leq t}(x_l-x_j)$. Then for $c_m=c_p$
the left hand side of (7.4) becomes
$$
\align
&
(-1)^m\Delta(c_0,\dotsc,c_{m-1},c_{m+1},\dotsc,c_p,\dotsc,c_n)(qc_m)_s(c_m)_{n-s}+
\\
&\ \ \ \ \ \ \ \ \ \ \ \ \ \ \ \ \ \ \ \
(-1)^p\Delta(c_0,\dotsc,c_{m},\dotsc,c_{p-1},c_{p+1},\dotsc,c_n)(qc_p)_s(c_p)_{n-s}.
\endalign
$$
However, since $c_m=c_p$, the second list of indeterminates on which $\Delta$ acts above 
is obtained from the first such list by moving $c_p$ across $p-m-1$ neighbors to its left. This
implies that the two $\Delta$-expressions differ by a multiplicative constant of $(-1)^{p-m-1}$,
and thus the above expression equals 0.

Regard the two sides of (7.4) as polynomials is $\Z[q][c_0,\dotsc,c_n]$.
As a consequence of our observation in the previous paragraph, the polynomial on the left is
divisible by the product $\prod_{0\leq j<l\leq n}(c_l-c_j)$. Furthermore, the degree of the polynomial
on the left is ${n \choose 2}+n$, which is the same as the degree of this product. It follows that
the two sides of (7.4) are equal up to a multiplicative factor in $\Z[q]$. 

To finish proving (7.4) it suffices to show that it holds for a specialization of the $c_j$'s. For
$c_j=j$, $j=0,\dotsc,n$, this amounts to
$$
\sum_{j=0}^n\frac{(-1)^{n-j}}{(n-j)!\,j!}(qj)_s\,(j)_{n-s}=q^s,\ \ \ 0\leq s\leq n.\tag7.5
$$

Take $f(x)=x^t$ and $u=0$ in the expression (4.16) for the $n$-th power of the divided difference 
operator $D$. We obtain
$$
\spreadmatrixlines{3\jot}
\sum_{j=0}^n\frac{(-1)^{n-j}}{(n-j)!\,j!}j^t
=
\left\{\matrix 
0,&{\text {\rm if $0\leq t<n$,}}\\
1,&{\text {\rm if $k=n$.}}
\endmatrix\right.\tag7.6
$$
Thus, only the term in $j^n$ in the expansion of
$$
\align
(qj)_s\,(j)_{n-s}&=qj(qj+1)\cdots(qj+s-1)j(j+1)\cdots(j+n-s-1)
\\&=q^sj^n+\ 
{\text{\rm terms in $j$ of degree $<n$}}
\endalign
$$
contributes to the left hand side of (7.5). This proves (7.5), and completes the proof of (7.3).

We now turn to proving (7.2), so assume $k<n$. By extracting the coefficients of $z^{k-2s}$,
(7.2)~becomes equivalent to
$$
(-1)^n(qc_0)_s (c_0)_{k-s}\prod_{0\leq j<l\leq n \atop j,l\neq 0}(c_l-c_j)+\cdots
(-1)^{0}(qc_n)_s (c_n)_{k-s}\prod_{0\leq j<l\leq 0 \atop j,l\neq n}(c_l-c_j)
=0,\tag7.7
$$
for $0\leq s\leq k$.

By the arguments we used in the case $k=n$, the left hand side of (7.7) is seen to be of the form
$\alpha\prod_{0\leq j<l\leq n}(c_l-c_j)$, with $\alpha\in\Z[q]$. To establish (7.7) it suffices to
show $\alpha=0$. The specialization that led to (7.5) shows that in order to deduce the latter it is 
enough to check that
$$
\sum_{j=0}^n\frac{(-1)^{n-j}}{(n-j)!\,j!}(qj)_s\,(j)_{k-s}=0,\ \ \ 0\leq s\leq k<n.
$$
This follows from (7.6) by the same reasoning we used to prove (7.5). \epf

\proclaim{Lemma 7.3} Let $q\in\Q$ with $3|1-q$ and consider a sequence $a_0,a_1,\dotsc$ of integers
so that $qa_j\in\Z$, $j\geq0$. Then for any $n\geq0$, in the power series expansion 
$$
{\Cal D}_x^n t^{-qx}(-1-t)^{-x}|_{x=a_0}=\sum_{k\geq0}q_k(t-\zeta)^k\tag7.8
$$
around $\zeta=e^{2\pi i/3}$, the first non-zero coefficient is 
$q_n=\frac{1}{n!}(\zeta-q\zeta^{-1})^n$.

\endproclaim

\pf By \cite{\Jor,(1),\,p.19}, for any function $f$ we have
$$
\spreadlines{2\jot}
\align
{\Cal D}^n f(a_0)=
\frac{f(a_0)}{(a_0-a_1)(a_0-a_2)\cdots(a_0-a_n)}&+
\frac{f(a_1)}{(a_1-a_0)(a_1-a_2)\cdots(a_1-a_n)}+\cdots\\
&\ \ \ \ \ \ 
+\frac{f(a_n)}{(a_n-a_0)(a_n-a_1)\cdots(a_n-a_{n-1})}.
\\
\tag7.9
\endalign
$$
Take $f(x,t)=t^{-qx}(-1-t)^{-x}$ in the above equality. Then the resulting expression will supply
the power series expansion (7.8), provided we work out the coefficients of the series expansion 
$$
f(a_j,t)=\sum_{k\geq0}\frac{1}{k!}\frac{\partial^k f(a_j,\zeta)}{\partial t^k}(t-\zeta)^k.
$$

Let $g(t)=t^{-qx}$ and $h(t)=(-1-t)^{-x}$. Their derivatives of order $l$ are given by
$$
\align
g^{(l)}(t)&=(-1)^l(qx)_lt^{-qx-l},\\
h^{(l)}(t)&=(x)_l(-1-t)^{-x-l}.
\endalign
$$
Thus, by Leibniz's formula for the derivative of a product,
$$
\align
\frac{\partial^k f(a_j,\zeta)}{\partial t^k}&=\sum_{l=0}^k{k \choose l}(-1)^l(qa_j)_l(a_j)_{k-l}
\zeta^{-qa_j-l}(-1-\zeta)^{-a_j-(k-l)}\\
&=\sum_{l=0}^k{k \choose l}(-1)^l(qa_j)_l(a_j)_{k-l}\zeta^{k-2l},
\endalign
$$
where at the last equality we used $-1-\zeta=\zeta^{-1}$ and $\zeta^{a_j(1-q)}=1$. Substituting this 
in (7.9) we obtain that the coefficient of $(t-\zeta)^k$ in the power series expansion 
of ${\Cal D}_x^n t^{-qx}(-1-t)^{-x}|_{x=a_0}$ around $t=\zeta$ is
$$
\spreadlines{2\jot}
\align
&
\frac{\sum_{j=0}^k\frac{\displaystyle{(-1)^j(qa_0)_j}}{\displaystyle{j!}}
\frac{\displaystyle{(a_0)_{k-j}}}{\displaystyle{(k-j)!}}\zeta^{k-2j}}
{(a_0-a_1)(a_0-a_2)\cdots(a_0-a_n)}
\\
&\ \ \ \ \ \ \ \ \ \ \ \ 
+
\frac{\sum_{j=0}^k\frac{\displaystyle{(-1)^j(qa_1)_j}}{\displaystyle{j!}}
\frac{\displaystyle{(a_1)_{k-j}}}{\displaystyle{(k-j)!}}\zeta^{k-2j}}
{(a_1-a_0)(a_1-a_2)\cdots(a_1-a_n)}
+\cdots
+\frac{\sum_{j=0}^k\frac{\displaystyle{(-1)^j(qa_n)_j}}{\displaystyle{j!}}
\frac{\displaystyle{(a_n)_{k-j}}}{\displaystyle{(k-j)!}}\zeta^{k-2j}}
{(a_n-a_0)(a_n-a_1)\cdots(a_n-a_{n-1})}.
\endalign
$$
Apply Lemma 7.2 to complete the proof. \epf

\proclaim{Lemma 7.4} Let $q\in\Q$ with $3|1-q$, and let $a_1,a_2,\dotsc$ and $b_1,b_2,\dotsc$ be
sequences of integers so that $qa_j,qb_j\in\Z$, $j\geq1$. 
Then for any $k,l\geq0$, in the power series expansion 
$$
{\Cal D}_y^l\left.\left\{{\Cal D}_x^k \,  t^{-q(x+y)}(-1-t)^{-(x+y)}|_{x=a_1}\right\}\right|_{y=b_1}
=\sum_{s\geq0}q_s(t-\zeta)^s
$$
around $\zeta=e^{2\pi i/3}$, the first non-zero coefficient is 
$q_{k+l}=\frac{1}{k!l!}(\zeta-q\zeta^{-1})^{k+l}$.

\endproclaim

\pf We have
$$
\spreadlines{2\jot}
\align
&
{\Cal D}_y^l\left.\left\{{\Cal D}_x^k \, t^{-q(x+y)}(-1-t)^{-(x+y)}|_{x=a_1}\right\}\right|_{y=b_1}
\\
&
\ \ \ \ \ \ \ \ \ \ \ \ \ \ \ \ \ \ \ \ 
=
{\Cal D}_y^l\left.\left\{ t^{-qy}(-1-t)^{-y} {\Cal D}_x^k \, t^{-qx}(-1-t)^{-x}|_{x=a_1}
\right\}\right|_{y=b_1}\\
&
\ \ \ \ \ \ \ \ \ \ \ \ \ \ \ \ \ \ \ \ 
=
{\Cal D}_y^l \, t^{-qy}(-1-t)^{-y}|_{y=b_1}\,
{\Cal D}_x^k \, t^{-qx}(-1-t)^{-x}|_{x=a_1}.
\endalign
$$
The statement follows therefore by Lemma 7.3. \epf

{\it Proof of Proposition 7.1.} Assume first that $u>0$. Then by (4.19) we obtain that
$$
\spreadlines{2\jot}
\align
P(-Ru+x+y+c,&-Rv+q(x+y)+d)\\
&=-\frac{i}{2\pi}\int_{e^{2\pi i/3}}^{e^{4\pi i/3}}
t^{Rv}(-1-t)^{Ru}t^{-q(x+y)-d-1}(-1-t)^{-(x+y)-c-1}dt\\
&=-\frac{i}{2\pi}
\left\{\int_{e^{2\pi i/3}}^{-1}e^{-Rp(t)}Q(t)dt - \int_{e^{4\pi i/3}}^{-1}e^{-Rp(t)}Q(t)dt\right\}
\endalign
$$
for all sufficiently large $R$, where 
$$
\align
p(t)&=-v\ln t -u\ln(-1-t),\tag7.10\\
Q(t)&=t^{-q(x+y)-d-1}(-1-t)^{-(x+y)-c-1}.
\endalign
$$
Applying ${\Cal D}_y^l {\Cal D}_x^k$ to both sides yields
$$
\spreadlines{2\jot}
\align
&
{\Cal D}_y^l\left.\left\{{\Cal D}_x^k P(-Ru+x+y+c,-Rv+q(x+y)+d)|_{x=a_1}\right\}\right|_{y=b_1}\\
&\ \ \ \ \ \ \ \ \ \ \ \ \ \ \ \ \ \ \ \ 
=-\frac{i}{2\pi}
\left\{\int_{e^{2\pi i/3}}^{-1}e^{-Rp(t)}\tilde{Q}(t)dt 
      - \int_{e^{4\pi i/3}}^{-1}e^{-Rp(t)}\tilde{Q}(t)dt\right\},\tag7.11
\endalign
$$
where $p(t)$ is given by (7.10) and
$$
\tilde{Q}(t)=t^{-d-1}(-1-t)^{-c-1}
{\Cal D}_y^l\left.\left\{{\Cal D}_x^k \, t^{-q(x+y)}(-1-t)^{-(x+y)}|_{x=a_1}\right\}\right|_{y=b_1}.
\tag7.12
$$

As in (4.20), the values of the two integrals in (7.11) are complex conjugates. Let
$$
I(R)=\int_{e^{2\pi i/3}}^{-1}e^{-Rp(t)}\tilde{Q}(t)dt.
$$
We find the asymptotics of $I(R)$ by the approach we used in the proof of Proposition 4.1---Laplace's
method for contour integrals as presented in \cite{\Ol,Theorem 6.1,\,p.125}. The integral $I(R)$ is
readily seen to satisfy conditions $(i)$--$(v)$ of \cite{\Ol,pp.121--122} ($(v)$ holds since we are
in the case $u>0$).

The first two coefficients of the power series expansion
$$
p(t)=p(\zeta)+p_0(t-\zeta)+p_1(t-\zeta)^2+\cdots
$$
of the function $p(t)$ of (7.10) are easily checked to satisfy $e^{-Rp(\zeta)}=\zeta^{-R(u-v)}=1$ (the
second equality since $u$ and $v$ are multiples of 3) and $p_0=p'(\zeta)=u\zeta-v\zeta^{-1}$. Since
$p_0\neq0$ it also follows that the value of $\mu$ when applying (4.21) for $I(R)$ is $\mu=1$.

The function $\tilde{Q}(t)$ given by (7.12) is precisely the function whose power series expansion is
considered in Lemma 7.4, multiplied by $t^{-d-1}(-1-t)^{-c-1}$. Since the first non-zero coefficient
in the power series expansion of $t^{-d-1}(-1-t)^{-c-1}$ around $t=\zeta$ is the constant term, and it
equals $\zeta^{-d-1}(-1-\zeta)^{-c-1}=\zeta^{c-d}$, we obtain by Lemma 7.4 that the 
power series expansion of $\tilde{Q}(t)$ has the form
$$
\tilde{Q}(t)
=\tilde{q}_0(t-\zeta)^{k+l}+\tilde{q}_1(t-\zeta)^{k+l+1}+\cdots
$$
with $\tilde{q}_0=\frac{\zeta^{c-d}}{k!\,l!}(\zeta-q\zeta^{-1})^{k+l}$.
Therefore, since $\tilde{q}_0\neq0$, the value of $\lambda$ when applying (4.21) for $I(R)$ is 
$\lambda=k+l+1$. Furthermore, the $a_0$ of (4.21) is now, by \cite{O,p.123}, equal to 
$$
\frac{\tilde{q}_0}{\mu p_0^{\lambda/\mu}}=
\frac{\zeta^{c-d}}{k!\,l!}\frac{(\zeta-q\zeta^{-1})^{k+l}}{(u\zeta-v\zeta^{-1})^{k+l+1}}.
$$
Thus, by applying (4.21) to $I(R)$ and looking just at the first term of the asymptotic series,
we obtain
$$
\spreadlines{2\jot}
\align
I(R)=\frac{I(R)}{e^{-Rp(\zeta)}}
&=\frac{\Gamma(k+l+1)}{R^{k+l+1}}
\frac{\zeta^{c-d}}{k!\,l!}\frac{(\zeta-q\zeta^{-1})^{k+l}}{(u\zeta-v\zeta^{-1})^{k+l+1}}
+O(R^{-k-l-2})\\
&=
{k+l \choose l}\zeta^{c-d-1}\frac{(1-q\zeta)^{k+l}}{(u-v\zeta)^{k+l+1}}+O(R^{-k-l-2}),\tag7.13
\endalign
$$
where at the last equality we used $\zeta^{-2}=\zeta$.

The asymptotics of the second integral in (7.11) is obtained by taking complex conjugation in~(7.13).
Replacing these asymptotics into (7.11) we obtain (7.1).

We now extend the proof to the case when $u$ is not necessarily positive. The same arguments we
employed above imply for $u>0$ the slightly more general asymptotics
$$
\spreadlines{3\jot}
\align
&
{\Cal D}_y^l\left.\left\{{\Cal D}_x^k P(-Ru+q'(x+y)+c,-Rv+q(x+y)+d)|_{x=a_1}\right\}\right|_{y=b_1}=
\\
&\! 
-\frac{i}{2\pi}{k+l \choose l}
\left[\zeta^{c-d}\frac{(q'\zeta-q\zeta^{-1})^{k+l}}{(u\zeta-v\zeta^{-1})^{k+l+1}}
-\zeta^{-c+d}\frac{(q'\zeta^{-1}-q\zeta)^{k+l}}{(u\zeta^{-1}-v\zeta)^{k+l+1}}\right]
\frac{1}{R^{k+l+1}}
\\
&
\ \ \ \ \ \ \ \ \ \ \ \ \ \ \ \ \ \ \ \ \ \ \ \ \ \ \ \ \ \ \ \ \ \ \ \ \ \ \ \ 
+O(R^{-k-l-2}),\tag7.14
\endalign
$$
for any $q'\in\Q$ with $3|q'-q$, and any integer multiples $u$ and $v$ of 3 at least one of which is
non-zero.

Suppose now $v>0$. By its definition (1.3) it is clear that 
$P(\alpha,\beta)=P(\beta,\alpha)$, so we can write
$$
P(-Ru+x+y+c,-Rv+q(x+y)+d)=P(-Rv+q(x+y)+d,-Ru+x+y+c).
$$
Since $v>0$, (7.14) applies to the right hand side above. Its asymptotics is thus
$$
\spreadlines{2\jot}
\align
&
-\frac{i}{2\pi}{k+l \choose l}
\left[\zeta^{d-c}\frac{(q\zeta-\zeta^{-1})^{k+l}}{(v\zeta-u\zeta^{-1})^{k+l+1}}
-\zeta^{-d+c}\frac{(q\zeta^{-1}-\zeta)^{k+l}}{(v\zeta^{-1}-u\zeta)^{k+l+1}}\right]
\frac{1}{R^{k+l+1}}\\
&
\ \ \ \ \ \ \ \ \ \ \ \ \ \ \ \ \ \ \ \ \ \ \ \ \ \ \ \ \ \ \ \ \ \ \ \ \ \ \ \ 
+O(R^{-k-l-2})
\\
=
&
-\frac{i}{2\pi}{k+l \choose l}
\left[-\zeta^{d-c}\frac{(\zeta^{-1}-q\zeta)^{k+l}}{(u\zeta^{-1}-v\zeta)^{k+l+1}}
+\zeta^{-d+c}\frac{(\zeta-q\zeta^{-1})^{k+l}}{(u\zeta-v\zeta^{-1})^{k+l+1}}\right]
\frac{1}{R^{k+l+1}}\\
&
\ \ \ \ \ \ \ \ \ \ \ \ \ \ \ \ \ \ \ \ \ \ \ \ \ \ \ \ \ \ \ \ \ \ \ \ \ \ \ \ 
+O(R^{-k-l-2}),
\endalign
$$
which agrees with (7.1).

If neither $u>0$ nor $v>0$ hold, then since by hypothesis $(u,v)\neq(0,0)$, we must have
$-u-v>0$. By \cite{\Kone}, $P$ also satisfies the symmetry 
$P(\alpha,\beta)=P(-\alpha-\beta-1,\alpha)$. Write therefore
$$
\align
&
P(-Ru+x+y+c,-Rv+q(x+y)+d)
\\
&
\ \ \ \ \ \ \ \ \ \ \ \ \ \ \ \ \ \ \ \
=P(-R(-u-v)+(-1-q)(x+y)-c-d-1,-Ru+x+y+c).
\endalign
$$

Relation (7.14) is applicable to the right had side above: $-u-v$ and $-v$ are multiples of 3 (since
$u$ and $v$ are assumed to be so), $(-1-q)-1=-3+1-q$ is divisible by 3 (since $3|1-q$), and 
$-u-v>0$. We obtain by (7.14) that its asymptotics is
$$
\spreadlines{2\jot}
\align
&
-\frac{i}{2\pi}{k+l \choose l}
\left[\zeta^{-2c-d-1}\frac{[(-1-q)\zeta-\zeta^{-1}]^{k+l}}{[(-u-v)\zeta-u\zeta^{-1}]^{k+l+1}}
\right.
\\
&
\ \ \ \ \ \ \ \ \ \ \ \ \ \ \ \ \ \ \ \ \ \ \ \ \ \ \ \ \ \ \
\left.
-\zeta^{2c+d+1}\frac{[(-1-q)\zeta^{-1}-\zeta]^{k+l}}{[(-u-v)\zeta^{-1}-u\zeta]^{k+l+1}}\right]
\frac{1}{R^{k+l+1}}
+O(R^{-k-l-2})
\\
&
=
-\frac{i}{2\pi}{k+l \choose l}
\left[\zeta^{c-d-1}\frac{(1-q\zeta)^{k+l}}{(u-v\zeta)^{k+l+1}}
\right.
\\
&
\ \ \ \ \ \ \ \ \ \ \ \ \ \ \ \ \ \ \ \ \ \ \ \ \ \ \ \ \ \ \
\left.
-\zeta^{-c+d+1}\frac{(1-q\zeta^{-1})^{k+l}}{(u-v\zeta^{-1})^{k+l+1}}\right]
\frac{1}{R^{k+l+1}}+O(R^{-k-l-2})
\endalign
$$
(we used $\zeta^{-2c}=\zeta^c$ and $-\zeta-\zeta^{-1}=1$). This is precisely the right hand
side of (7.1), and the proof is complete. \epf

\mysec{8. The evaluation of $\det(M'')$}

We evaluate the determinant of the matrix $M''$ given by (5.11)--(5.15) by the method of factor
exhaustion. It turns out that this evaluation holds for any $\zeta$, not just for the specific
$\zeta=e^{2\pi i/3}$ from the statement of Proposition 5.3. Therefore we will treat $\zeta$ as an
indeterminate in Sections~8--12 and in the Appendix.

The proof of the evaluation of $\det(M'')$ involves several technical steps, presented in 
Sections 9--12 and in the Appendix.
In the current section we show how to put together those steps to deduce the evaluation.

\proclaim{Theorem 8.1} Let $S=\sum_{i=1}^m s_i$, $T=\sum_{j=1}^n t_j$, and assume $S\geq T$. 
Then the determinant of the matrix $M''$ defined in $(5.11)$--$(5.15)$, with $\zeta$, $q$,
$x_i$, $y_i$, $w_j$, $z_j$, $i=1,\dotsc,m$, $j=1,\dotsc,n$ being 
indeterminates, is given by
$$
\spreadlines{2\jot}
\align
\det(M'')=(\zeta^2-\zeta^{-2})^{2S}
&[(q-\zeta)(q-\zeta^{-1})]^{\sum_{i=1}^m {s_i \choose 2}+\sum_{j=1}^n {t_j \choose 2}}
\\
\times
\prod_{1\leq i<j\leq m}&[((x_i-x_j)-\zeta(y_i-y_j))((x_i-x_j)-\zeta^{-1}(y_i-y_j))]^{s_is_j}
\\
\times
\prod_{1\leq i<j\leq n}&[((z_i-z_j)-\zeta(w_i-w_j))((z_i-z_j)-\zeta^{-1}(w_i-w_j))]^{t_it_j}
\\
\times
\prod_{i=1}^m\prod_{j=1}^n&[((x_i-z_j)-\zeta(y_i-w_j))((x_i-z_j)-\zeta^{-1}(y_i-w_j))]^{-s_it_j}.
\tag8.1
\endalign
$$

\endproclaim

\pf Regard both sides of (8.1) as polynomials in $q$ with coefficients in the field of rational
functions $\Q(x_1,\dotsc,x_m,y_1,\dotsc,y_m,z_1,\dotsc,z_n,w_1,\dotsc,w_n,\zeta)$. 

It is not hard to show that the degree in $q$ of the left hand side of (8.1) is at 
most $\sum_{i=1}^m {s_i \choose 2}+\sum_{j=1}^n {t_j \choose 2}$. Indeed, multiplying the first
$2T$ columns of $M''$ from left to right by 
$$
\align
q^0,q^0,q^{-1},q^{-1},\dotsc,q^{t_1-1},q^{t_1-1},
q^0,q^0,&\,q^{-1},q^{-1},\dotsc,q^{t_2-1},q^{t_2-1},\dotsc,
\\
&
q^0,q^0,q^{-1},q^{-1},\dotsc,q^{t_n-1},q^{t_n-1},\tag8.2
\endalign
$$
respectively, the degrees in $q$ of the entries become constant along the rows (except for some 0 
entries generated by (5.15) when $l<k$). Namely, from top to bottom, these common degrees along the 
rows are
$$
0,0,1,1,\dotsc,s_1-1,s_1-1,
0,0,1,1,\dotsc,s_2-1,s_2-1,\dotsc,
0,0,1,1,\dotsc,s_m-1,s_m-1,
$$
respectively. Thus, the degree in $q$ of any element in the expansion of this scaled-column 
modification of $M''$ is either 0 or 
$$
2\sum_{i=0}^{s_1-1} i +\cdots +2\sum_{i=0}^{s_m-1} i =\sum_{i=1}^m {s_i \choose 2}.
$$
Taking into account the factors (8.2) we multiplied the columns of $M''$ by, it follows that the
degree in $q$ of the left hand side of (8.1) is at most 
$\sum_{i=1}^m {s_i \choose 2}+\sum_{j=1}^n {t_j \choose 2}$.

On the other hand, by Corollary 9.2, the left hand side of (8.1) is 
divisible by 
$$
[(q-\zeta)(q-\zeta^{-1})]^{\sum_{i=1}^m {s_i \choose 2}+\sum_{j=1}^n {t_j \choose 2}},
$$
which is, up to a scalar multiple, the polynomial on the right hand side of (8.1).
It follows that the two polynomials on the left and right in (8.1) must be the same up to 
multiplication by a factor independent of $q$. In particular, to prove (8.1) it suffices to 
establish it for $q=0$, i.e., to show that
$$
\spreadlines{2\jot}
\align
&
\det(M''|_{q=0})
\\
&\ \ \ \ \ \ \ \
=(\zeta^2-\zeta^{-2})^{2S}
\prod_{1\leq i<j\leq m}[((x_i-x_j)-\zeta(y_i-y_j))((x_i-x_j)-\zeta^{-1}(y_i-y_j))]^{s_is_j}
\\
&
\ \ \ \ \ \ \ \ \ \ \ \ \ \ \ \ \ \ \ \ \ \ \ \ 
\times
\prod_{1\leq i<j\leq n}[((z_i-z_j)-\zeta(w_i-w_j))((z_i-z_j)-\zeta^{-1}(w_i-w_j))]^{t_it_j}
\\
&
\ \ \ \ \ \ \ \ \ \ \ \ \ \ \ \ \ \ \ \ \ \ \ \ 
\times
\prod_{i=1}^m\prod_{j=1}^n[((x_i-z_j)-\zeta(y_i-w_j))((x_i-z_j)-\zeta^{-1}(y_i-w_j))]^{-s_it_j}.
\tag8.3
\endalign
$$

Taking $h=0$ in Proposition 10.1 it follows that
$$
\spreadlines{2\jot}
\align
\det(M''|_{q=0})=c
&\prod_{1\leq i<j\leq m}[((x_i-x_j)-\zeta(y_i-y_j))((x_i-x_j)-\zeta^{-1}(y_i-y_j))]^{s_is_j}
\\
\times
&\prod_{1\leq i<j\leq n}[((z_i-z_j)-\zeta(w_i-w_j))((z_i-z_j)-\zeta^{-1}(w_i-w_j))]^{t_it_j}
\\
\times
&\prod_{i=1}^m\prod_{j=1}^n[((x_i-z_j)-\zeta(y_i-w_j))((x_i-z_j)-\zeta^{-1}(y_i-w_j))]^{-s_it_j},
\tag8.4
\endalign
$$
with $c\in\Z[\zeta,\zeta^{-1}]$. 

To determine $c$, specialize in (8.4) $y_i=0$, $w_j=0$, $i=1,\dotsc,m$, $j=1,\dotsc,n$. Then, given
that we also have $q=0$, formulas (5.14)--(5.15) become equivalent to
$$
A^{\left\{\matrix {\scriptstyle 11}\\{\scriptstyle 12}\\{\scriptstyle 21}\endmatrix\right\}}(k,l,u,v)
=\frac{{k+l \choose k}}{u^{k+l+1}}
\left\{\matrix \zeta^{-1}-\zeta\\\zeta^{-3}-\zeta^3\\\zeta-\zeta^{-1}\endmatrix\right\}\tag8.5
$$
and
$$
B_l^{\left\{\matrix {\scriptstyle 11}\\{\scriptstyle 12}\\{\scriptstyle 21}\endmatrix\right\}}(k,u,v)
={l \choose k}u^{l-k}
\left\{\matrix \zeta^{-1}-\zeta\\\zeta^{-3}-\zeta^3\\\zeta-\zeta^{-1}\endmatrix\right\}.\tag8.6
$$
Denote by $M''_0$ the matrix given by (5.11)--(5.13) with definitions (8.5)--(8.6). 
Equality~(8.4)~becomes under our specialization
$$
\spreadlines{2\jot}
\align
\det(M''_0)=c\,
\frac{\prod_{1\leq i<j\leq m}(x_i-x_j)^{2s_is_j}\prod_{1\leq i<j\leq n}(z_i-z_j)^{2t_it_j}}
{\prod_{i=1}^m\prod_{j=1}^n(x_i-z_j)^{2s_it_j}}.
\tag8.7
\endalign
$$

Permute the rows of $M''_0$ by listing them in the order $1,3,\dotsc,2S-1,2,4,\dotsc,2S$, and then
apply this same permutation to the columns of the resulting matrix. Denote by $M_1$ the matrix obtained
permuting this way the rows and columns of $M''_0$.

Note that when using definitions (8.5) and (8.6) in (5.12) and (5.13), in all $2\times2$ blocks 
the entries in positions $(2,1)$ and $(2,2)$ are negatives of each other. 
The consequence of this for the matrix $M_1$ is that it has the form
$$
M_1=\left[\matrix C&D\\-C&C\endmatrix\right],
$$
where $C$ and $D$ are $S\times S$ matrices. We obtain
$$
\det(M''_0)=\det(M_1)=\det\left[\matrix C+D&D\\0&C\endmatrix\right].\tag8.8
$$
Furthermore, (8.5) and (8.6) imply that we have 
$$
\align
C&=(\zeta^{-1}-\zeta)N'\tag8.9\\
C+D&=(\zeta^{-1}-\zeta+\zeta^{-3}-\zeta^3)N'=(\zeta^{-1}-\zeta)(\zeta^{-1}+\zeta)^2 N',\tag8.10
\endalign
$$
where $N'$ is the matrix obtained from the matrix $N$ in Theorem A.1 by replacing $z_j$ by $-z_j$, 
$j=1,\dotsc,n$. By (8.8)--(8.10) we obtain
$$
\det(M''_0)=(\zeta^{-1}-\zeta)^{2S}(\zeta^{-1}+\zeta)^{2S}(\det N')^2=
(\zeta^{2}-\zeta^{-2})^{2S}(\det N')^2.\tag8.11
$$
By Theorem A.1 the above equality implies
$$
\det(M''_0)=(\zeta^{2}-\zeta^{-2})^{2S}\,
\frac{\prod_{1\leq i<j\leq m}(x_i-x_j)^{2s_is_j}\prod_{1\leq i<j\leq n}(z_i-z_j)^{2t_it_j}}
{\prod_{i=1}^m\prod_{j=1}^n(x_i-z_j)^{2s_it_j}}.
$$
Comparing this to (8.7) we obtain that $c=(\zeta^{2}-\zeta^{-2})^{2S}$. Replacing this in (8.4) implies
(8.3). This completes the proof. \epf

\mysec{9. Divisibility of $\det(M'')$ by the powers of $q-\zeta$ and $q-\zeta^{-1}$}

It turns out that rather than proving directly the divisibility of $\det(M'')$ by the factors 
involving $q$ on the right hand side of (8.1), it is more convenient to deduce this as a special
case of a more general result. 

More precisely, deform definitions (5.14)--(5.15) by introducing a new parameter $h$ as follows:
$$
\spreadlines{3\jot}
\align
&
A^{\left\{\matrix {\scriptstyle{11}}\\{\scriptstyle{12}}\\{\scriptstyle{21}}
\endmatrix\right\}}(k,l,u,v)={k+l \choose k}
\left[
\left\{\matrix{\scriptstyle{1}}\\{\scriptstyle{\zeta^{-2}}}\\{\scriptstyle{\zeta^2}}\endmatrix\right\}
\zeta^{-1}\frac{(-\zeta)^{k+l}(q-\zeta^{-1})_{k,h}(q-\zeta^{-1})_{l,h}}{(u-v\zeta)^{k+l+1}}
\right.
\\
&
\ \ \ \ \ \ \ \ \ \ \ \ \ \ \ \ \ \ \ \ \ \ \ \ \ \ \ \ \ \ \ \ \ \ \ \ \ \ \ \ \ \ \ \ \ 
-
\left.
\left\{\matrix{\scriptstyle{1}}\\{\scriptstyle{\zeta^{2}}}\\{\scriptstyle{\zeta^{-2}}}\endmatrix\right\}
\zeta\frac{(-\zeta^{-1})^{k+l}(q-\zeta)_{k,h}(q-\zeta)_{l,h}}{(u-v\zeta^{-1})^{k+l+1}}
\right],
\tag9.1
\endalign
$$
and
$$
\spreadlines{3\jot}
\align
&
B_{l}^{\left\{\matrix {\scriptstyle{11}}\\{\scriptstyle{12}}\\{\scriptstyle{21}}
\endmatrix\right\}}
(k,u,v)={l \choose k}
\left[
\left\{\matrix{\scriptstyle{1}}\\{\scriptstyle{\zeta^{-2}}}\\{\scriptstyle{\zeta^{2}}}\endmatrix\right\}
\zeta^{-1}(-\zeta)^k (q-\zeta^{-1})_{k,h} (u-v\zeta)^{l-k}
\right.
\\
&
\ \ \ \ \ \ \ \ \ \ \ \ \ \ \ \ \ \ \ \ \ \ \ \ \ \ \ \ \ \ \ \ \ \ \ \ \ \ \ \ \ \ \ \ \ 
-
\left.
\left\{\matrix{\scriptstyle{1}}\\{\scriptstyle{\zeta^{2}}}\\{\scriptstyle{\zeta^{-2}}}\endmatrix\right\}
\zeta (-\zeta^{-1})^k (q-\zeta)_{k,h} (u-v\zeta^{-1})^{l-k}
\right],
\tag9.2
\endalign
$$
where 
$$
(a)_{n,h}:=a(a+h)(a+2h)\cdots(a+(n-1)h).
$$
Clearly, formulas (5.14)--(5.15) correspond to the special case $h=0$ in the above expressions.

\proclaim{Proposition 9.1} Regarded as a polynomial in $q$ with coefficients in
$$
\Q(x_1,\dotsc,x_m,y_1,\dotsc,y_m,z_1,\dotsc,z_n,w_1,\dotsc,w_n,\zeta)[h],
$$ 
the determinant of the
matrix $(5.11)$--$(5.13)$ with definitions $(9.1)$--$(9.2)$ admits $q=\zeta^{\pm1}-\lambda h$ as a
root of multiplicity 
$$
\sum_{i=1}^m \max(s_i-(\lambda+1),0)+\sum_{j=1}^n \max(t_j-(\lambda+1),0),
$$
for all $\lambda\geq0$.

\endproclaim

\proclaim{Corollary 9.2} Regard the determinant of the matrix $M''$ given by $(5.11)$--$(5.15)$
as a polynomial in $q$ with coefficients in 
$\Q(x_1,\dotsc,x_m,y_1,\dotsc,y_m,z_1,\dotsc,z_n,w_1,\dotsc,w_n,\zeta)$. Then $\det(M'')$ is
divisible by
$$
[(q-\zeta)(q-\zeta^{-1})]^{\sum_{i=1}^m {s_i \choose 2}+
\sum_{j=1}^n {t_j \choose 2}}.
$$

\endproclaim

\pf Take $h=0$ in Proposition 9.1 and use that
$$
\align
&
\sum_{\lambda\geq0}
\left\{\sum_{i=1}^m \max(s_i-(\lambda+1),0)+\sum_{j=1}^n \max(t_j-(\lambda+1),0)\right\}\\
&\ \ \ \ \ \ \ \ \ \
=\sum_{i=1}^m [(s_i-1)+(s_i-2)+\cdots+1]+\sum_{j=1}^n [(t_j-1)+(t_j-2)+\cdots+1]\\
&\ \ \ \ \ \ \ \ \ \
=\sum_{i=1}^m {s_i \choose 2}+\sum_{j=1}^n {t_j \choose 2}.
\endalign
$$
\epfmath

{\it Proof of Proposition 9.1.} 
Recall the terminology we defined after the proof of Lemma 5.2 to describe the structure of the
matrix (5.11)--(5.13): the blocks outlined by large rectangles form block-rows and block-columns,
while the $2\times2$ blocks form bi-rows and bi-columns. Refer to portion (5.12) as the $A$-part
and to portion (5.13) as the $B$-part of the matrix.
In block-row $i$, bi-row $k+1$ consists of the juxtaposition of the $n+1$ matrices
$$
\spreadmatrixlines{2\jot}
\left[\matrix 
{\scriptscriptstyle A^{11}(k,0,z_j-x_i,w_j-y_i)}\!\!&
{\scriptscriptstyle A^{12}(k,0,z_j-x_i,w_j-y_i)}&\ \ 
{\scriptscriptstyle \dotsc}&\ \ 
{\scriptscriptstyle A^{11}(k,t_j-1,z_j-x_i,w_j-y_i)}\!\!&
{\scriptscriptstyle A^{12}(k,t_j-1,z_j-x_i,w_j-y_i)}
\\
{\scriptscriptstyle A^{21}(k,0,z_j-x_i,w_j-y_i)}\!\!&
{\scriptscriptstyle A^{11}(k,0,z_j-x_i,w_j-y_i)}&\ \ 
{\scriptscriptstyle \dotsc}&\ \ 
{\scriptscriptstyle A^{21}(k,t_j-1,z_j-x_i,w_j-y_i)}\!\!&
{\scriptscriptstyle A^{11}(k,t_j-1,z_j-x_i,w_j-y_i)}
\endmatrix\right]\!\!,
$$
for $j=1,\dotsc,n$, and
$$
\spreadmatrixlines{2\jot}
\left[\matrix 
{\scriptstyle B_0^{11}(k,x_i,y_i)}\!\!&{\scriptstyle B_0^{12}(k,x_i,y_i)}&\ \ 
{\scriptstyle \dotsc}&\ \ 
{\scriptstyle B_{S-T-1}^{11}(k,x_i,y_i)}\!\!&{\scriptstyle B_{S-T-1}^{12}(k,x_i,y_i)}
\\
{\scriptstyle B_0^{21}(k,x_i,y_i)}\!\!&{\scriptstyle B_0^{11}(k,x_i,y_i)}&\ \ 
{\scriptstyle \dotsc}&\ \ 
{\scriptstyle B_{S-T-1}^{21}(k,x_i,y_i)}\!\!&{\scriptstyle B_{S-T-1}^{11}(k,x_i,y_i)}
\endmatrix\right].
$$
We consider first the case of the roots of type $q=\zeta-\lambda h$.
Perform the elementary row operation of adding the top half of this bi-row, multiplied by $-\zeta^2$,
to its bottom half:
$$
R_{2k+2}^{(i)}:=R_{2k+2}^{(i)}-\zeta^2 R_{2k+1}^{(i)}\tag9.3
$$
($R_p^{(i)}$ stands for row $p$ of the block-row $i$ in matrix (5.11)--(5.13)).
By (9.1), the modified entries in bi-column $l+1$ of block-column $j$ are
$$
\align
A^{21}(k,l,z_j-x_i,w_j-y_i)-&\zeta^2 A^{11}(k,l,z_j-x_i,w_j-y_i)=
\\
&
\!\!\!\!\!\!\!\!
-{k+l \choose k}
\left[\zeta(\zeta^{-2}-\zeta^2)\frac{(-\zeta^{-1})^{k+l}(q-\zeta)_{k,h}(q-\zeta)_{l,h}}
{[(z_j-x_i)-\zeta(w_j-y_i)]^{k+l+1}}\right]\tag9.4
\endalign
$$
and
$$
\align
A^{11}(k,l,z_j-x_i,w_j-y_i)-&\zeta^2 A^{12}(k,l,z_j-x_i,w_j-y_i)=
\\
&
-{k+l \choose k}
\left[\zeta(1-\zeta^4)\frac{(-\zeta^{-1})^{k+l}(q-\zeta)_{k,h}(q-\zeta)_{l,h}}
{[(z_j-x_i)-\zeta(w_j-y_i)]^{k+l+1}}\right],\tag9.5
\endalign
$$
while by (9.2) the modified entries in bi-column $l+1$ of the $B$-part are
$$
B_l^{21}(k,x_i,y_i)-\zeta^2 B_l^{11}(k,x_i,y_i)=-{l \choose k}
\left[\zeta(\zeta^{-2}-\zeta^2)(-\zeta^{-1})^k (q-\zeta)_{k,h} (u-v\zeta^{-1})^{l-k}\right]
$$
and
$$
B_l^{11}(k,x_i,y_i)-\zeta^2 B_l^{12}(k,x_i,y_i)=-{l \choose k}
\left[\zeta(1-\zeta^4)(-\zeta^{-1})^k (q-\zeta)_{k,h} (u-v\zeta^{-1})^{l-k}\right].
$$

Note that
$$
q-(\zeta-\lambda h)\ |\ (q-\zeta)_{k,h}=(q-\zeta)(q-(\zeta-h))\cdots(q-(\zeta-(k-1))h)\tag9.6
$$
provided $\lambda\leq k-1$. The above expressions 
show therefore that whenever $k\geq \lambda+1$,
operation (9.3) produces a row along which all entries are divisible by $q-(\zeta-\lambda h)$.

Let $D$ be the matrix obtained from matrix (5.11)--(5.13), (9.1)--(9.2) by applying row operations
(9.3) for all $1\leq i\leq m$ and $k\geq \lambda+1$. Let $E$ be the matrix obtained from $D$ by
performing the column operations
$$
C_{2l+1}^{(j)}:=C_{2l+1}^{(j)}-\zeta^2 C_{2l+2}^{(j)},\ \ \ 1\leq j\leq n,\ l\geq \lambda+1\tag9.7
$$
($C_p^{(j)}$ stands for column $p$ of the block-column $j$ in matrix~$D$). 

Then to finish the proof, it suffices to show that all entries of $E$ along its columns produced
by (9.7) are divisible by $q-(\zeta-\lambda h)$, and that those of them situated
in bottom halves of bi-rows of index $\geq\lambda+1$ are in fact divisible by 
$(q-(\zeta-\lambda h))^2$.

To verify this, consider an arbitrary entry $\alpha$ of a column of type (9.7) of $E$.

{\it Case 1.} The entry $\alpha$ is in the top half of bi-row $k+1$ of block-row $i$. 

Then the entries of $D$ on which (9.7) acts are unmodified by the row operations (9.3). Formulas
(9.1)--(9.2) imply that 
$$
\alpha=-{k+l \choose k}\zeta(1-\zeta^4)
\frac{(-\zeta^{-1})^{k+l}(q-\zeta)_{k,h}(q-\zeta)_{l,h}}{[(z_j-x_i)-\zeta(w_j-y_i)]^{k+l+1}}.
$$
Since $l\geq \lambda+1$, (9.6) implies that this is divisible by $q-(\zeta-\lambda h)$.

{\it Case 2.} The entry $\alpha$ is in the bottom half of bi-row $k+1$ of block-row $i$. 

Then the entries of $D$ on which (9.7) acts are given by (9.4) and (9.5). It follows that
$$
\alpha=-{k+l \choose k}\zeta[-(\zeta^{-2}-\zeta^2)+\zeta^2(1-\zeta^4)]
\frac{(-\zeta^{-1})^{k+l}(q-\zeta)_{k,h}(q-\zeta)_{l,h}}{[(z_j-x_i)-\zeta(w_j-y_i)]^{k+l+1}}.
$$
By (9.6), this is divisible by $q-(\zeta-\lambda h)$, since $l\geq \lambda+1$. If in addition
$k\geq \lambda+1$, (9.6) implies that $\alpha$ is divisible by $(q-(\zeta-\lambda h))^2$.

This verifies our statement about the entries of $E$ along the columns produced 
by (9.7), and completes the proof of the statement of the Proposition for the roots of type 
$q=\zeta-\lambda h$. 

To obtain the corresponding statement about the roots of type $q=\zeta^{-1}-\lambda h$, 
replace $\zeta$ by $\zeta^{-1}$ in the statement we just proved, and use the fact that
this has the effect of replacing the quantities (9.1)--(9.2) by their negatives.
\epf

\mysec{10. The case $q=0$ of Theorem 8.1, up to a constant multiple}

Proposition 9.1 reduces the proof of Theorem 8.1 to the case $q=0$ (see Section 8). 
It turns out that in order to 
handle this case it is convenient to deform the $q=0$ specialization of definitions (5.14)--(5.15)
by introducing a new parameter $h$ as follows:
$$
\align
A^{\left\{\matrix {\scriptstyle{11}}\\{\scriptstyle{12}}\\{\scriptstyle{21}}
\endmatrix\right\}}(k,l,u,v)=
{k+l \choose k}
\left[
\left\{\!\!\matrix{\scriptstyle{1}}\\{\scriptstyle{\zeta^{-2}}}\\{\scriptstyle{\zeta^{2}}}
\endmatrix\!\!\right\}
\zeta^{-1}\frac{1}{(u-v\zeta)_{k+l+1,h}}
-
\left\{\!\!\matrix{\scriptstyle{1}}\\{\scriptstyle{\zeta^{2}}}\\{\scriptstyle{\zeta^{-2}}}
\endmatrix\!\!\right\}
\zeta\frac{1}{(u-v\zeta^{-1})_{k+l+1,h}}
\right]\!,
\\
\tag10.1
\endalign
$$
and
$$
\align
B_l^{\left\{\matrix {\scriptstyle{11}}\\{\scriptstyle{12}}\\{\scriptstyle{21}}
\endmatrix\right\}}
(k,u,v)=
{l \choose k}\!\!
\left[
\left\{\matrix{\scriptstyle{1}}\\{\scriptstyle{\zeta^{-2}}}\\{\scriptstyle{\zeta^2}}\endmatrix\right\}
\zeta^{-1} (u-v\zeta)_{l-k,h}
-
\left\{\matrix{\scriptstyle{1}}\\{\scriptstyle{\zeta^2}}\\{\scriptstyle{\zeta^{-2}}}\endmatrix\right\}
\zeta  (u-v\zeta^{-1})_{l-k,h}
\right],
\tag10.2
\endalign
$$
where as in the previous section $(a)_{n,h}:=a(a+h)(a+2h)\cdots(a+(n-1)h)$. Clearly, setting $h=0$ 
above yields the specialization $q=0$ of the definitions (5.14)--(5.15).

\proclaim{Proposition 10.1}  
Let $M_0$ be the $2S\times 2S$ matrix given by {\rm (5.11)--(5.13)} and {\rm (10.1)--(10.2)}. 
Then we have
$$
\spreadlines{3\jot}
\align
&
\det(M_0)=
\\
&
c
\prod_{1\leq i<j\leq m}\prod_{k=0}^{s_i-1}\prod_{l=0}^{s_j-1}
[(x_i-x_j)-\zeta(y_i-y_j)-(k-l)h][(x_i-x_j)-\zeta^{-1}(y_i-y_j)-(k-l)h]\\
&
\times\!\!\!
\prod_{1\leq i<j\leq n}\prod_{k=0}^{t_i-1}\prod_{l=0}^{t_j-1}
[(z_i-z_j)-\zeta(w_i-w_j)-(k-l)h][(z_i-z_j)-\zeta^{-1}(w_i-w_j)-(k-l)h]\\
&
\times
\prod_{i=1}^m\prod_{j=1}^n\prod_{k=0}^{s_i-1}\prod_{l=0}^{t_j-1}
\left\{
[(x_i-z_j)-\zeta(y_i-w_j)-(k+l)h]^{-1}
\right.
\\
&
\left.
\ \ \ \ \ \ \ \ \ \ \ \ \ \ \ \ \ \ \ \ \ \ \ \ \ \ \ \ \ \ \ \ \ \ \ \ \ \
\times
[(x_i-z_j)-\zeta^{-1}(y_i-w_j)-(k+l)h]^{-1}
\right\}
,\tag10.3
\endalign
$$
where $c\in\Z[\zeta,\zeta^{-1}]$.

\endproclaim

\pf 
Let
$$
d=\prod_{i=1}^m\prod_{j=1}^n\prod_{a=0}^{s_i+t_j-2}
[(x_i-z_j)-\zeta(y_i-w_j)-ah][(x_i-z_j)-\zeta^{-1}(y_i-w_j)-ah].\tag10.4
$$
It follows from the definition of $M_0$ that when multiplying it by $d$, all its entries become 
polynomials in the
variables $h,x_1,\dotsc,x_m,y_1,\dotsc,y_m,z_1,\dotsc,z_n,w_1,\dotsc,w_n$ with coefficients in
$\Z[\zeta,\zeta^{-1}]$.

For $a\in\Z$ and integers $s,t\geq1$ define
$$
N_1(a,s,t)=|\{(k,l): k-l=a, 0\leq k\leq s-1, 0\leq l\leq t-1\}|\tag10.5
$$
and
$$
N_2(a,s,t)=|\{(k,l): k+l=a, 0\leq k\leq s-1, 0\leq l\leq t-1\}|.\tag10.6
$$
Then, multiplying all rows of $M_0$ by $d$, (10.3) becomes equivalent to
$$
\align
&
\det(dM_0)=
\\
&\ 
c\ 
\prod_{1\leq i<j\leq m}\prod_{a=-(s_j-1)}^{s_i-1}
\left\{
[(x_i-x_j)-\zeta(y_i-y_j)-ah]^{N_1(a,s_i,s_j)}
\right.
\\
&\ \ \ \ \ \ \ \ \ \ \ \ \ \ \ \ \ \ \ \ \ \ \ \ \ \ \ \ \ \ \ \ \ \ \ \ \ \ \ \ \ \ \ \ \ \ \ \ \ \ \ \
\left.
\times
[(x_i-x_j)-\zeta^{-1}(y_i-y_j)-ah]^{N_1(a,s_i,s_j)}\right\}
\\
&
\times
\prod_{1\leq i<j\leq n}\prod_{a=-(t_j-1)}^{t_i-1}
\left\{
[(z_i-z_j)-\zeta(w_i-w_j)-ah]^{N_1(a,t_i,t_j)}
\right.
\\
&\ \ \ \ \ \ \ \ \ \ \ \ \ \ \ \ \ \ \ \ \ \ \ \ \ \ \ \ \ \ \ \ \ \ \ \ \ \ \ \ \ \ \ \ \ \ \ \ \ \ \ \
\left.
\times
[(z_i-z_j)-\zeta^{-1}(w_i-w_j)-ah]^{N_1(a,t_i,t_j)}\right\}
\\
&
\times
\prod_{i=1}^m\prod_{j=1}^n\prod_{a=0}^{s_i+t_j-2}
\{[(x_i-z_j)-\zeta(y_i-w_j)-ah]^{2S-N_2(a,s_i,t_j)}
\\
&\ \ \ \ \ \ \ \ \ \ \ \ \ \ \ \ \ \ \ \ \ \ \ \ \ \ \ \ \ \ \ \ \ \ \ \ \ \  
\times
[(x_i-z_j)-\zeta^{-1}(y_i-w_j)-ah]^{2S-N_2(a,s_i,t_j)}\}.\tag10.7
\endalign
$$
Regard both sides of (10.7) as polynomials in the
variables 
$$
h,x_1,\dotsc,x_m,y_1,\dotsc,y_m,z_1,\dotsc,z_n,w_1,\dotsc,w_n,
$$ 
with coefficients in $\Z[\zeta,\zeta^{-1}]$. 

By Corollary 11.3 and Proposition 12.1, $\det(dM_0)$ is divisible by the full product expression on 
the right hand side of (10.7). It follows from definitions (10.5) and (10.6) that 
$$
\sum_{a=-(t-1)}^{s-1}N_1(a,s,t)=\sum_{a=0}^{s+t-2}N_2(a,s,t)=st.
$$
By (10.4), the degree of $d$ is
$$
\deg(d)=2\sum_{i=1}^m\sum_{j=1}^n(s_i+t_j-1).
$$
Therefore, jointly in its variables, the polynomial on the right hand side of (10.7) has degree
$$
2S\deg(d)+2\sum_{1\leq i<j\leq m}s_is_j+2\sum_{1\leq i<j\leq n}t_it_j
-2\sum_{i=1}^m\sum_{j=1}^n s_it_j.\tag10.8
$$
To complete the proof, it suffices to show that the degree of $\det(dM_0)$ is less or equal than
the value given by this expression. In turn, this is clearly equivalent to the degree\footnote{
The degree of a rational function is the difference between the degrees of its numerator and 
de\-no\-mi\-na\-tor.}
of $\det(M_0)$ being at most the value given by the three sum-terms in (10.8). 
However, this follows by the argument we used
in the proof of Proposition 5.3 to factor out the powers of $R$ from the determinant of the matrix
with blocks (5.16)--(5.17): Indeed, the degree in $R$ of each entry of that matrix is equal to the
joint degree in its variables of the corresponding entry of $M_0$.\epf

\medskip
\flushpar
{\smc Remark 10.2.} Setting $h=0$ in (10.3) and using (8.3) (which is a special case of Theorem 8.1) 
we obtain the explicit value 
$
c=(\zeta^{-2}-\zeta^2)^{2S}
$ 
for the multiplicative constant in Proposition 10.1. This provides a more general deformation of Cauchy's
determinant than the one presented in Theorem A.1.

\bigskip
\medskip
\bigskip
\centerline{\bf 11. Divisibility of $\det(dM_0)$ by the powers of
$(x_i-x_j)-\zeta^{\pm1}(y_i-y_j)-ah$}
\medskip
\centerline{\bf and $(z_i-w_j)-\zeta^{\pm1}(z_i-w_j)-ah$}

\medskip
\medskip
\bigskip
For any integers $s\geq1$ and $0\leq i\leq s-1$, define the $2s$-vector ${\bold v}_i^s$ by
$$
\spreadlines{3\jot}
\align
{\bold v}_i^s&=[v_{i0},-\zeta^2 v_{i0},v_{i1},-\zeta^2 v_{i1},\dotsc,
v_{s-1},-\zeta^2 v_{i,s-1}],\\
v_{ij}&=(-1)^{j}j!{i \choose j}h^{j},\ \ j=0,1,\dotsc,s-1.\tag11.1
\endalign
$$

Denote by ${\bold 0}_r$ the $r$-vector with all coordinates equal to 0.

\proclaim{Lemma 11.1} Let $k$ and $l$ be integers, $0\leq k\leq s_i-1$, $0\leq l\leq s_j-1$, $i<j$.
Then, if $x_i=x_j+\zeta(y_i-y_j)+(k-l)h$, the linear combination of the rows of matrix $M_0$ 
specified by the~$2S$-vector
$$
[{\bold 0}_{2s_1},\cdots,{\bold 0}_{2s_{i-1}},-{\bold v}_k^{s_i},{\bold 0}_{2s_{i+1}},\cdots,
{\bold 0}_{2s_{j-1}},{\bold v}_l^{s_j},{\bold 0}_{2s_{j+1}},\cdots,{\bold 0}_{2s_{m}}]\tag11.2
$$
is a vanishing row combination. Furthermore, for any fixed $a\in\Z$, the vectors of the above type
satisfying $k-l=a$ are linearly independent. 

\endproclaim

\pf Note that in any subset of the set of vectors with $k-l=a$ fixed 
there is a unique vector with a rightmost non-zero coordinate. Indeed, if
two such vectors have their rightmost non-zero coordinates in the same position, the values of $l$
they correspond to are equal, but then $k-l=a$ implies that the $k$-values they correspond to
are also equal, so the two vectors are identical. This implies the linear independence claimed in the
statement of the Lemma.

Due to the form of $M_0$, the statement that the specified row combination is vanishing 
amounts to four separate identities. Namely, we need to check
that the indicated row combination produces 0 along the even and odd indexed columns of portion (5.12) 
of $M_0$, and also along the even and odd indexed columns of portion (5.13) of $M_0$. 
It is convenient to refer to these two portions as the $A$-part and $B$-part of $M_0$.

Along column $2r+1$ of block-column $b$ (cf. our terminology related to the structure of (5.12)--(5.13))
of the $A$-part of $M_0$, the vanishing of the indicated row combination amounts to
$$
\align
-\sum_{\alpha=0}^{s_i-1} 
&v_{k\alpha}[A^{11}(\alpha,r,z_b-x_i,w_b-y_i)-\zeta^2A^{21}(\alpha,r,z_b-x_i,w_b-y_i)]
\\
&
+\sum_{\alpha=0}^{s_j-1} 
v_{l\alpha}[A^{11}(\alpha,r,z_b-x_j,w_b-y_j)-\zeta^2A^{21}(\alpha,r,z_b-x_j,w_b-y_j)]=0,\tag11.3
\endalign
$$
provided $x_i=x_j+\zeta(y_i-y_j)+(k-l)h$.

By (11.1) and (10.1)--(10.2), the left hand side above equals
$$
\align
-\sum_{\alpha=0}^k&(-h)^\alpha\frac{k!}{(k-\alpha)!}{\alpha+r \choose \alpha}\frac{1-\zeta^4}{\zeta}
\frac{1}{((z_b-x_i)-\zeta(w_b-y_i))_{\alpha+r+1,h}}\\
&
+\sum_{\alpha=0}^l(-h)^\alpha\frac{l!}{(l-\alpha)!}{\alpha+r \choose \alpha}\frac{1-\zeta^4}{\zeta}
\frac{1}{((z_b-x_j)-\zeta(w_b-y_j))_{\alpha+r+1,h}}.\tag11.4
\endalign
$$
When $x_i=x_j+\zeta(y_i-y_j)+(k-l)h$, we have 
$(z_b-x_j)-\zeta(w_b-y_j)=(z_b-x_i)-\zeta(w_b-y_i)+(k-l)h$. Denoting $c=(z_b-x_i)-\zeta(w_b-y_i)$,
we obtain that (11.3) is equivalent to
$$
\spreadlines{2\jot}
\align
&
\sum_{\alpha=0}^k (-h)^\alpha\frac{k!}{(k-\alpha)!}{\alpha+r \choose \alpha}\frac{1}{(c)_{\alpha+r+1,h}}
\\
&
\ \ \ \ \ \ \ \ \ \ \ \ \ \ \ \ \ \ \ \ \ \ \ \ \ \
=
\sum_{\alpha=0}^l (-h)^\alpha\frac{l!}{(l-\alpha)!}{\alpha+r \choose \alpha}
\frac{1}{(c+(k-l)h)_{\alpha+r+1,h}}.\tag11.5
\endalign
$$

We claim that the left hand side of (11.5) evaluates as
$$
\sum_{\alpha=0}^k (-h)^\alpha\frac{k!}{(k-\alpha)!}{\alpha+r \choose \alpha}\frac{1}{(c)_{\alpha+r+1,h}}
=
\frac{1}{h^{r+1}}\frac{1}{(\frac{c}{h}+k)_{r+1}}.\tag11.6
$$
Indeed, let $c'=c/h$. The left hand side of (11.6) can then be written as
$$
\align
&\!\!\!\!\!\!\!\!\!\!\!\!
\sum_{\alpha=0}^k (-h)^\alpha\frac{k!}{(k-\alpha)!}{\alpha+r \choose \alpha}
\frac{1}{(c')_{\alpha+r+1}h^{\alpha+r+1}}
\\
&=
\frac{1}{h^{r+1}}\sum_{\alpha=0}^k (-k)_{\alpha}\frac{(r+1)_{\alpha}}{\alpha !}
\frac{1}{(c')_{r+1}(c'+r+1)_{\alpha}}
\\
&=
\frac{1}{(c')_{r+1}h^{r+1}}{}_2 F_1\!\!\left[\matrix -k,r+1\\ c'+r+1\endmatrix;1\right].
\endalign
$$
Then (11.6) follows by Lemma 11.2.

By (11.6), the right hand side of (11.5) equals
$$
\frac{1}{h^{r+1}}\frac{1}{(\frac{c+(k-l)h}{h}+l)_{r+1}}.
$$
This proves (11.5), and therefore (11.3). 

Next, we verify that the row combination (11.2) is vanishing along the 
column $2r+2$ of block-column $b$. This amounts to proving a close analog
of (11.3); namely, that
$$
\align
-\sum_{\alpha=0}^{s_i-1} 
&v_{k\alpha}[A^{12}(\alpha,r,z_b-x_i,w_b-y_i)-\zeta^2A^{11}(\alpha,r,z_b-x_i,w_b-y_i)]
\\
&
+\sum_{\alpha=0}^{s_j-1} 
v_{l\alpha}[A^{12}(\alpha,r,z_b-x_j,w_b-y_j)-\zeta^2A^{11}(\alpha,r,z_b-x_j,w_b-y_j)]=0,\tag11.7 
\endalign
$$
provided $x_i=x_j+\zeta(y_i-y_j)+(k-l)h$. By (11.1) and (10.1)--(10.2), the left hand side of (11.7)
is nearly the same as the left hand side of (11.3)---the only difference is that the factor $1-\zeta^4$
at the numerator gets now replaced by $\zeta^{-2}-\zeta^2$. Therefore (11.7) follows by (11.3).

To complete the proof we need to check that row combination (11.2) vanishes along the columns of the
$B$-part of $M_0$. 

Along column $2r+1$ of the $B$-part of $M_0$, the vanishing of row combination (11.2) amounts to 
showing that for $x_i=x_j+\zeta(y_i-y_j)+(k-l)h$ one has
$$
\align
-\sum_{\alpha=0}^{s_i-1} 
&v_{k\alpha}[B^{11}(\alpha,r,x_i,y_i)-\zeta^2B^{12}(\alpha,r,x_i,y_i)]
\\
&
+\sum_{\alpha=0}^{s_j-1} 
v_{l\alpha}[B^{11}(\alpha,r,x_j,y_j)-\zeta^2B^{12}(\alpha,r,x_j,y_j)]=0.\tag11.8 
\endalign
$$
By (11.1) and formulas (10.1)--(10.2), the left hand side of (11.8) becomes
$$
\align
&\!\!\!\!\!\!\!\!\!\!\!\!\!\!\!\!
\frac{1-\zeta^4}{\zeta}
\left\{
-\sum_{\alpha=0}^k
(-h)^\alpha\frac{k!}{(k-\alpha)!}{r \choose \alpha}
(x_i-\zeta y_i)_{r-\alpha,h}
\right.
\\
&\ \ \ \ \ \ \ \ \ \ \ \ \ \ \ \ \ \ \ \ \ \ \ \ \ \ \ \
\left.
+\sum_{\alpha=0}^l(-h)^\alpha\frac{l!}{(l-\alpha)!}{r \choose \alpha}
(x_j-\zeta y_j)_{r-\alpha,h}
\right\}.\tag11.9
\endalign
$$
Letting $c=(x_i-\zeta y_i)/h$, the first sum above can be written as
$$
\align
&\!\!\!\!\!\!\!\!\!\!\!\!
\sum_{\alpha=0}^k (-h)^\alpha\frac{k!}{(k-\alpha)!}{r \choose \alpha}
(c)_{r-\alpha}h^{r-\alpha} 
\\
&=
h^r\sum_{\alpha=0}^k (-k)_{\alpha}\frac{(-r)_{\alpha}}{\alpha !}
\frac{(c)_{r}}{(-c-r+1)_{\alpha}}
\\
&=
(c)_{r} h^r {}_2 F_1\!\!\left[\matrix -k,-r\\ -c-r+1\endmatrix;1\right].
\endalign
$$
Replacing $r$ by $-r-1$ and $c$ by $-c+1$, Lemma 11.2 implies, after simplifications, that
$$
{}_2 F_1\!\!\left[\matrix -k,-r\\ -c-r+1\endmatrix;1\right]=\frac{(c-k)_r}{(c)_r}.
$$
Therefore, the first sum in (11.9) equals $h^r(c-k)_r$. But the second sum in (11.9) has the same form,
except $(x_j-\zeta y_j)/h=(x_i-\zeta y_i)/h-(k-l)=c-(k-l)$, since we are assuming 
$x_i=x_j+\zeta(y_i-y_j)+(k-l)h$.
It follows that the second sum equals $h^r((c-(k-l))-l)_r$, and thus cancels the first sum. This proves
(11.8). 

The final identity to check is that row combination (11.2) is vanishing along column $2r+2$ of the 
$B$-part of $M_0$. This is equivalent to the following counterpart of (11.8):
$$
\align
-\sum_{\alpha=0}^{s_i-1} 
&v_{k\alpha}[B^{12}(\alpha,r,x_i,y_i)-\zeta^2B^{11}(\alpha,r,x_i,y_i)]
\\
&
+\sum_{\alpha=0}^{s_j-1} 
v_{l\alpha}[B^{12}(\alpha,r,x_j,y_j)-\zeta^2B^{11}(\alpha,r,x_j,y_j)]=0.\tag11.10 
\endalign
$$
Replacing (11.1) and (10.1)--(10.2) in this shows that, just as it was the case for (11.3) and (11.7),
the expressions on the left hand sides of (11.10) and (11.8) differ only in that the factor
$1-\zeta^4$ in the former gets replaced by $\zeta^{-2}-\zeta^2$ in the latter. Therefore (11.10) follows
from (11.8), and the proof is complete. \epf

\proclaim{Lemma 11.2} Let $0\leq k\in\Z$, $r\in\Z$. Then if $c$ is an indeterminate one has
$$
\spreadmatrixlines{2\jot}
{}_2 F_1\!\!\left[\matrix -k,r+1\\ c+r+1\endmatrix;1\right]=
\left\{
\matrix
\frac{ \displaystyle (c)_{r+1} }{ \displaystyle (c+k)_{r+1} },\ \ \ \ \ \ \ \ \ \ \ \ \ \,
{\text {\rm if $r+1\geq0$,}}\\
\frac{\displaystyle (c+k+r+1)_{-r-1}}{\displaystyle (c+r+1)_{-r-1}},\ \ {\text {\rm if $r+1<0$}}.
\endmatrix
\right.
$$

\endproclaim

\pf Since $k\geq0$, the hypergeometric series is terminating and the equality in the statement is
equivalent to a polynomial identity in $c$. By Gauss' formula (4.9), for any large enough integer $c$
we have
$$
{}_2 F_1\!\!\left[\matrix -k,r+1\\ c+r+1\endmatrix;1\right]=
\frac{\Gamma(c+r+1)\Gamma(c+k)}{\Gamma(c)\Gamma(c+k+r+1)}.
$$
The expression on the right hand side readily checks to simplify to the ones in the statement.
This proves that the rational functions in $c$ on the two sides of the equality in the statement 
agree on infinitely many values of $c$. They must therefore be identical.~\epf

\proclaim{Corollary 11.3} Regarded as a polynomial with coefficients in $\Z[\zeta,\zeta^{-1}]$ in the 
variables 
$h$, $x_1,\dotsc,x_m,y_1,\dotsc,y_m,z_1,\dotsc,z_n,w_1,\dotsc,w_n$, 
$\det(dM_0)$ is divisible by
$$
\spreadlines{1\jot}
\align
&
\prod_{1\leq i<j\leq m}\prod_{a=-(s_j-1)}^{s_i-1}
\left\{
[(x_i-x_j)-\zeta(y_i-y_j)-ah]^{N_1(a,s_i,s_j)}
\right.
\\
&\ \ \ \ \ \ \ \ \ \ \ \ \ \ \ \ \ \ \ \ \ \ \ \ \ \ \ \ \ \ \ \ \ \ \ \ \ \ \ \ \ \
\left.
\times
[(x_i-x_j)-\zeta^{-1}(y_i-y_j)-ah]^{N_1(a,s_i,s_j)}\right\}
\\
\times
&
\prod_{1\leq i<j\leq n}\prod_{a=-(t_j-1)}^{t_i-1}
\left\{
[(z_i-z_j)-\zeta(w_i-w_j)-ah]^{N_1(a,t_i,t_j)}
\right.
\\
&\ \ \ \ \ \ \ \ \ \ \ \ \ \ \ \ \ \ \ \ \ \ \ \ \ \ \ \ \ \ \ \ \ \ \ \ \ \ \ \ \ \
\left.
\times
[(z_i-z_j)-\zeta^{-1}(w_i-w_j)-ah]^{N_1(a,t_i,t_j)}\right\}.
\\
\tag11.11
\endalign
$$

\endproclaim

\pf By Lemma 11.1, $x_i=x_j+\zeta(y_i-y_j)+ah$ is a root of multiplicity $N_1(a,s_i,s_j)$ of 
$\det(dM_0)$, when the latter is regarded as a polynomial in the single variable $x_i$,
with coefficients in $\Z[\zeta,\zeta^{-1},h,x_1,\dotsc,x_{i-1},x_{i+1},\dotsc,x_m,
y_1,\dotsc,y_m,z_1,\dotsc,z_n,w_1,\dotsc,w_n]$. By B\'ezout's theorem, it follows that 
$$
[(x_i-x_j)-\zeta(y_i-y_j)-ah]^{N_1(a,s_i,s_j)}\ |\ \det(dM_0).\tag11.12
$$
By the form of the expressions (10.1)--(10.2) and (10.4), the effect of replacing $\zeta$ by 
$\zeta^{-1}$ in $dM_0$ is to replace each entry by its negative. Therefore (11.12) implies that
we also have
$$
[(x_i-x_j)-\zeta^{-1}(y_i-y_j)-ah]^{N_1(a,s_i,s_j)}\ |\ \det(dM_0).
$$

To obtain the divisibility by the factors on the second line of (11.11), note that Lemma 11.1 admits
a perfect analog for column combinations. The proof of this analog is identical to the part of the proof
of Lemma 11.1 that pertains to verifying that combination (11.2) is vanishing along the columns of the
$A$-part of $M_0$. This is due to the symmetry of formula (10.1) under interchanging $k$ and $l$.~\epf

\medskip
\mysec{12. Divisibility of $\det(dM_0)$ by the powers of $(x_i-z_j)-\zeta^{\pm1}(y_i-w_j)$}

\medskip
\proclaim{Proposition 12.1} Regarded as a polynomial with coefficients in $\Z[\zeta,\zeta^{-1}]$ 
in the variables $h$, $x_1,\dotsc,x_m,y_1,\dotsc,y_m,z_1,\dotsc,z_n,w_1,\dotsc,w_n$, 
$\det(dM_0)$ is divisible by
$$
\spreadlines{1\jot}
\align
\prod_{i=1}^{m}\prod_{j=1}^{n}\prod_{a=0}^{s_i+t_j-2}
[(x_i-z_j)-\zeta&(y_i-w_j)-ah]^{2S-N_2(a,s_i,t_j)}
\\
\times
&[(x_i-z_j)-\zeta^{-1}(y_i-w_j)-ah]^{2S-N_2(a,s_i,t_j)}.
\endalign
$$

\endproclaim

Our proof is based on finding suitable sets of independent linear 
combinations of the rows and columns of $dM_0$ that vanish when 
$(x_i-z_j)-\zeta^{\pm1}(y_i-w_j)-ah=0$. These are presented in
the following four preliminary results.

\proclaim{Lemma 12.2} Each column of $dM_0$ outside of block-column $j$ becomes zero when
$(x_i-z_j)-\zeta^{\pm1}(y_i-w_j)-ah=0$.

\endproclaim

\pf This holds on account of the multiplicative factor $d$---outside block-column $j$ the
factors $(x_i-z_j)-\zeta^{\pm1}(y_i-w_j)-ah$ of $d$ do not get canceled by factors in the 
denominators of expressions~(10.1)--(10.2). \epf

\proclaim{Lemma 12.3} Let $a\in\{0,1,\dotsc,s_i+t_j-2\}$ be fixed, and assume
$(x_i-z_j)-\zeta^{\pm1}(y_i-w_j)-ah=0$. Then the column combination
$$
[{\bold 0}_{2t_1},\dotsc,{\bold 0}_{2t_{j-1}},{\bold 0}_{2l},1,-\zeta^{\pm2},{\bold 0}_{2t_j-2l-2},
{\bold 0}_{2t_{j+1}},\dotsc,{\bold 0}_{2t_n},{\bold 0}_{2S-2T}]
$$
in $dM_0$ is vanishing, for $l=0,\dotsc,t_j-1$.

\endproclaim

\pf The vanishing of the given column combination along row $2r+1$ of block-row $i$ of 
$dM_0$ amounts to
$$
d^{ij}A^{11}(r,l,z_j-x_i,w_j-y_i)-\zeta^{\pm2}d^{ij}A^{12}(r,l,z_j-x_i,w_j-y_i)=0,\tag12.1
$$
where 
$$
d^{ij}=\prod_{a=0}^{s_i+t_j-2}
[(x_i-z_j)-\zeta(y_i-w_j)-ah][(x_i-z_j)-\zeta^{-1}(y_i-w_j)-ah]\tag12.2
$$
is the part of $d$ involving precisely indices $i$ and $j$. By definition (10.1), (12.1) 
is equivalent to
$$
\spreadlines{2\jot}
\align
&
{r+l \choose r}
\\
&
\times
\left[
\zeta^{-1}(1-\zeta^{\pm2}\zeta^{-2})
\frac{((z_j-x_i)-\zeta(w_j-y_i))_{s_i+t_j-1,h}((z_j-x_i)
-\zeta^{-1}(w_j-y_i))_{s_i+t_j-1,h}}
{((z_j-x_i)-\zeta(w_j-y_i))_{r+l+1,h}}
\right.
\\
&\ \ \ \ \ 
-
\left.
\zeta(1-\zeta^{\pm2}\zeta^2)
\frac{((z_j-x_i)-\zeta(w_j-y_i))_{s_i+t_j-1,h}((z_j-x_i)-\zeta^{-1}(w_j-y_i))_{s_i+t_j-1,h}}
{((z_j-x_i)-\zeta^{-1}(w_j-y_i))_{r+l+1,h}}
\right]
\\
&\ \ \ \ \ \ \ \ \ 
=0.
\\
\tag12.3
\endalign
$$
Suppose $+2$ is chosen at the exponent in (12.3). Then the first term in the square brackets is 
clearly 0. The second term is 0 due to the factor of  $(z_j-x_i)-\zeta(w_j-y_i)+ah$ in the numerator,
which clearly does not get canceled by any factor of the denominator. The case when $-2$ is chosen 
at the exponent in (12.3) follows similarly.

The vanishing of the indicated column combination along row $2r+2$ of block-row $i$ of 
$dM_0$ amounts to
$$
d^{ij}A^{21}(r,l,z_j-x_i,w_j-y_i)-\zeta^{\pm2}d^{ij}A^{11}(r,l,z_j-x_i,w_j-y_i)=0.
$$
This is obtained from (12.1) by dividing by $-\zeta^{\pm2}$. \epf

\proclaim{Lemma 12.4} Let $a=s_i+k$, $k\in\{0,\dotsc,t_j-2\}$, be fixed.
Then when $(x_i-z_j)-\zeta^{\pm1}(y_i-w_j)-ah=0$, columns $2,4,6,\dotsc,2k+2$ of 
block-column $j$ of $dM_0$ become identically zero.

\endproclaim

\pf For $0\leq l \leq k$ and $0\leq r\leq s_i-1$, the entry in column $2l+2$ of block-column $j$ of 
$dM_0$ that is situated in row $2r+1$ of block-row $i$ is
$$
d^{ij}A^{12}(r,l,z_j-x_i,w_j-y_i),
$$
where $d^{ij}$ is given by (12.2). By (10.1) and (12.2) the explicit value of this is
$$
\spreadlines{2\jot}
\align
&
{r+l \choose r}
\left[
\zeta^{-3}
\frac{((z_j-x_i)-\zeta(w_j-y_i))_{s_i+t_j-1,h}((z_j-x_i)-\zeta^{-1}(w_j-y_i))_{s_i+t_j-1,h}}
{((z_j-x_i)-\zeta(w_j-y_i))_{r+l+1,h}}
\right.
\\
&\ \ \ \ \ \ \ \ \ \ \ \ \ \ \ 
-
\left.
\zeta^3
\frac{((z_j-x_i)-\zeta(w_j-y_i))_{s_i+t_j-1,h}((z_j-x_i)-\zeta^{-1}(w_j-y_i))_{s_i+t_j-1,h}}
{((z_j-x_i)-\zeta^{-1}(w_j-y_i))_{r+l+1,h}}
\right].
\endalign
$$
The factors $(x_i-z_j)-\zeta^{\pm1}(y_i-w_j)-ah$ in the numerators are not canceled by any of the 
factors in the denominators, as the latter are of the form $(x_i-z_j)-\zeta^{\pm1}(y_i-w_j)-bh$
with $0\leq b\leq r+l < s_i+k=a$. Thus the expression above is 0 for 
$(x_i-z_j)-\zeta^{\pm1}(y_i-w_j)-ah=0$.

The entry in column $2l+2$ of block-column $j$ of $dM_0$ which is situated in row $2r+2$ of 
block-row $i$ is seen to be 0 by the same argument. \epf

\proclaim{Lemma 12.5} Let $a\in\{0,\dotsc,t_j-2\}$ be fixed. For $0\leq k\leq t_j-2-a$, consider
the vectors
$$
{\bold u}_k=\frac{(-1)^k(t_j-1)!}{k!(t_j-2-a-k)!}
\left[
\left(0,\frac{(-h)^\alpha}{(a-\alpha)!(a-\alpha+1+k)}\right)_{0 \leq \alpha \leq a}
\right]
$$
and
$$
{\bold v}_k=
\left[
{\bold 0}_{2k},0,\frac{(t_j-1)!}{(t_j-2-a-k)!}(-h)^{1+a+k},{\bold 0}_{2(t_j-2-a-k)}
\right].
$$
Then if $(x_i-z_j)-\zeta^{\pm1}(y_i-w_j)-ah=0$, the column combinations of $dM_0$ with coefficients
$$
[{\bold 0}_{2t_1},\dotsc,{\bold 0}_{2t_{j-1}},{\bold u}_k,{\bold v}_k,
{\bold 0}_{2t_{j+1}},\dotsc,{\bold 0}_{2t_{n}}]\tag12.4
$$
are vanishing for all $k=0,1,\dotsc,t_j-2-a$.

\endproclaim

\pf The fact that the indicated column combination is vanishing along row $2r+1$ in block-row $i$ is
equivalent to
$$
\spreadlines{2\jot}
\align
\frac{(-1)^k(t_j-1)!}{k!(t_j-2-a-k)!}
\sum_{\alpha=0}^a &\frac{(-h)^\alpha}{(a-\alpha)!(k+1+a-\alpha)}dA^{12}(r,\alpha,z_j-x_i,w_j-y_i)
\\
&\!\!\!\!
=
-\frac{(t_j-1)!}{(t_j-2-a-k)!}(-h)^{a+k+1}dA^{12}(r,a+k+1,z_j-x_i,w_j-y_i),\tag12.5
\endalign
$$
provided $(x_i-z_j)-\zeta^{\pm1}(y_i-w_j)-ah=0$. 

Assume $(x_i-z_j)-\zeta(y_i-w_j)-ah=0$.
Using (10.1), (10.4), the fact that $(a)_{n,h}/(a)_{k,h}=(a+kh)_{n-k,h}$ and dividing through by
all the factors of $d$ not present in (12.2), the above equality becomes
$$
\spreadlines{3\jot}
\align
&
\frac{(-1)^k(t_j-1)!}{k!(t_j-2-a-k)!}
\sum_{\alpha=0}^a \frac{(-h)^\alpha}{(a-\alpha)!(k+1+a-\alpha)}{r+\alpha \choose \alpha}
\\
&\ \ \ \ \ \ \ \ \ \ \ 
\times
\left[
\zeta^{-3}(-ah+(r+\alpha+1)h)_{s_i+t_j-r-\alpha,h}(u-v\zeta^{-1})_{s_i+t_j+1,h}
\right.
\\
&\ \ \ \ \ \ \ \ \ \ \ \ \ \ \ \ \ \ \ \ \ \ \ \ \ \ \ \ \ \ \ \ \ \ \ \ \ \
\left.
-
\zeta^{3}(-ah)_{s_i+t_j+1,h}(u-v\zeta^{-1}+(r+\alpha+1)h)_{s_i+t_j-r-\alpha,h}
\right]
\\
&
=
-\frac{(t_j-1)!}{(t_j-2-a-k)!}(-h)^{a+k+1}{r+a+k+1 \choose r}
\\
&\ \ \ \ \ \ \ \ \ \ \ 
\times
\left[
\zeta^{-3}(-ah+(r+a+k+2)h)_{s_i+t_j-r-a-k-1,h}(u-v\zeta^{-1})_{s_i+t_j+1,h}
\right.
\\
&\ \ \ \ \ \ \ \ \ \ \ \ \ \ \ \ \ \ \ \ \ \ \ \ \ \ \ \
\left.
-
\zeta^{3}(-ah)_{s_i+t_j+1,h}(u-v\zeta^{-1}+(r+a+k+2)h)_{s_i+t_j-r-a-k-1,h}
\right],
\endalign
$$
where $u=z_j-x_i$ and $v=w_j-y_i$. Using that $(-ah)_{s_i+t_j+1,h}=0$, the above simplifies to
$$
\spreadlines{3\jot}
\align
\sum_{\alpha=0}^a \frac{(-1)^\alpha}{(a-\alpha)!(k+1+a-\alpha)}&{r+\alpha \choose \alpha}
(-a+r+\alpha+1)_{s_i+t_j-r-\alpha}
\\
&
=
(-1)^{a}k!{r+a+k+1 \choose r}(r+k+2)_{s_i+t_j-r-a-k-1}.\tag12.6
\endalign
$$
Write for notational simplicity $s=s_i+t_j$. Then after some manipulation, the left hand side of 
(12.6) becomes
$$
-\frac{(s-a)!}{a!\Gamma(r+1)(-a-k-1)}\sum_{\alpha=0}^a
\frac{(-a)_\alpha(-a-k-1)_\alpha(r+1)_\alpha}{\alpha!(-a-k)_\alpha}
\frac{(r+1-a)_a}{(r+1-a)_\alpha},
$$
an expression that is well-defined for any $0\leq a,k\in\Z$, $a\leq s\in\Z$ and $r\in\C$. 
Furthermore, when $r\notin\Z$, the above expression can be written as
$$
\spreadlines{3\jot}
\align
-\frac{(s-a)!(r+1-a)_a}{a!\Gamma(r+1)(-a-k-1)}&\sum_{\alpha=0}^a
\frac{(-a)_\alpha(-a-k-1)_\alpha(r+1)_\alpha}{\alpha!(-a-k)_\alpha(r+1-a)_\alpha}
\\
&
=
-\frac{(s-a)!(r+1-a)_a}{a!\Gamma(r+1)(-a-k-1)}
{}_3 F_2\!\!\left[\matrix -a,-a-k-1,r+1\\ r+1-a,-a-k\endmatrix;1\right].
\\
\tag12.7
\endalign
$$
The ${}_3 F_2$ above can be evaluated by choosing $n=a$, $x=-a-k-1$, $y=r+1$, $z=r+1-a$ in the 
Pfaff-Saalsch\"utz summation formula (found e.g. in \cite{\Sl, (2.3.1.3),\,Appendix (III.2)})
$$
{}_3 F_2\!\!\left[\matrix -n,x,y\\ z,1+x+y-z-n\endmatrix;1\right]
=
\frac{(z-x)_n(z-y)_n}{(z)_n(z-x-y)_n}.
$$
Thus, by (12.7) we obtain that for $r\notin\Z$ the left hand side of (12.6) equals
$$
-\frac{(s-a)!(r+1-a)_a}{a!\Gamma(r+1)(-a-k-1)}\frac{(r+k+2)_a(-a)_a}{(r+1-a)_a(k+1)_a}.
$$
However, this is readily seen to agree with the right hand side of (12.6) (which, by writing 
$(r+n)_{m-r}=\Gamma(m+n)/\Gamma(r+n)$, is defined for any $r\in\C\setminus\Z$, $m,n\in\Z$).
This proves equality (12.6) for $r\in\C\setminus\Z$. Since the two sides
of (12.6) are analytic functions of $r$ that coincide on $\C\setminus\Z$, they must be identical.
This proves (12.6) and therefore (12.5). 

The vanishing of column combination (12.4) along row $2r+2$ in block-row $i$ is
equivalent to the equality obtained from (12.5) by replacing the $A^{12}$'s by $A^{11}$'s. This
is readily seen to be equivalent to (12.6) as well (the only difference from the case of $A^{12}$'s
is that the factors $\zeta^{-3}$ and $\zeta^{3}$ in the formula displayed above (12.6) get replaced
by $\zeta^{-1}$ and $\zeta$, respectively). This completes the proof of the statement of the lemma
for the case $(x_i-z_j)-\zeta(y_i-w_j)-ah=0$.

The case when $(x_i-z_j)-\zeta^{-1}(y_i-w_j)-ah=0$ follows by replacing $\zeta$ by $\zeta^{-1}$
in the case we proved above, and using the fact that the right hand side of (10.1) simply
changes sign under this substitution. \epf

{\it Proof of Proposition 12.1.} We treat first the case $s_i\geq t_j$.

For each $a\in\{0,\dotsc,s_i+t_j-2\}$, Lemmas 12.2 and 12.3 provide together 
a set of $2S-2t_j+t_j=2S-t_j$ 
independent column combinations that are vanishing when $(x_i-z_j)-\zeta^{\pm1}(y_i-w_j)-ah=0$. To
complete the proof it suffices to enlarge this set by $t_j-N_2(a,s_i,t_j)$ more vanishing column 
combinations, so that the enlarged set of column combinations is linearly independent.

If $0\leq a\leq t_j-2$, Lemma 12.5 provides $t_j-1-a$ such additional column combinations. However, 
it follows by definition (10.6) that $N_2(a,s_i,t_j)=a+1$ for $0\leq a\leq t_j-2$. 
Therefore these can be taken as the additional $t_j-N_2(a,s_i,t_j)$ column combinations that we need.

For $t_j-1\leq a\leq s_i-1$, (10.6) implies that $N_2(a,s_i,t_j)=t_j$, so no vanishing column
combinations are needed in addition to the ones supplied by Lemmas 12.2 and 12.3.

If on the other hand $s_i\leq a\leq s_i+t_j-2$, then $N_2(a,s_i,t_j)=s_i+t_j-a-1$, and the needed
additional vanishing column combinations are provided by Lemma 12.4. 

Similar arguments, based on the counterparts of Lemmas 12.2--12.5 presented below, imply the
same conclusion if $s_i\leq t_j$. \epf

\proclaim{Lemma 12.2'} Each row of $dM_0$ outside of block-row $i$ becomes zero when
$(x_i-z_j)-\zeta^{\pm1}(y_i-w_j)-ah=0$.

\endproclaim

\pf This follows by the same argument as Lemma 12.2 \epf

\proclaim{Lemma 12.3'} Let $a\in\{0,1,\dotsc,s_i+t_j-2\}$ be fixed, and assume
$(x_i-z_j)-\zeta^{\pm1}(y_i-w_j)-ah=0$. Then the row combination
$$
[{\bold 0}_{2s_1},\dotsc,{\bold 0}_{2s_{i-1}},{\bold 0}_{2r},1,-\zeta^{\pm2},{\bold 0}_{2s_i-2r-2},
{\bold 0}_{2s_{i+1}},\dotsc,{\bold 0}_{2s_m}]
$$
in $dM_0$ is vanishing, for $r=0,\dotsc,s_i-1$.

\endproclaim

\pf The vanishing of the given column combination along column $2l+1$ of block-row $i$ 
of $dM_0$ amounts to
$$
d^{ij}A^{11}(r,l,z_j-x_i,w_j-y_i)-\zeta^{\pm2}d^{ij}A^{21}(r,l,z_j-x_i,w_j-y_i)=0.
$$
with $d^{ij}$ given by (12.2). This follows by the argument used in the proof of Lemma 12.3. The case
of column $2l+2$ follows by dividing the above equality by $\zeta^{\pm2}$.

In addition, we need to verify the vanishing of the indicated row combination for the $B$-part of
$dM_0$. But this holds trivially since the entire $B$-part of $dM_0$ is zero when 
$(x_i-z_j)-\zeta^{\pm1}(y_i-w_j)-ah=0$. \epf

\proclaim{Lemma 12.4'} Let $a=t_j+l$, $l\in\{0,\dotsc,s_i-2\}$, be fixed.
Then when $(x_i-z_j)-\zeta^{\pm1}(y_i-w_j)-ah=0$, rows $2,4,6,\dotsc,2l+2$ of 
block-row $i$ of $dM_0$ become identically zero.

\endproclaim

\pf This follows by an argument perfectly analogous to the one used in the proof of Lemma~12.4. \epf

\proclaim{Lemma 12.5'} Let $a\in\{0,\dotsc,s_i-2\}$ be fixed. For $0\leq k\leq s_i-2-a$, consider
the vectors
$$
{\bold u}_k=\frac{(-1)^k(s_i-1)!}{k!(s_i-2-a-k)!}
\left[
\left(0,\frac{(-h)^\alpha}{(a-\alpha)!(a-\alpha+1+k)}\right)_{0 \leq \alpha \leq a}
\right]
$$
and
$$
{\bold v}_k=
\left[
{\bold 0}_{2k},0,\frac{(s_i-1)!}{(s_i-2-a-k)!}(-h)^{1+a+k},{\bold 0}_{2(s_i-2-a-k)}
\right].
$$
Then if $(x_i-z_j)-\zeta^{\pm1}(y_i-w_j)-ah=0$, the row combinations of $dM_0$ with coefficients
$$
[{\bold 0}_{2s_1},\dotsc,{\bold 0}_{2s_{i-1}},{\bold u}_k,{\bold v}_k,
{\bold 0}_{2s_{i+1}},\dotsc,{\bold 0}_{2s_{m}}]
$$
are vanishing for all $k=0,1,\dotsc,s_i-2-a$.

\endproclaim

\pf The vanishing of the indicated row combination on the $A$-part of $dM_0$ follows by the 
calculations in the proof of Lemma 12.5. Along the $B$ part of $dM_0$ this holds trivially, 
as explained in the proof of Lemma 12.3'.\epf

%

\mysec{13. The proofs of Theorem 2.1 and Proposition 2.2}

We are now ready to present the proofs of the two results from which we deduced Theorem~1.2 in
Section 2.

{\it Proof of Theorem 2.1.} By reflecting the lattice and the holes about a vertical lattice line
if necessary, we may assume $S=\sum_{k=1}^m s_k\geq \sum_{l=1}^n t_l = T$.

The statement follows then directly by Proposition 5.1, Proposition 5.3 and
Theorem 8.1, using that for $\zeta=e^{2\pi i/3}$ one has
$$
\zeta^{-2}-\zeta^2=i\sqrt{3}\tag13.1
$$
and
$$
(u-\zeta v)(u-\zeta^{-1}v)=u^2+uv+v^2.\tag13.2
$$
\epf

{\it Proof of Proposition 2.2.} By Proposition 5.1 we have
$$
\hat{\omega}(E(a_1,qa_1),\dotsc,E(a_s,qa_s))=|\det(M_1)|,\tag13.3
$$
where $M_1$ is the matrix
$$
\left[\matrix
{\scriptstyle U_0(a_1,qa_1)}&\!\!{\scriptstyle U_0(a_1-1,qa_1+1)}&\cdots&
{\scriptstyle U_{s-1}(a_1,qa_1)}&\!\!{\scriptstyle U_{s-1}(a_1-1,qa_1+1)}\\
{\scriptstyle U_0(a_1+1,qa_1-1)}&\!\!{\scriptstyle U_0(a_1,qa_1)}&\cdots&
{\scriptstyle U_{s-1}(a_1+1,qa_1-1)}&\!\!{\scriptstyle U_{s-1}(a_1,qa_1)}\\
\cdot&\!\!\cdot&\cdots&\cdot&\!\!\cdot\\
\cdot&\!\!\cdot&\cdots&\cdot&\!\!\cdot\\
\cdot&\!\!\cdot&\cdots&\cdot&\!\!\cdot\\
{\scriptstyle U_0(a_s,qa_s)}&\!\!{\scriptstyle U_0(a_s-1,qa_s+1)}&\cdots&
{\scriptstyle U_{s-1}(a_s,qa_s)}&\!\!{\scriptstyle U_{s-1}(a_s-1,qa_s+1)}\\
{\scriptstyle U_0(a_s+1,qa_s-1)}&\!\!{\scriptstyle U_0(a_s,qa_s)}&\cdots&
{\scriptstyle U_{s-1}(a_s+1,qa_s-1)}&\!\!{\scriptstyle U_{s-1}(a_s,qa_s)}
\endmatrix\right].
$$
Apply the determinant-preserving operator (5.4) to $M_1$ twice, first along the rows of odd index
and then along the rows of even index. By the arguments in the proof of Proposition 5.3 we obtain
that
$$
\det(M_1)=\prod_{1\leq k<l\leq s}(a_k-a_l)^2\det(M_2),\tag13.4
$$
where $M_2$ is the matrix
$$
\left[\matrix
{\scriptscriptstyle {\Cal D}^0U_0(a_1,qa_1)}&{\scriptscriptstyle \!\!{\Cal D}^0U_0(a_1-1,qa_1+1)}&\cdots&
{\scriptscriptstyle {\Cal D}^0U_{s-1}(a_1,qa_1)}&{\scriptscriptstyle \!\!{\Cal D}^0U_{s-1}(a_1-1,qa_1+1)}\\
{\scriptscriptstyle {\Cal D}^0U_0(a_1+1,qa_1-1)}&{\scriptscriptstyle \!\!{\Cal D}^0U_0(a_1,qa_1)}&\cdots&
{\scriptscriptstyle {\Cal D}^0U_{s-1}(a_1+1,qa_1-1)}&{\scriptscriptstyle \!\!{\Cal D}^0U_{s-1}(a_1,qa_1)}\\
\cdot&\!\!\cdot&\cdots&\cdot&\!\!\cdot\\
\cdot&\!\!\cdot&\cdots&\cdot&\!\!\cdot\\
\cdot&\!\!\cdot&\cdots&\cdot&\!\!\cdot\\
{\scriptscriptstyle {\Cal D}^{s-1}U_0(a_s,qa_s)}&{\scriptscriptstyle \!\!{\Cal D}^{s-1}U_0(a_s-1,qa_s+1)}&\cdots&
{\scriptscriptstyle {\Cal D}^{s-1}U_{s-1}(a_s,qa_s)}&{\scriptscriptstyle \!\!{\Cal D}^{s-1}U_{s-1}(a_s-1,qa_s+1)}\\
{\scriptscriptstyle {\Cal D}^{s-1}U_0(a_s+1,qa_s-1)}&{\scriptscriptstyle \!\!{\Cal D}^{s-1}U_0(a_s,qa_s)}&\cdots&
{\scriptscriptstyle {\Cal D}^{s-1}U_{s-1}(a_s+1,qa_s-1)}&{\scriptscriptstyle \!\!{\Cal D}^{s-1}U_{s-1}(a_s,qa_s)}
\endmatrix\right]\!\!.
$$
We have seen in the proof of Proposition 6.1 that
$$
\spreadlines{2\jot}
\align
U_l(a,b)=&-\frac{i}{2\pi}[\zeta^{a-b-1}(a-b\zeta)^l-\zeta^{-a+b+1}(a-b\zeta^{-1})^l]\\
&+\ {\text{\rm monomials in $a$ and $b$ of joint degree $<l$}}.
\endalign
$$
This implies that, viewed as a block-matrix consisting of $2\times 2$ blocks, $M_2$ is 
block-upper-triangular. Furthermore, using Lemma 6.4 and the assumption $3|1-q$, the $(k+1)$-st diagonal 
$2\times 2$ block of $M_2$ is seen to be
$$
\spreadmatrixlines{2\jot}
-\frac{i}{2\pi}
\left[\matrix
\zeta^{-1}(1-q\zeta)^k-\zeta(1-q\zeta^{-1})^k&\zeta^{-3}(1-q\zeta)^k-\zeta^3(1-q\zeta^{-1})^k\\
\zeta(1-q\zeta)^k-\zeta^{-1}(1-q\zeta^{-1})^k&\zeta^{-1}(1-q\zeta)^k-\zeta(1-q\zeta^{-1})^k
\endmatrix\right].
$$
Since the determinant of this block equals
$$
-\frac{1}{4\pi^2}(\zeta^{-2}-\zeta^2)^2(q-\zeta)^k(q-\zeta^{-1})^k,
$$
equality (2.2) follows by (13.3) and (13.4), using (13.1)--(13.2). \epf

\mysec{14. The case of arbitrary slopes}

Let us now consider the pure linear multiholes of not necessarily equal slopes 
$E^{q_1}_{{\bold a}_1},\dotsc,E^{q_m}_{{\bold a}_m}$ and 
$W^{q'_1}_{{\bold b}_1},\dotsc,W^{q'_n}_{{\bold b}_n}$. Then their correlation is still given by 
Lemma 5.1, but with the following change in the definition of the block matrices $C_j$ and $D$ of matrix $M$:
 Modify the blocks of (5.2) so that the top left entries of the $2\times2$ matrices
change from
$$
\!\!\!\!\!\!
\spreadmatrixlines{3\jot}
\matrix
{\scriptscriptstyle P(-R(z_j-x_i)+a_{i1}-b_{j1}-1,-R(w_j-y_i)+q(a_{i1}-b_{j1})-1)}\!\!\!\!\!\!&&
{\scriptstyle \cdots}\!\!\!\!\!\!&&
{\scriptscriptstyle P(-R(z_j-x_i)+a_{i1}-b_{j,t_j}-1,-R(w_j-y_i)+q(a_{i1}-b_{j,t_j})-1)}&&
\\
{\scriptscriptstyle P(-R(z_j-x_i)+a_{i2}-b_{j1}-1,-R(w_j-y_i)+q(a_{i2}-b_{j1})-1)}\!\!\!\!\!\!&&
{\scriptstyle \cdots}\!\!\!\!\!\!&&
{\scriptscriptstyle P(-R(z_j-x_i)+a_{i2}-b_{j,t_j}-1,-R(w_j-y_i)+q(a_{i2}-b_{j,t_j})-1)}&&
\\
{\scriptstyle \vdots}\!\!\!\!\!\!&&\ \!\!\!\!\!\!&&{\scriptstyle \vdots}
\\
{\scriptscriptstyle P(-R(z_j-x_i)+a_{i,s_i}-b_{j1}-1,-R(w_j-y_i)+q(a_{i,s_i}-b_{j1})-1)}\!\!\!\!\!\!&&
{\scriptstyle \cdots}\!\!\!\!\!\!&&
{\scriptscriptstyle P(-R(z_j-x_i)+a_{i,s_i}-b_{j,t_j}-1,-R(w_j-y_i)+q(a_{i,s_i}-b_{j,t_j})-1)}&&
\endmatrix
$$
to
$$
\!\!\!\!\!\!\!\!\!
\spreadmatrixlines{3\jot}
\matrix
{\scriptscriptstyle P(-R(z_j-x_i)+a_{i1}-b_{j1}-1,-R(w_j-y_i)+q_ia_{i1}-q'_jb_{j1}-1)}\!\!\!\!\!\!&&
{\scriptstyle \cdots}\!\!\!\!\!\!&&
{\scriptscriptstyle P(-R(z_j-x_i)+a_{i1}-b_{j,t_j}-1,-R(w_j-y_i)+q_ia_{i1}-q'_jb_{j,t_j}-1)}&&
\\
{\scriptscriptstyle P(-R(z_j-x_i)+a_{i2}-b_{j1}-1,-R(w_j-y_i)+q_ia_{i2}-q'_jb_{j1}-1)}\!\!\!\!\!\!&&
{\scriptstyle \cdots}\!\!\!\!\!\!&&
{\scriptscriptstyle P(-R(z_j-x_i)+a_{i2}-b_{j,t_j}-1,-R(w_j-y_i)+q_ia_{i2}-q'_jb_{j,t_j}-1)}&&
\\
{\scriptstyle \vdots}\!\!\!\!\!\!&&\ \!\!\!\!\!\!&&{\scriptstyle \vdots}
\\
{\scriptscriptstyle P(-R(z_j-x_i)+a_{i,s_i}-b_{j1}-1,-R(w_j-y_i)+q_ia_{i,s_i}-q'_jb_{j1}-1)}\!\!\!\!\!\!&&
{\scriptstyle \cdots}\!\!\!\!\!\!&&
{\scriptscriptstyle P(-R(z_j-x_i)+a_{i,s_i}-b_{j,t_j}-1,-R(w_j-y_i)+q_ia_{i,s_i}-q'_jb_{j,t_j}-1)}&&
\endmatrix
$$
and the blocks of (5.3) so that the top left entries of the $2\times2$ matrices change from
$$
\spreadmatrixlines{3\jot}
\matrix
{\scriptscriptstyle U_0(Rx_i+a_{i1},Ry_j+qa_{i1})}&&
{\scriptstyle \cdots}&&
{\scriptscriptstyle U_{S-T-1}(Rx_i+a_{i1},Ry_j+qa_{i1})}&&
\\
{\scriptscriptstyle U_0(Rx_i+a_{i2},Ry_j+qa_{i2})}&&
{\scriptstyle \cdots}&&
{\scriptscriptstyle U_{S-T-1}(Rx_i+a_{i2},Ry_j+qa_{i2})}&&
\\
{\scriptstyle \vdots}&&\ &&{\scriptstyle \vdots}
\\
{\scriptscriptstyle U_0(Rx_i+a_{i,s_i},Ry_j+qa_{i,s_i})}&&
{\scriptstyle \cdots}&&
{\scriptscriptstyle U_{S-T-1}(Rx_i+a_{i,s_i},Ry_j+qa_{i,s_i})}
\endmatrix
$$
to
$$
\spreadmatrixlines{3\jot}
\matrix
{\scriptscriptstyle U_0(Rx_i+a_{i1},Ry_j+q_ia_{i1})}&&
{\scriptstyle \cdots}&&
{\scriptscriptstyle U_{S-T-1}(Rx_i+a_{i1},Ry_j+q_ia_{i1})}&&
\\
{\scriptscriptstyle U_0(Rx_i+a_{i2},Ry_j+q_ia_{i2})}&&
{\scriptstyle \cdots}&&
{\scriptscriptstyle U_{S-T-1}(Rx_i+a_{i2},Ry_j+q_ia_{i2})}&&
\\
{\scriptstyle \vdots}&&\ &&{\scriptstyle \vdots}
\\
{\scriptscriptstyle U_0(Rx_i+a_{i,s_i},Ry_j+q_ia_{i,s_i})}&&
{\scriptstyle \cdots}&&
{\scriptscriptstyle U_{S-T-1}(Rx_i+a_{i,s_i},Ry_j+q_ia_{i,s_i})}
\endmatrix
$$
It is apparent from the above that in this more general version of the matrix $M$ we have the same 
$2m+2n$ opportunities to apply operation (5.4) (or its analog for columns) as for the original matrix $M$
of Section 5, obtained from the above when $q_1=\cdots=q'_n=q$. Indeed, 
the arguments of Section 5 apply in this more general case of arbitrary slopes, and give that
the determinant of the above modification of the matrix $M$ is equal to the determinant of the deformation
of the matrix $M'$ of (5.6)--(5.9) in which the quantities $qa_{ik}$ are replaced by $q_ia_{ik}$'s and the
quantities $qb_{jl}$ by $q'_jb_{jl}$'s. The asymptotics of the thus modified entries in (5.9) is given by
Proposition 6.1. For the modified entries of (5.6) we need the following slight extension of Proposition 7.1,
whose proof follows by the arguments of Section 7.

\proclaim{Proposition 7.1'} 
Let $q$ and $q'$ be rational numbers so that $3|1-q$ and $3|1-q'$.
Then for any fixed $c,d\in\Z$, $u,v\in3\Z$, $(u,v)\neq(0,0)$, and $0\leq k,l\in\Z$, we have
$$
\spreadlines{1\jot}
\align
&
{\Cal D}_y^l\left.\left\{{\Cal D}_x^k P(-Ru+x+y+c,-Rv+qx+q'y+d)|_{x=a_1}\right\}\right|_{y=b_1}=
\\
&\ \ 
-\frac{i}{2\pi}{k+l \choose l}\left[\zeta^{c-d-1}\frac{(1-q\zeta)^{k}(1-q'\zeta)^{l}}{(u-v\zeta)^{k+l+1}}
\right.
\\
&\ \ \ \ \ \ \ \ \ \ \ \ \ \ \ \ \ \ \ \ \ \ \ \ \ \ \ \ \ \ \ \ \ 
\left. 
-\zeta^{-c+d+1}\frac{(1-q\zeta^{-1})^{k}(1-q'\zeta^{-1})^{l}}{(u-v\zeta^{-1})^{k+l+1}}\right]\frac{1}{R^{k+l+1}}
+O(R^{-k-l-2}).
\endalign
$$

\endproclaim

The arguments that proved Proposition 5.3 yield then that
$$
\spreadlines{3\jot}
\align
\!\!\!\!\!\!\!\!\!\!\!\!\!\!\!\!\!\!
&
\hat\omega(E_{{\bold a}_1}^{q_1}(Rx_1,Ry_1),\dotsc,E_{{\bold a}_m}^{q_m}(Rx_m,Ry_m),
W_{{\bold b}_1}^{q'_1}(Rz_1,Rw_1),\dotsc,W_{{\bold b}_n}^{q'_n}(Rz_n,Rw_n))
\\
&\ \ \ \ \ \ \ \ \ \ \ \ 
=
\left(\frac{1}{2\pi}\right)^{2S}\prod_{i=1}^m \ \prod_{1\leq j<k\leq s_i} (a_{ij}-a_{ik})^2
\prod_{i=1}^n \ \prod_{1\leq j<k\leq t_i} (b_{ij}-b_{ik})^2 
\\
&\ \ \ \ \ \ \ \ \ \ \ \ \ \ \ \ \ \ 
\times
\left|\det(\bar{M}'')\right|\,\,
R^{2\{\sum_{1\leq i<j\leq m}s_is_j+\sum_{1\leq i<j\leq n}t_it_j-\sum_{i=1}^m\sum_{j=1}^ns_it_j \}}
\\
&\ \ \ \ \ \ \ \ \ \ \ \ \ \ \ \ \ \ 
+O(R^{2\{\sum_{1\leq i<j\leq m}s_is_j+\sum_{1\leq i<j\leq n}t_it_j-\sum_{i=1}^m\sum_{j=1}^ns_it_j\} 
-1}),\tag14.1
\endalign
$$
where
$$
\bar{M}''=\left[\matrix \bar{C}''&&\bar{D}''\endmatrix\right],\tag14.2
$$
$$
\bar{C}''=\left[\matrix
\ &&\vdots &&\ 
\\
\\
\cdots&&C''_{ij}&&\cdots
\\
\\
\ &&\vdots &&\ 
\endmatrix\right],\tag14.3
$$
$$
\spreadlines{3\jot}
\spreadmatrixlines{2\jot}
\align
&
C''_{ij}=
\\
&\!\!\!\!\!\!\!\!\!\!
\left[\matrix 
{\scriptscriptstyle A_{ij}^{11}(0,0,z_j-x_i,w_j-y_i)}\!\!\!\!\!\!\!\!&&
{\scriptscriptstyle A_{ij}^{12}(0,0,z_j-x_i,w_j-y_i)}\!\!\!\!\!\!&&
{\scriptstyle \cdots}\!\!\!\!\!\!&&
{\scriptscriptstyle A_{ij}^{11}(0,t_j-1,z_j-x_i,w_j-y_i)}\!\!\!\!\!\!\!\!&&
{\scriptscriptstyle A_{ij}^{12}(0,t_j-1,z_j-x_i,w_j-y_i)}
\\
{\scriptscriptstyle A_{ij}^{21}(0,0,z_j-x_i,w_j-y_i)}\!\!\!\!\!\!\!\!&&
{\scriptscriptstyle A_{ij}^{11}(0,0,z_j-x_i,w_j-y_i)}\!\!\!\!\!\!&&
{\scriptstyle \cdots}\!\!\!\!\!\!&&
{\scriptscriptstyle A_{ij}^{21}(0,t_j-1,z_j-x_i,w_j-y_i)}\!\!\!\!\!\!\!\!&&
{\scriptscriptstyle A_{ij}^{11}(0,t_j-1,z_j-x_i,w_j-y_i)}
\\
\\
{\scriptstyle \vdots}&&{\scriptstyle \vdots}&&\ &&{\scriptstyle \vdots}&&{\scriptstyle \vdots}
\\
\\
{\scriptscriptstyle A_{ij}^{11}(s_i-1,0,z_j-x_i,w_j-y_i)}\!\!\!\!\!\!\!\!&&
{\scriptscriptstyle A_{ij}^{12}(s_i-1,0,z_j-x_i,w_j-y_i)}\!\!\!\!\!\!&&
{\scriptstyle \cdots}\!\!\!\!\!\!&&
{\scriptscriptstyle A_{ij}^{11}(s_i-1,t_j-1,z_j-x_i,w_j-y_i)}\!\!\!\!\!\!\!\!&&
{\scriptscriptstyle A_{ij}^{12}(s_i-1,t_j-1,z_j-x_i,w_j-y_i)}
\\
{\scriptscriptstyle A_{ij}^{21}(s_i-1,0,z_j-x_i,w_j-y_i)}\!\!\!\!\!\!\!\!&&
{\scriptscriptstyle A_{ij}^{11}(s_i-1,0,z_j-x_i,w_j-y_i)}\!\!\!\!\!\!&&
{\scriptstyle \cdots}\!\!\!\!\!\!&&
{\scriptscriptstyle A_{ij}^{21}(s_i-1,t_j-1,z_j-x_i,w_j-y_i)}\!\!\!\!\!\!\!\!&&
{\scriptscriptstyle A_{ij}^{11}(s_i-1,t_j-1,z_j-x_i,w_j-y_i)}
\endmatrix\right],
\\
\tag14.4
\endalign
$$
$$
\bar{D}''=\left[\matrix
{\scriptstyle \vdots}
\\
\\
D_i''
\\
\\
{\scriptstyle \vdots}
\endmatrix\right],\tag14.5
$$
$$
D_i''=\left[\matrix
{\scriptscriptstyle B_{0,i}^{11}(0,x_i,y_i)}\!\!\!\!\!\!&&
{\scriptscriptstyle B_{0,i}^{12}(0,x_i,y_i)}\!\!\!\!\!\!&&
{\scriptstyle \cdots}\!\!\!\!\!\!&&
{\scriptscriptstyle B_{S-T-1,i}^{11}(0,x_i,y_i)}\!\!\!\!\!\!&&
{\scriptscriptstyle B_{S-T-1,i}^{12}(0,x_i,y_i)}
\\
{\scriptscriptstyle B_{0,i}^{21}(0,x_i,y_i)}\!\!\!\!\!\!&&
{\scriptscriptstyle B_{0,i}^{22}(0,x_i,y_i)}\!\!\!\!\!\!&&
{\scriptstyle \cdots}\!\!\!\!\!\!&&
{\scriptscriptstyle B_{S-T-1,i}^{21}(0,x_i,y_i)}\!\!\!\!\!\!&&
{\scriptscriptstyle B_{S-T-1,i}^{11}(0,x_i,y_i)}
\\
\\
{\scriptstyle \vdots}&&{\scriptstyle \vdots}&&\ &&{\scriptstyle \vdots}&&{\scriptstyle \vdots}
\\
\\
{\scriptscriptstyle B_{0,i}^{11}(s_i-1,x_i,y_i)}\!\!\!\!\!\!&&
{\scriptscriptstyle B_{0,i}^{12}(s_i-1,x_i,y_i)}\!\!\!\!\!\!&&
{\scriptstyle \cdots}\!\!\!\!\!\!&&
{\scriptscriptstyle B_{S-T-1,i}^{11}(s_i-1,x_i,y_i)}\!\!\!\!\!\!&&
{\scriptscriptstyle B_{S-T-1,i}^{12}(s_i-1,x_i,y_i)}
\\
{\scriptscriptstyle B_{0,i}^{21}(s_i-1,x_i,y_i)}\!\!\!\!\!\!&&
{\scriptscriptstyle B_{0,i}^{22}(s_i-1,x_i,y_i)}\!\!\!\!\!\!&&
{\scriptstyle \cdots}\!\!\!\!\!\!&&
{\scriptscriptstyle B_{S-T-1,i}^{21}(s_i-1,x_i,y_i)}\!\!\!\!\!\!&&
{\scriptscriptstyle B_{S-T-1,i}^{11}(s_i-1,x_i,y_i)}
\endmatrix\right],\tag14.6
$$
and
$$
\align
&
A_{ij}^{\left\{\matrix {\scriptstyle 11}\\{\scriptstyle 12}\\{\scriptstyle 21}\endmatrix\right\}}(k,l,u,v)=
{k+l\choose k}
\left[
\left\{\matrix {\scriptstyle 1}\\{\scriptstyle \zeta^{-2}}\\{\scriptstyle \zeta^{2}}\endmatrix\right\}
\zeta^{-1}\frac{(1-q_i\zeta)^k(1-q'_j\zeta)^l}{(u-v\zeta)^{k+l+1}}
\right.
\\
&\ \ \ \ \ \ \ \ \ \ \ \ \ \ \ \ \ \ \ \ \ \ \ \ \ \ \ \ \ \ \ \ \ \ \ \ \ \ \ \ \ \ \ \ \ \ 
\left.
-
\left\{\matrix {\scriptstyle 1}\\{\scriptstyle \zeta^{2}}\\{\scriptstyle \zeta^{-2}}\endmatrix\right\}
\zeta\frac{(1-q_i\zeta^{-1})^k(1-q'_j\zeta^{-1})^l}{(u-v\zeta^{-1})^{k+l+1}}
\right],\tag14.7
\endalign
$$
$$
\align
&
B_{l,i}^{\left\{\matrix {\scriptstyle 11}\\{\scriptstyle 12}\\{\scriptstyle 21}\endmatrix\right\}}(k,u,v)=
{l\choose k}
\left[
\left\{\matrix {\scriptstyle 1}\\{\scriptstyle \zeta^{-2}}\\{\scriptstyle \zeta^{2}}\endmatrix\right\}
\zeta^{-1}(1-q_i\zeta)^k(u-v\zeta)^{l-k}
\right.
\\
&\ \ \ \ \ \ \ \ \ \ \ \ \ \ \ \ \ \ \ \ \ \ \ \ \ \ \ \ \ \ \ \ \ \ \ \ \ \ \ \ \ \ \ \
\left.
-
\left\{\matrix {\scriptstyle 1}\\{\scriptstyle \zeta^{2}}\\{\scriptstyle \zeta^{-2}}\endmatrix\right\}
\zeta(1-q_i\zeta^{-1})^k(u-v\zeta^{-1})^{l-k}
\right].\tag14.8
\endalign
$$
Theorem 3.1 --- the asymptotics of the correlation in the case of arbitrary slopes --- follows by (14.1) and 
the following extension of Theorem 8.1.

\proclaim{Theorem 14.1} Let $S=\sum_{i=1}^m s_i$, $T=\sum_{j=1}^n t_j$, and assume $S\geq T$. 
Then the determinant of the matrix $\bar{M}''$ defined in $(14.2)$--$(14.8)$, with $\zeta$, $q$,
$x_i$, $y_i$, $w_j$, $z_j$, $i=1,\dotsc,m$, $j=1,\dotsc,n$ being 
indeterminates, is given by
$$
\spreadlines{2\jot}
\align
\det(\bar{M}'')=(\zeta^2-\zeta^{-2})^{2S}
&\prod_{i=1}^m[(q_i-\zeta)(q_i-\zeta^{-1})]^{s_i \choose 2}
\prod_{j=1}^n[(q'_j-\zeta)(q'_j-\zeta^{-1})]^{t_j \choose 2}
\\
\times
\prod_{1\leq i<j\leq m}&[((x_i-x_j)-\zeta(y_i-y_j))((x_i-x_j)-\zeta^{-1}(y_i-y_j))]^{s_is_j}
\\
\times
\prod_{1\leq i<j\leq n}&[((z_i-z_j)-\zeta(w_i-w_j))((z_i-z_j)-\zeta^{-1}(w_i-w_j))]^{t_it_j}
\\
\times
\prod_{i=1}^m\prod_{j=1}^n&[((x_i-z_j)-\zeta(y_i-w_j))((x_i-z_j)-\zeta^{-1}(y_i-w_j))]^{-s_it_j}.
\tag14.9
\endalign
$$

\endproclaim

\pf  Theorem 8.1 was proved in Sections 8--12. The only part of the proof affected by the generalization from
the matrix $M''$ to $\bar{M}''$ is Section 9, as we now need to show that $\det\bar{M}''$ is divisible by
$(q_i-\zeta^{\pm1})^{s_i\choose 2}$ and $(q'_j-\zeta^{\pm1})^{t_j\choose 2}$. As in Section 9, deform
the definitions (14.7) and (14.8), by introducing a new parameter $h$, to
$$
\spreadlines{3\jot}
\align
&
A_{ij}^{\left\{\matrix {\scriptstyle{11}}\\{\scriptstyle{12}}\\{\scriptstyle{21}}
\endmatrix\right\}}(k,l,u,v)={k+l \choose k}
\left[
\left\{\matrix{\scriptstyle{1}}\\{\scriptstyle{\zeta^{-2}}}\\{\scriptstyle{\zeta^2}}\endmatrix\right\}
\zeta^{-1}\frac{(-\zeta)^{k+l}(q_i-\zeta^{-1})_{k,h}(q'_j-\zeta^{-1})_{l,h}}{(u-v\zeta)^{k+l+1}}
\right.
\\
&
\ \ \ \ \ \ \ \ \ \ \ \ \ \ \ \ \ \ \ \ \ \ \ \ \ \ \ \ \ \ \ \ \ \ \ \ \
-
\left.
\left\{\matrix{\scriptstyle{1}}\\{\scriptstyle{\zeta^{2}}}\\{\scriptstyle{\zeta^{-2}}}\endmatrix\right\}
\zeta\frac{(-\zeta^{-1})^{k+l}(q_i-\zeta)_{k,h}(q'_j-\zeta)_{l,h}}{(u-v\zeta^{-1})^{k+l+1}}
\right]
\tag14.10
\endalign
$$
and
$$
\spreadlines{3\jot}
\align
&
B_{l,i}^{\left\{\matrix {\scriptstyle{11}}\\{\scriptstyle{12}}\\{\scriptstyle{21}}
\endmatrix\right\}}
(k,u,v)={l \choose k}
\left[
\left\{\matrix{\scriptstyle{1}}\\{\scriptstyle{\zeta^{-2}}}\\{\scriptstyle{\zeta^{2}}}\endmatrix\right\}
\zeta^{-1}(-\zeta)^k (q_i-\zeta^{-1})_{k,h} (u-v\zeta)^{l-k}
\right.
\\
&
\ \ \ \ \ \ \ \ \ \ \ \ \ \ \ \ \ \ \ \ \ \ \ \ \ \ \ \ \ \ \ \ \ 
-
\left.
\left\{\matrix{\scriptstyle{1}}\\{\scriptstyle{\zeta^{2}}}\\{\scriptstyle{\zeta^{-2}}}\endmatrix\right\}
\zeta (-\zeta^{-1})^k (q_i-\zeta)_{k,h} (u-v\zeta^{-1})^{l-k}
\right],
\tag14.11
\endalign
$$
where $(a)_{n,h}:=a(a+h)(a+2h)\cdots(a+(n-1)h)$.
Running the arguments of the proof of Proposition 9.1 with the sole change that the row operations (9.3)
and (9.7) are now applied only for a fixed value of $i$ and a fixed value of $j$ (as opposed to all 
$1\leq i\leq m$
and all $1\leq j\leq n$), we obtain that for any integer $\lambda$ the left hand side of (14.9), when
regarded as a polynomial in $q_i$, admits
$q_i=\zeta^{\pm1}-\lambda h$ as a root of multiplicity $\max(s_i-(\lambda+1),0)$, and as a polynomial in $q'_j$ 
admits $q'_j=\zeta^{\pm1}-\lambda h$ as a root of multiplicity $\max(t_j-(\lambda+1),0)$. Setting $h=0$ we
obtain the desired divisibility of $\det \bar{M}''$ by the powers of $q_i-\zeta^{\pm1}$ and $q'_j-\zeta^{\pm1}$,
and the proof is complete. \epf

\mysec{15. Random covering surfaces and physical interpretation}


Let $q_1,\dotsc,q_n$ be fixed even integers, and assume
$2q=\sum_{i=1}^n q_i\geq0$. Let $U$ be a fixed rhombus on the triangular lattice, centered at the origin
and having side $l$, with $l$ very large.
Consider the lattice triangular holes of side two $W(3r_1,0),\dotsc,W(3r_q,0)$, where $r_1$ is
vastly larger than the side of $U$ and $r_i$ is vastly larger than $r_{i-1}$, for $i=2,\dotsc,q$. Let
$S$ be the lattice rhombus $rU$ with opposite sides identified so as to make a torus,
where $r$ is vastly larger than $r_q$.

Let $\Cal C$ be the collection of all non-overlapping placements $C=\{Q_1,\dotsc,Q_n\}$ of 
lattice triangular holes inside $S$, where $Q_i$ has side $q_i$, for $i=1,\dotsc,n$.

Given an integer $x\geq 1$, a collection of holes on the triangular lattice $T$ may also be viewed as 
residing on the $x$-fold refinement $T_x$ of the triangular lattice. Regarded so, their correlation is
still well-defined; denote it by $\hat\omega_x$.

Let 
$$
{\Cal S}_x=\bigcup_{\{Q_1,\dotsc,Q_n\}\in{\Cal C}}
{\Cal T}_x(S\setminus Q_1\cup\cdots\cup Q_n\cup W(3r_1,0)\cdots\cup W(3r_q,0)),
$$ 
where ${\Cal T}_x(R)$ is the set of all 
lozenge tilings of the region $R$ when regarded as residing on the $x$-fold refinement of the triangular
lattice. 

Then Theorem 1.1 implies that when sampling uniformly at random from ${\Cal S}_x$,
the ratio of the probabilities $P$ that $Q_1,\dotsc,Q_n$ are at mutual
distances $d_{ij}$ and $P'$ that $Q_1,\dotsc,Q_n$ are at mutual
distances $d'_{ij}$ is, in the limit $1<<l<<r_1<<\dotsc<<r_q<<r$ and for large separations between the
holes, given by
$$
\frac{P}{P'}\sim
\frac
{ e^{-\frac{x^2}{2} \sum_{1\leq i\leq j\leq n}q_iq_j(-\ln d_{ij})   }   } 
{ e^{-\frac{x^2}{2} \sum_{1\leq i\leq j\leq n}q_iq_j(-\ln d'_{ij})   }   }.\tag15.1
$$

On the other hand, the Fundamental Theorem of Statistical Physics (see e.g. \cite{\Fone,\S 40-3})
implies that, given a physical system of two dimensional charges of magnitudes $q_1q_e,\dotsc,q_nq_e$ 
(where $q_e$ is the elementary charge), the relative probability that the charges
are at mutual distances $d_{ij}$ versus $d'_{ij}$ is given by 
$$
\frac
{ e^{-\frac{q_e^2}{2\pi\epsilon_0 kT} \sum_{1\leq i\leq j\leq n}q_iq_j(-\ln d_{ij})   }   } 
{ e^{-\frac{q_e^2}{2\pi\epsilon_0 kT} \sum_{1\leq i\leq j\leq n}q_iq_j(-\ln d'_{ij})   }   },\tag15.2
$$
where $k$ is Boltzmann's constant (see e.g. \cite{\Fone,\S39-4}), $T$ is absolute temperature, and
$\epsilon_0$ is the permitivity of (two dimensional) empty space.

Thus physical electrostatic interaction in two dimensions is perfectly paralleled by our model,
with charges corresponding to holes, charge magnitude to the difference between the number of 
right- and left-pointing unit triangles in the hole, and temperature to $c/x^2$, where $x$ is the 
lattice refinement parameter and $c$ is a constant.

We note that while there is evidence that the superposition principle of Theorem 1.1 is independent of
the type of the background lattice (see \cite{\sc,\S14}), the specific relationship obtained between 
$T$ and $x$ when doing the above analysis does depend on the lattice. We also note that there is an
analogous three dimensional model one can consider (see \cite{\sc,\S15}), and recent numerical 
simulations suggests that the parallel to electrostatics also holds in three dimensions 
(see \cite{\HKMS}). In the version presented in \cite{\sc,\S15} the relationship between physical
temperature and the lattice refinement parameter is one of direct proportionality.

We conclude this section by providing an equivalent interpretation of our result in terms of covering
surfaces.

View the plane with our triangular lattice $T$ as being the $(x,y)$-coordinate plane in $\R^3$.
Consider a copy of the lattice $L=\Z^3$ in $\R^3$ placed in such a way that the image of the
lattice points of $L$ under the orthogonal projection $p$ on the $(x,y)$-coordinate plane consists precisely of 
the lattice points of $T$. 

Then the triangular lattice lifts under $p:L\mapsto T$ to the cubic lattice $L$. Furthermore, it is
well known that the lozenge tilings of a simply connected region on $T$ lift under $p$ to lattice 
surfaces of $L$ (more precisely, to minimal lattice surfaces---unions $S$ of unit squares of $L$ so that
the restriction of $p$ to $S$ is one-to-one).

\topinsert
\twoline{\mypic{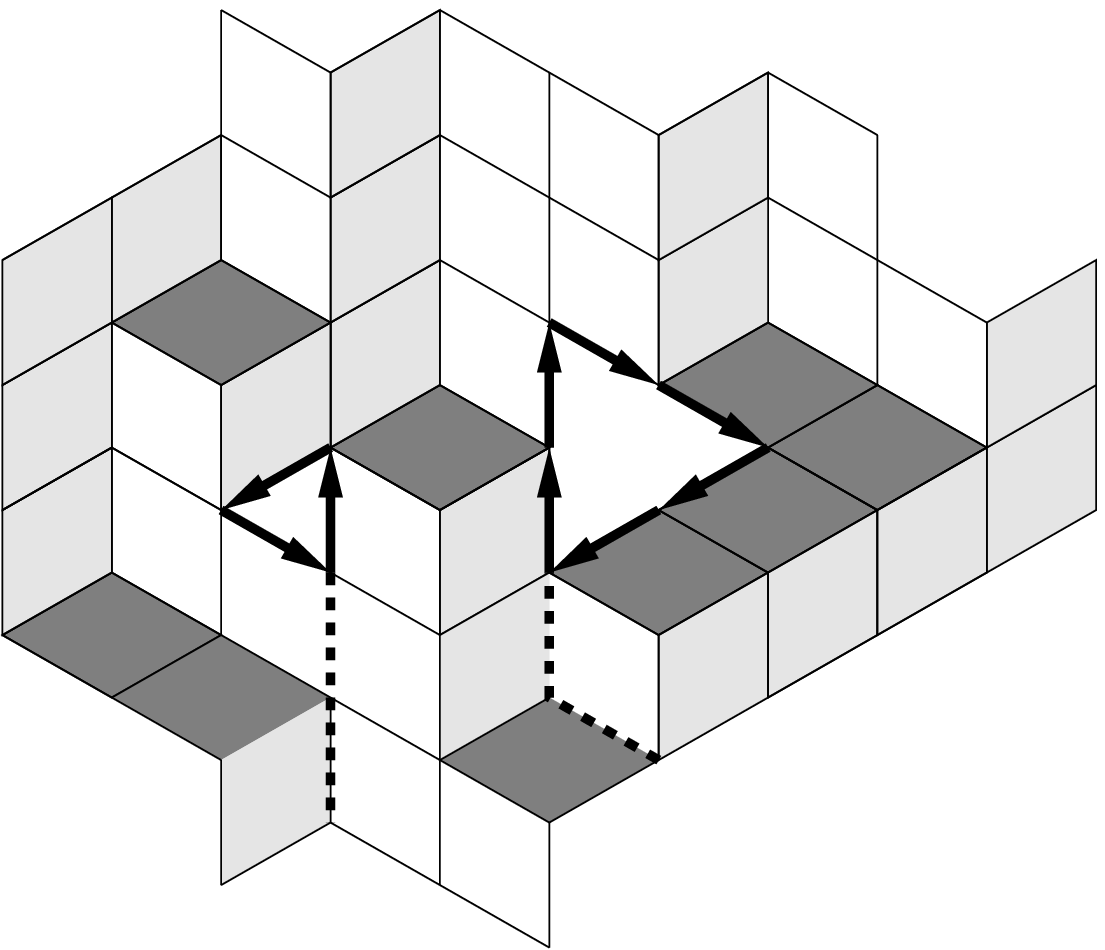}}{\mypic{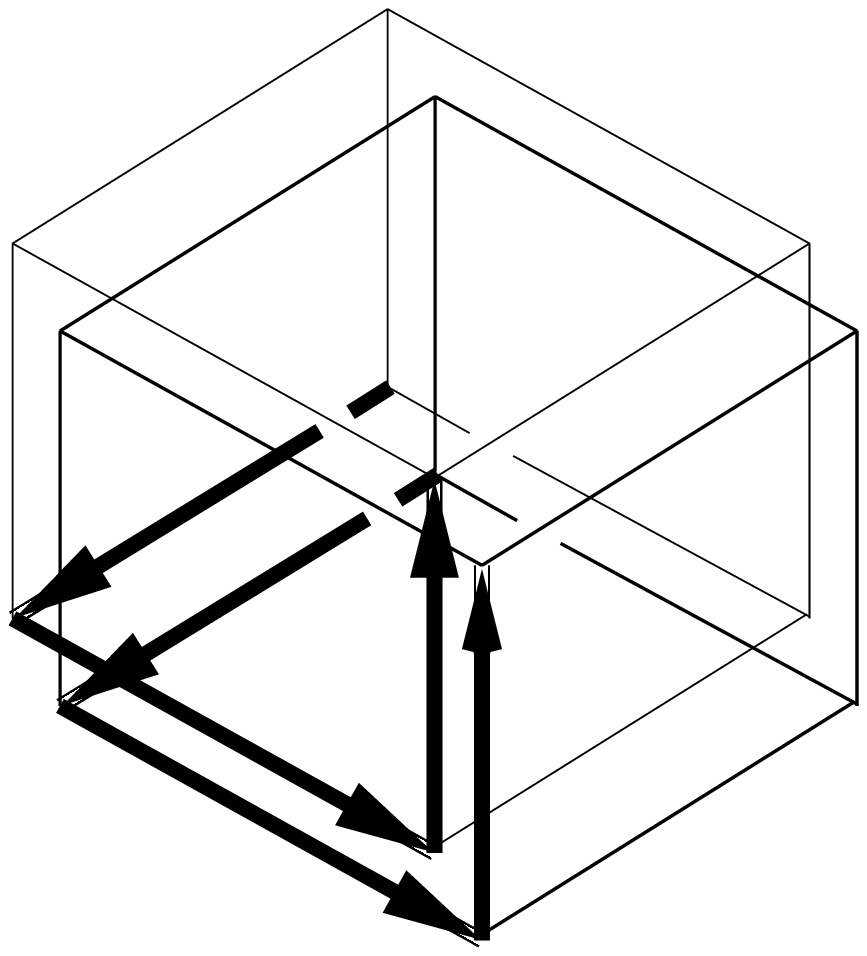}}
\twoline{Figure~15.1. {\rm Lifting with cuts.\ \ \ }}
{Figure~15.2. {\rm The boundary of a hole becomes a spiral.}}
\endinsert

\topinsert
\vskip-0.5in
\twoline{\mypic{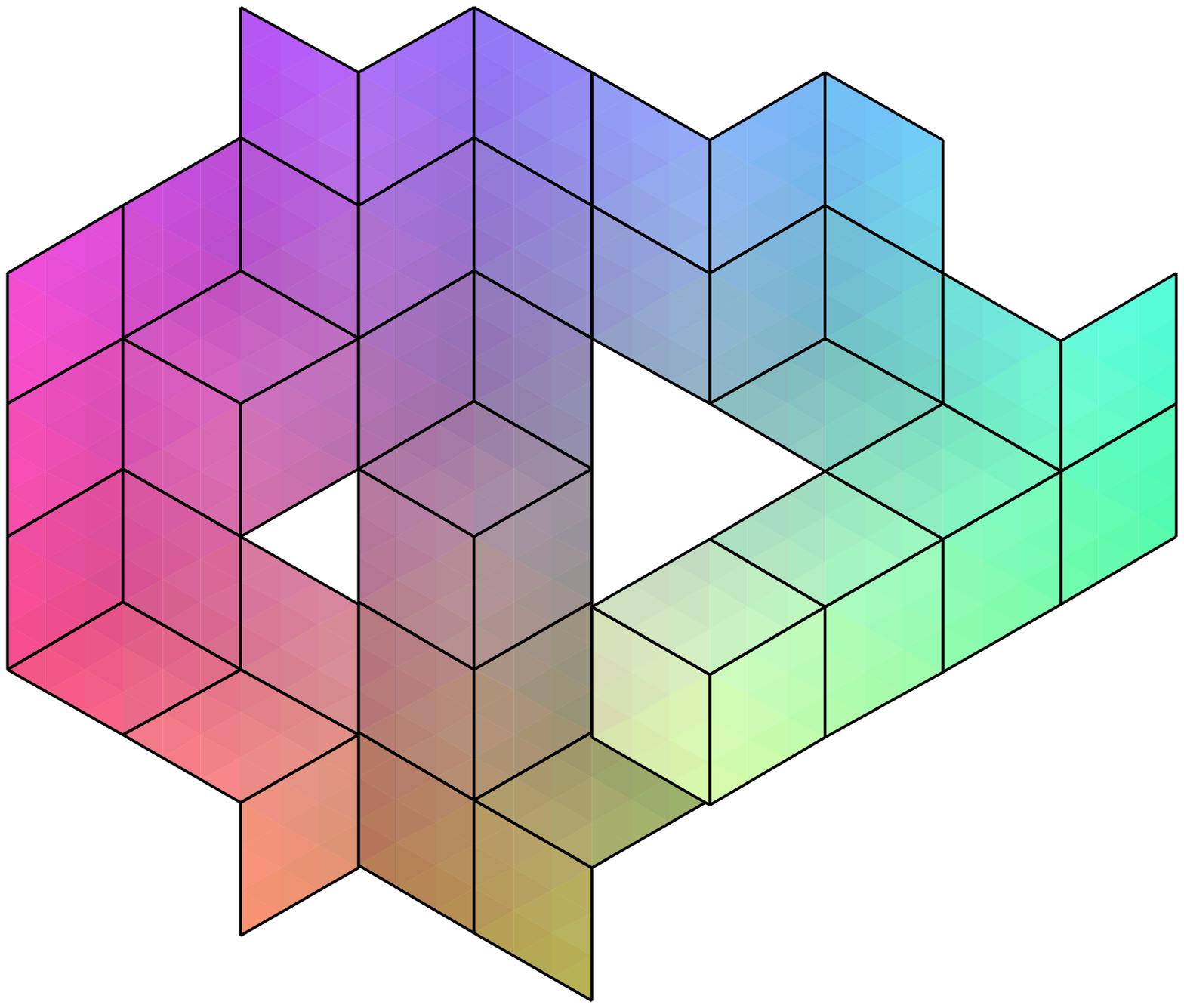}}{\mypic{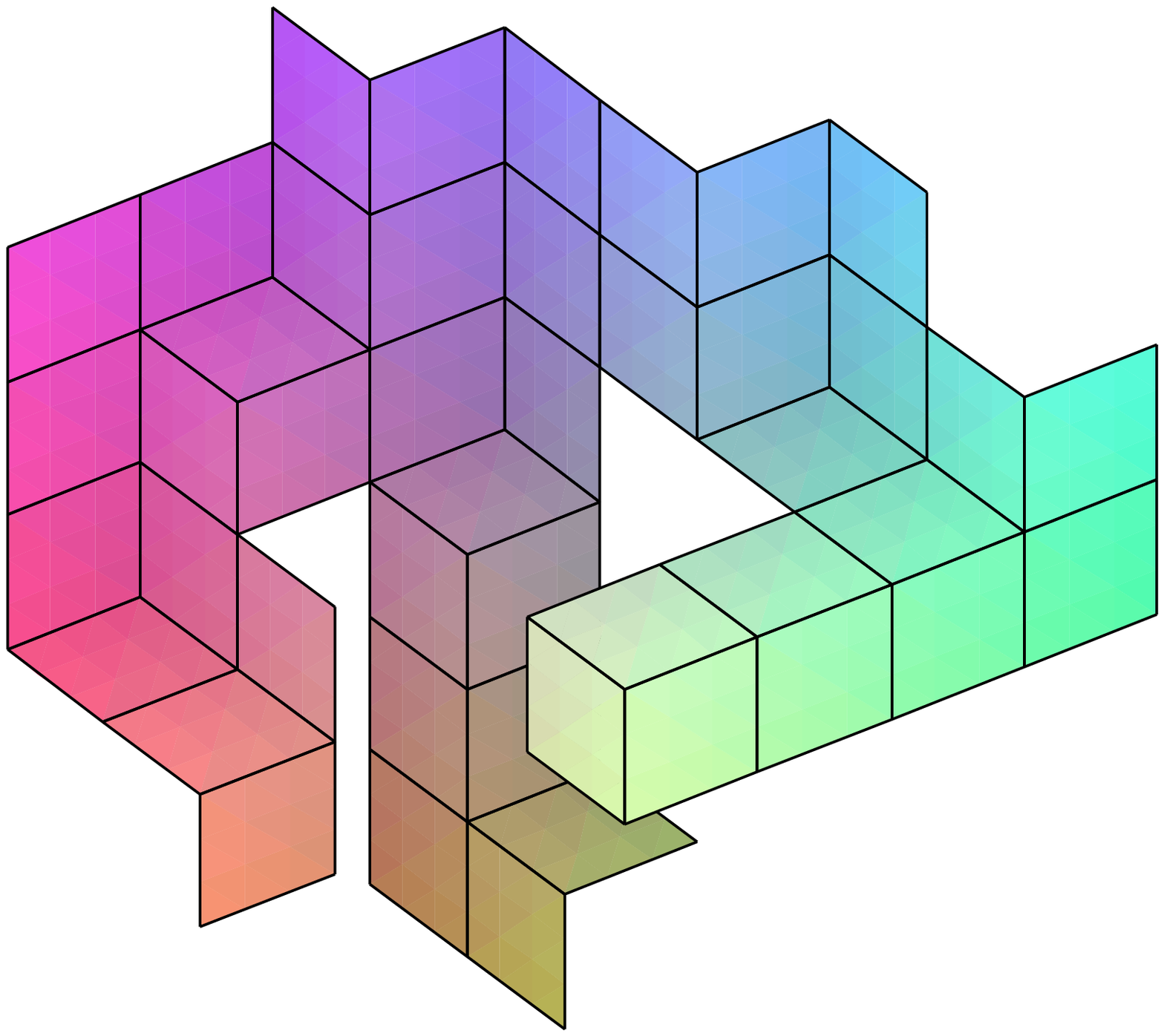}}

\vskip-0.68in
\twoline{\mypic{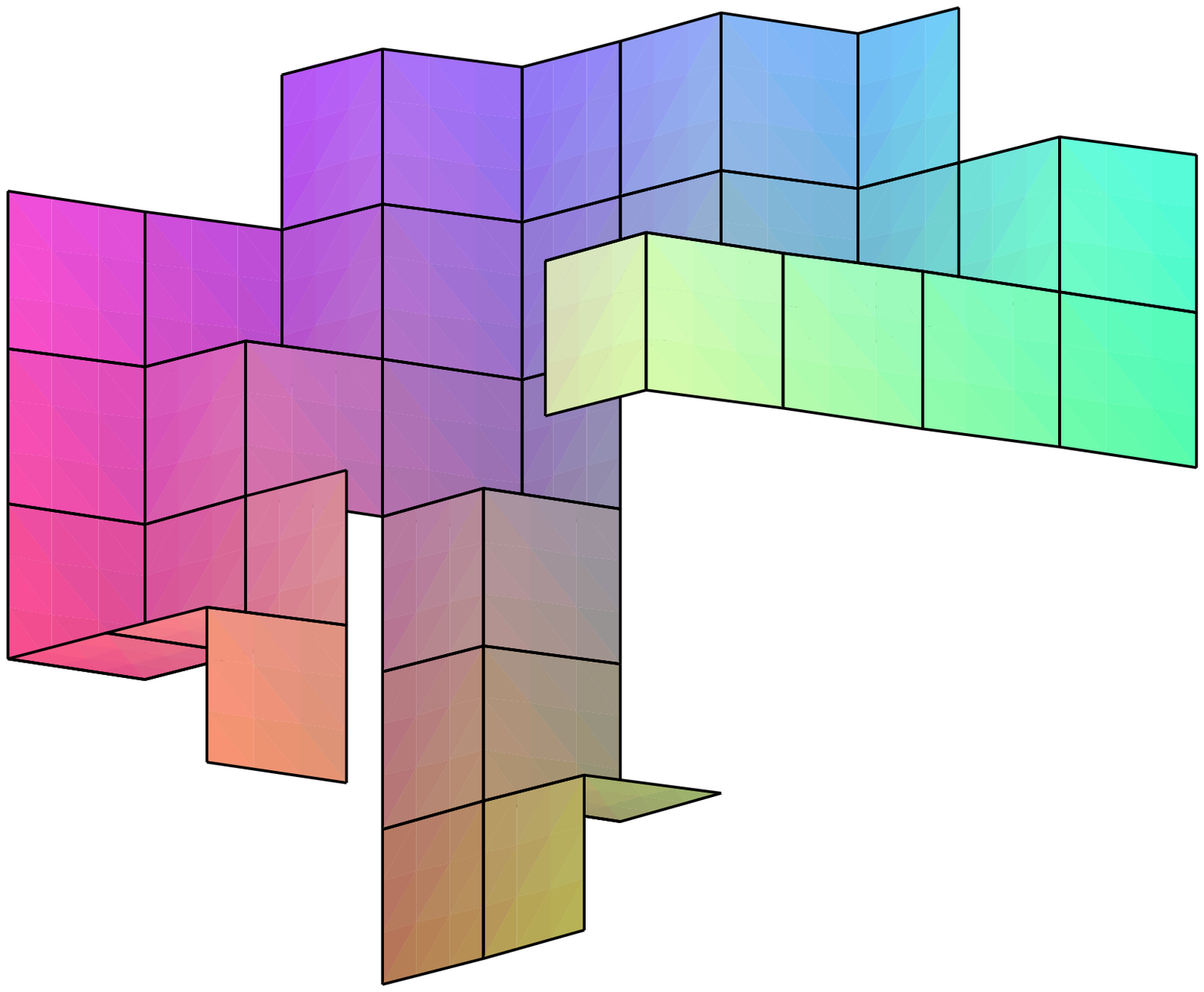}}{\mypic{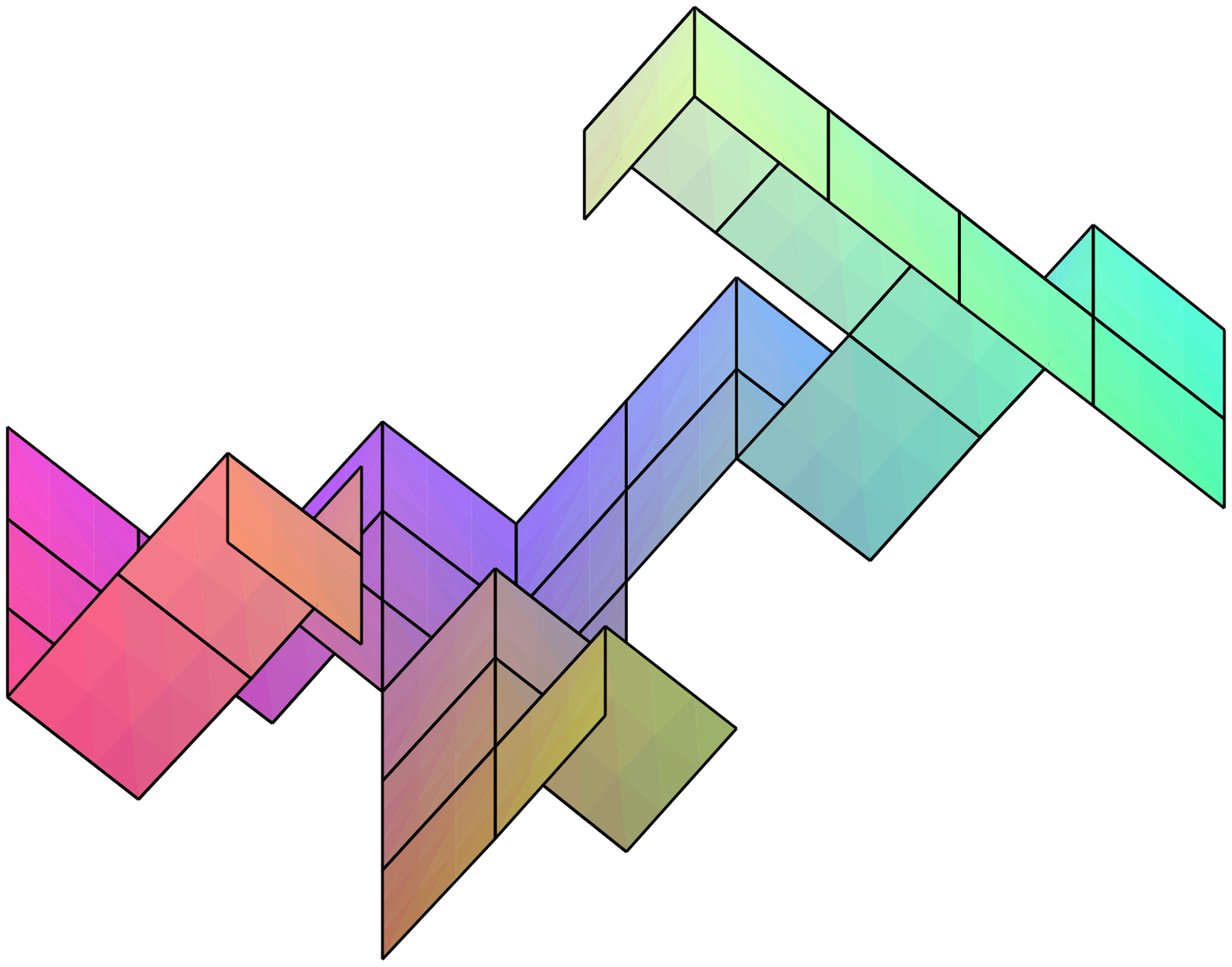}}
\vskip-0.15in
\centerline{Figure~15.3. {\rm Four views of the surface lifting the tiling of Figure 15.1.}}
\endinsert

\topinsert
\vskip-0.5in
\twoline{\!\!\!\!\!\!\!\!\!\!\!\!\!\!\!\!\!\!\!\!\!\mypic{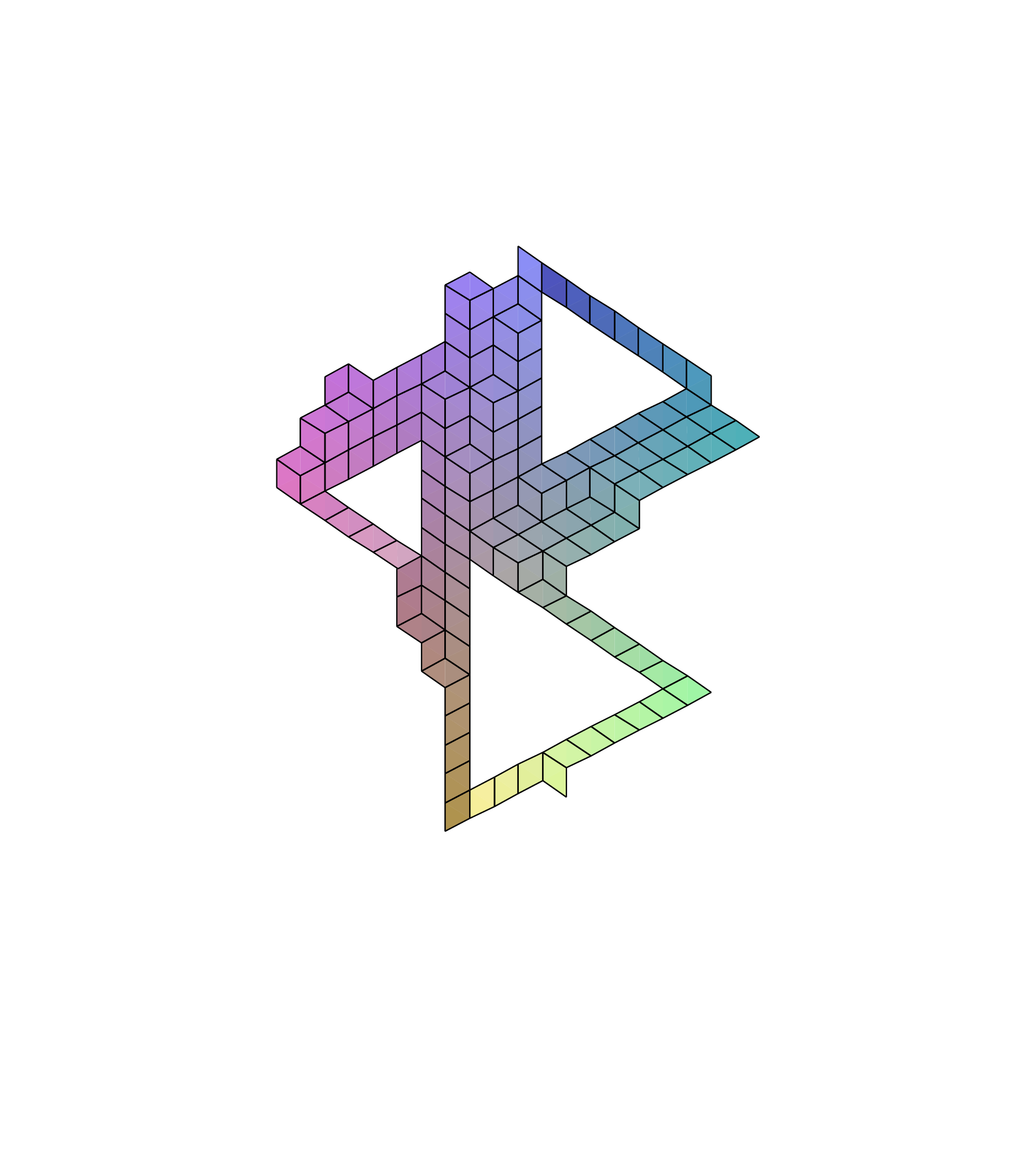}}
{\!\!\!\!\!\!\!\!\!\!\!\!\!\!\!\!\!\!\!\!\!\!\!\!\!\!\!\!\!\!\!\!\!\mypic{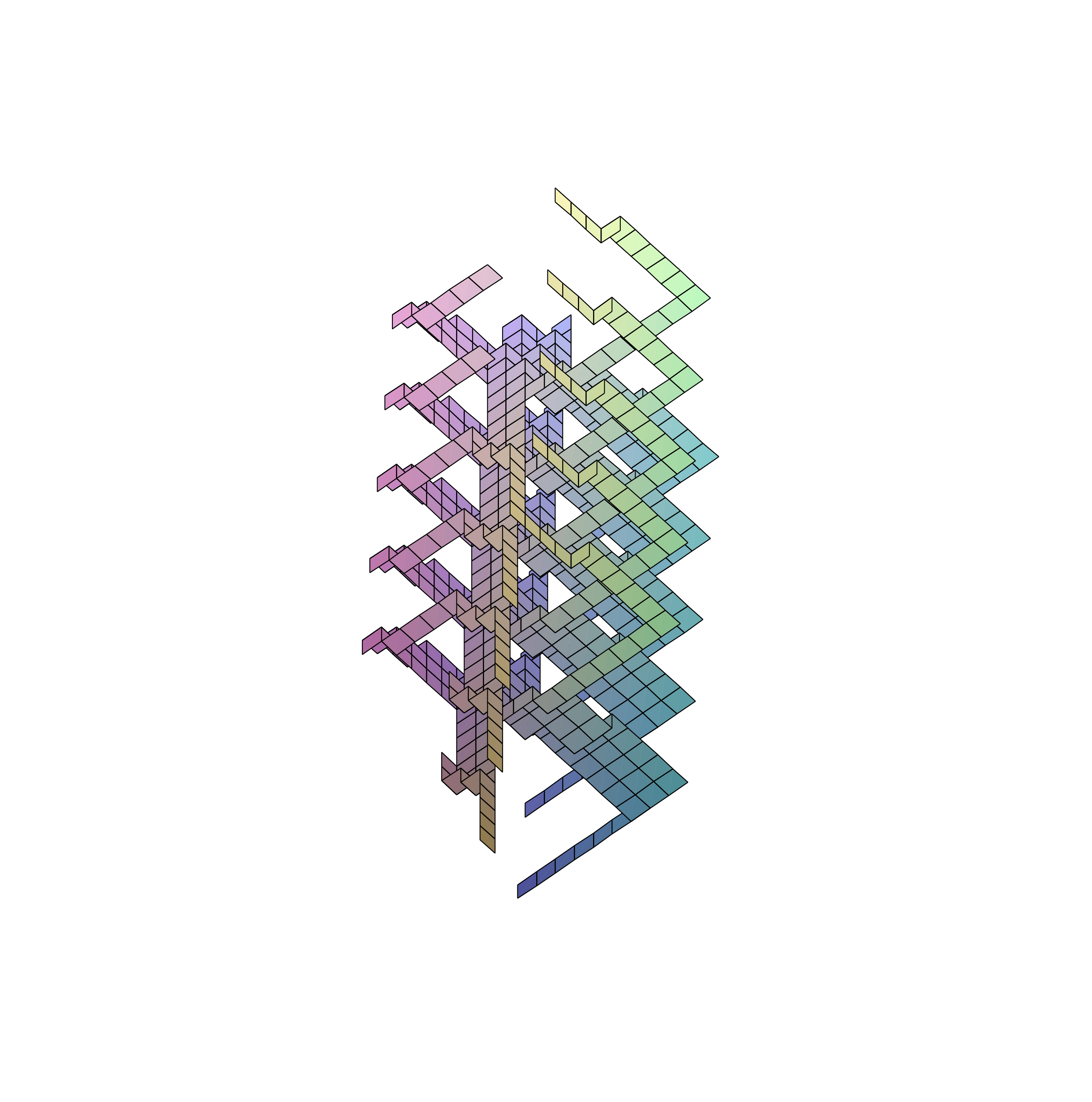}}
\vskip-0.4in
\centerline{Figure~15.4. {\rm Two views of the surface lifting a tiling of the }}
\centerline{{\rm complement of the holes in Figure 2.1.}}
\endinsert

This bijection breaks down if the lattice region has holes. However, it turns out that tilings of 
regions with holes can be lifted to covering surfaces under $p$. Indeed, consider for instance the 
region with two holes shown in Figure 15.1, and consider a tiling $\mu$ of it 
(also indicated in the figure). Cut the region along 
non-intersecting lattice paths leading from each hole to the boundary (these cuts are indicated by 
dotted lines in Figure 15.1). The resulting
region is simply connected, hence $\mu$ can be lifted to a lattice surface $S_0$ in $L$ (four different views
of the surface corresponding to the tiling and cuts in Figure 15.1 are shown in Figure 15.3; the view from
the top---with no perspective---looks just like the tiling; a further instance is shown in Figure 15.4.).
For $n\in\Z$, let $S_n$ be the translation of $S_0$ by the vector $nv$, where $v$ is the up-pointing 
large diagonal of a unit cube of $L$. Then $S=\cup_{n\in\Z}S_n$ is independent of the choice of the
cuts, and is a covering surface under the projection $p$ for the tiling $\mu$.

The boundary of a hole $Q$ lifts under $p:S_n\mapsto T$ to a lattice path on $L$ that in general is not 
closed: the vector that separates its endpoints is precisely $\ch(Q)v$.
The full image under $p^{-1}$ of the boundary of the hole---call it
a {\it tubular hole} of $S$---becomes a collection of $|\ch(Q)|$ identical interspersed spirals of 
step $\ch(Q)$ (for $\ch(Q)=-1$ this is illustrated in Figure 15.2). 
This step size is known as the local {\it holonomy} of the covering surface. The collection of step sizes
associated to all the holes form what we call the {\it holonomy type} of the covering surface.

By this correspondence, the parallel discussed at the beginning of this section becomes a parallel 
between relative probabilities of two dimensional electrostatic charges being at specified mutual
distances and relative probabilities of random covering surfaces of given holonomy type (and appropriate
periodic boundary conditions) to have their
tubular holes at specified mutual distances. The parameter $x$ that accounts for temperature controls
then the ``resolution'' of the covering surfaces.

From this perspective, the laws of two dimensional electrostatics arise by
averaging over all possible discrete geometries of the covering surfaces.

The joint correlation of the holes is a real number that arises by averaging over all the covering surfaces that
lift tilings of the complement of the holes. In this paper we studied the asymptotic behavior of this number
as the holes move away from one another (and found that this asymptotic behavior is governed by Coulomb's law for
the electrostatic potential energy). A complementary question, imperative from the point of view of the literature 
on surface models (see e.g. \cite{\CEP}, \cite{\CLP} and \cite{\CKP})), is the following. 
Keep the holes fixed, and move around a ``probing point,'' which
at each position records the (suitably defined) average height of the covering surfaces compatible with the
fixed position of the holes. What can we say about this average surface in the scaling limit?


More precisely, let us refine the triangular lattice more and more, keeping the holes of fixed side lengths {\it in
the refined lattice} --- this will cause the holes to shrink to points as the lattice spacing approaches zero. Then
if we arrange for the holes to shrink to points chosen beforehand, does the limit of the normalized\footnote{ To obtain
the normalized height function, as we refine the triangular lattice say $R$-fold, we replace the lattice $\Z^3$
in the surface interpretation of tilings presented earlier in this section by the lattice $\left(\frac1R\Z\right)^3$.}
average multivalued height function exist? If so, what is this limit?

\topinsert
\centerline{\mypic{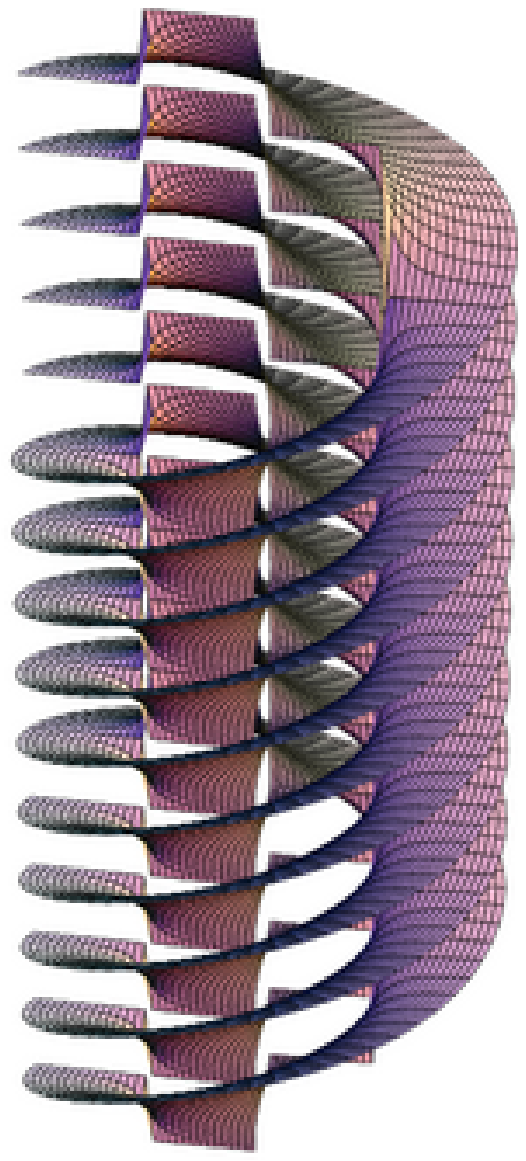}}
\vskip-0.10in
\centerline{Figure~15.5. {\rm The limit of the average over all surfaces of the type shown in Figure 15.4.}}
\endinsert

It turns out that the above described limit surface exists, and is just the horizontal coordinate plane with the 
corresponding points removed. Thus, unlike in the explicit limit surface results of Cohn et al. 
\cite{\CEP}\cite{\CLP} and the related results of \cite{\CKP}\cite{\KOS}\cite{\SheffRS}, 
to capture the interesting behavior we need to consider the {\it un-normalized} average height function. This
occurs also in Kenyon \cite{\Kthree}, where it is shown that the (un-normalized) average height function of Temperleyan 
domains on the square lattice converges in the scaling limit to a harmonic function.

We show in \cite{\ef} that the scaling limit of the latter is a sum of ``refined helicoids'' based at the points to
which the holes shrink, and with steps proportional to the charge of the corresponding holes (if for a helicoid $H$ 
the points on the same vertical are a distance $d$ apart, the $s$-refinement of $H$ is obtained by taking $s-1$ more
copies of $H$, all vertical translates of it, and interspersing them so that the points on their union that are on
the same vertical are a distance $d/s$ apart). For instance, the scaling limit of the average over all the surfaces 
of the kind shown in Figure 15.4 is of the type illustrated in Figure 15.5.


We conclude by mentioning that this limit surface result turns out to be equivalent to determining the scaling limit
of the discrete vector field defined by the average orientation of a lozenge covering a fixed unit triangle in
the complement of the holes (see \cite{\ef} for details). In this phrasing, the above result states that the
scaling limit of this discrete field is precisely the two dimensional electrostatic field obtained when regarding
the holes as charges of magnitude given by the difference between the number of right and left monomers in them. 
Thus the
two complementary viewpoints of studying the behavior of the correlation when the holes move around, and studying
the average height function when the holes are fixed, lead to the emergence of two complementary aspects of 
electrostatics: the electrostatic potential energy and the electric field.

\newpage
\mysec{{\smc Appendix.} A determinant evaluation}

The object of this section is to prove the following result.

\proclaim{Theorem A.1} Let $s_1\geq \cdots\geq s_m\geq1$ and $t_1\geq \cdots\geq t_n\geq1$
be integers. Denote $S=\sum_{i=1}^m s_i$, $T=\sum_{j=1}^n t_j$, and assume $S\geq T$. 
Let $x_1,\dotsc,x_m$ and $z_1,\dotsc,z_n$ be indeterminates. Define $N$ to be the $S\times S$ matrix
$$
N=\left[\matrix A&B \endmatrix\right]\tag A.1
$$
whose blocks are given by
$$
\spreadmatrixlines{1\jot}
\align
&
A=
\\
&\!\!\!\!\!\!\!\!\!\!\!\!
\left[\!
\matrix
{\scriptscriptstyle \frac{{0 \choose 0}}{-x_1-z_1}}\!\!\!\!&
{\scriptscriptstyle \frac{{1 \choose 0}}{(-x_1-z_1)^2}}\!\!\!\!&{\scriptscriptstyle \cdots}\!\!\!\!&
{\scriptscriptstyle \frac{{t_1-1 \choose 0}}{(-x_1-z_1)^{t_1}}}&
\ &
{\scriptscriptstyle \frac{{0 \choose 0}}{-x_1-z_n}}\!\!\!\!&
{\scriptscriptstyle \frac{{1 \choose 0}}{(-x_1-z_n)^2}}\!\!\!\!&{\scriptscriptstyle \cdots}\!\!\!\!&
{\scriptscriptstyle \frac{{t_n-1 \choose 0}}{(-x_1-z_n)^{t_n}}}
\\
{\scriptscriptstyle \frac{{1 \choose 1}}{(-x_1-z_1)^2}}\!\!\!\!&
{\scriptscriptstyle \frac{{2 \choose 1}}{(-x_1-z_1)^3}}\!\!\!\!&{\scriptscriptstyle \cdots}\!\!\!\!&
{\scriptscriptstyle \frac{{t_1 \choose 1}}{(-x_1-z_1)^{t_1+1}}}&
\ &
{\scriptscriptstyle \frac{{1 \choose 1}}{(-x_1-z_n)^2}}\!\!\!\!&
{\scriptscriptstyle \frac{{2 \choose 1}}{(-x_1-z_n)^3}}\!\!\!\!&{\scriptscriptstyle \cdots}\!\!\!\!&
{\scriptscriptstyle \frac{{t_n \choose 1}}{(-x_1-z_n)^{t_n+1}}}
\\
{\scriptstyle\cdot}\!\!\!\!&
{\scriptstyle\cdot}\!\!\!\!&\ \!\!\!\!&
{\scriptstyle\cdot}&
\!\!\!\!\cdots\!\!\! &
{\scriptstyle\cdot}\!\!\!\!&
{\scriptstyle\cdot}\!\!\!\!&\ \!\!\!\!&
{\scriptstyle\cdot}
\\
{\scriptstyle\cdot}\!\!\!\!&
{\scriptstyle\cdot}\!\!\!\!&\ \!\!\!\!&
{\scriptstyle\cdot}&
\ &
{\scriptstyle\cdot}\!\!\!\!&
{\scriptstyle\cdot}\!\!\!\!&\ \!\!\!\!&
{\scriptstyle\cdot}
\\
{\scriptstyle\cdot}\!\!\!\!&
{\scriptstyle\cdot}\!\!\!\!&\ \!\!\!\!&
{\scriptstyle\cdot}&
\ &
{\scriptstyle\cdot}\!\!\!\!&
{\scriptstyle\cdot}\!\!\!\!&\ \!\!\!\!&
{\scriptstyle\cdot}
\\
{\scriptscriptstyle \frac{{s_1-1 \choose s_1-1}}{(-x_1-z_1)^{s_1}}}\!\!\!\!&
{\scriptscriptstyle \frac{{s_1 \choose s_1-1}}{(-x_1-z_1)^{s_1+1}}}\!\!\!\!&{\scriptscriptstyle \cdots}\!\!\!\!&
{\scriptscriptstyle \frac{{s_1+t_1-2 \choose s_1-1}}{(-x_1-z_1)^{s_1+t_1-1}}}&
\ &
{\scriptscriptstyle \frac{{s_1-1 \choose s_1-1}}{(-x_1-z_n)^{s_1}}}\!\!\!\!&
{\scriptscriptstyle \frac{{s_1 \choose s_1-1}}{(-x_1-z_n)^{s_1+1}}}\!\!\!\!&{\scriptscriptstyle \cdots}\!\!\!\!&
{\scriptscriptstyle \frac{{s_1+t_n-2 \choose s_1-1}}{(-x_1-z_n)^{s_1+t_n-1}}}
\\
\ \!\!\!\!&
\ \!\!\!\!&\cdot\!\!\!\!&
\ &
\ &
\ \!\!\!\!&
\ \!\!\!\!&\cdot\!\!\!\!&
\ 
\\
\ \!\!\!\!&
\ \!\!\!\!&\cdot\!\!\!\!&
\ &
\ &
\ \!\!\!\!&
\ \!\!\!\!&\cdot\!\!\!\!&
\ 
\\
\ \!\!\!\!&
\ \!\!\!\!&\cdot\!\!\!\!&
\ &
\ &
\ \!\!\!\!&
\ \!\!\!\!&\cdot\!\!\!\!&
\ 
\\
{\scriptscriptstyle \frac{{0 \choose 0}}{-x_m-z_1}}\!\!\!\!&
{\scriptscriptstyle \frac{{1 \choose 0}}{(-x_m-z_1)^2}}\!\!\!\!&{\scriptscriptstyle \cdots}\!\!\!\!&
{\scriptscriptstyle \frac{{t_1-1 \choose 0}}{(-x_m-z_1)^{t_1}}}&
\ &
{\scriptscriptstyle \frac{{0 \choose 0}}{-x_m-z_n}}\!\!\!\!&
{\scriptscriptstyle \frac{{1 \choose 0}}{(-x_m-z_n)^2}}\!\!\!\!&{\scriptscriptstyle \cdots}\!\!\!\!&
{\scriptscriptstyle \frac{{t_n-1 \choose 0}}{(-x_m-z_n)^{t_n}}}
\\
{\scriptscriptstyle \frac{{1 \choose 1}}{(-x_m-z_1)^2}}\!\!\!\!&
{\scriptscriptstyle \frac{{2 \choose 1}}{(-x_m-z_1)^3}}\!\!\!\!&{\scriptscriptstyle \cdots}\!\!\!\!&
{\scriptscriptstyle \frac{{t_1 \choose 1}}{(-x_m-z_1)^{t_1+1}}}&
\ &
{\scriptscriptstyle \frac{{1 \choose 1}}{(-x_m-z_n)^2}}\!\!\!\!&
{\scriptscriptstyle \frac{{2 \choose 1}}{(-x_m-z_n)^3}}\!\!\!\!&{\scriptscriptstyle \cdots}\!\!\!\!&
{\scriptscriptstyle \frac{{t_n \choose 1}}{(-x_m-z_n)^{t_n+1}}}
\\
{\scriptstyle\cdot}\!\!\!\!&
{\scriptstyle\cdot}\!\!\!\!&\ \!\!\!\!&
{\scriptstyle\cdot}&
\!\!\!\!\cdots\!\!\! &
{\scriptstyle\cdot}\!\!\!\!&
{\scriptstyle\cdot}\!\!\!\!&\ \!\!\!\!&
{\scriptstyle\cdot}
\\
{\scriptstyle\cdot}\!\!\!\!&
{\scriptstyle\cdot}\!\!\!\!&\ \!\!\!\!&
{\scriptstyle\cdot}&
\ &
{\scriptstyle\cdot}\!\!\!\!&
{\scriptstyle\cdot}\!\!\!\!&\ \!\!\!\!&
{\scriptstyle\cdot}
\\
{\scriptstyle\cdot}\!\!\!\!&
{\scriptstyle\cdot}\!\!\!\!&\ \!\!\!\!&
{\scriptstyle\cdot}&
\ &
{\scriptstyle\cdot}\!\!\!\!&
{\scriptstyle\cdot}\!\!\!\!&\ \!\!\!\!&
{\scriptstyle\cdot}
\\
{\scriptscriptstyle \frac{{s_m-1 \choose s_m-1}}{(-x_m-z_1)^{s_m}}}\!\!\!\!&
{\scriptscriptstyle \frac{{s_m \choose s_m-1}}{(-x_m-z_1)^{s_m+1}}}\!\!\!\!&{\scriptscriptstyle \cdots}\!\!\!\!&
{\scriptscriptstyle \frac{{s_m+t_1-2 \choose s_m-1}}{(-x_m-z_1)^{s_m+t_1-1}}}&
\ &
{\scriptscriptstyle \frac{{s_m-1 \choose s_m-1}}{(-x_m-z_n)^{s_m}}}\!\!\!\!&
{\scriptscriptstyle \frac{{s_m \choose s_m-1}}{(-x_m-z_n)^{s_m+1}}}\!\!\!\!&{\scriptscriptstyle \cdots}\!\!\!\!&
{\scriptscriptstyle \frac{{s_m+t_n-2 \choose s_m-1}}{(-x_m-z_n)^{s_m+t_n-1}}}
\endmatrix
\!\right]
\\
\tag A.2
\endalign
$$

\flushpar
and

$$
\spreadmatrixlines{1\jot}
B=
\left[\matrix
{\scriptstyle {0 \choose 0}x_1^0}\!\!\!&
{\scriptstyle {1 \choose 0}x_1}\!\!\!&{\scriptstyle \cdots}\!\!\!&
{\scriptstyle {S-T-1 \choose 0}x_1^{S-T-1}}
\\
{\scriptstyle {0 \choose 1}x_1^{-1}}\!\!\!&
{\scriptstyle {1 \choose 1}x_1^0}\!\!\!&{\scriptstyle \cdots}\!\!\!&
{\scriptstyle {S-T-1 \choose 1}x_1^{S-T-2}}
\\
{\scriptstyle \cdot}\!\!\!&
{\scriptstyle \cdot}\!\!\!&{\scriptstyle \ }\!\!\!&
{\scriptstyle \cdot}
\\
{\scriptstyle \cdot}\!\!\!&
{\scriptstyle \cdot}\!\!\!&{\scriptstyle \ }\!\!\!&
{\scriptstyle \cdot}
\\
{\scriptstyle \cdot}\!\!\!&
{\scriptstyle \cdot}\!\!\!&{\scriptstyle \ }\!\!\!&
{\scriptstyle \cdot}
\\
{\scriptstyle {0 \choose s_1-1}x_1^{1-s_1}}\!\!\!&
{\scriptstyle {1 \choose s_1-1}x_1^{2-s_1}}\!\!\!&{\scriptstyle \cdots}\!\!\!&
{\scriptstyle {S-T-1 \choose s_1-1}x_1^{S-T-s_1}}
\\
{\scriptstyle \ }\!\!\!&
{\scriptstyle \ }\!\!\!&\cdot \!\!\!&
{\scriptstyle \ }
\\
{\scriptstyle \ }\!\!\!&
{\scriptstyle \ }\!\!\!&\cdot \!\!\!&
{\scriptstyle \ }
\\
{\scriptstyle \ }\!\!\!&
{\scriptstyle \ }\!\!\!&\cdot \!\!\!&
{\scriptstyle \ }
\\
{\scriptstyle {0 \choose 0}x_m^0}\!\!\!&
{\scriptstyle {1 \choose 0}x_m}\!\!\!&{\scriptstyle \cdots}\!\!\!&
{\scriptstyle {S-T-1 \choose 0}x_m^{S-T-1}}
\\
{\scriptstyle {0 \choose 1}x_m^{-1}}\!\!\!&
{\scriptstyle {1 \choose 1}x_m^0}\!\!\!&{\scriptstyle \cdots}\!\!\!&
{\scriptstyle {S-T-1 \choose 1}x_m^{S-T-2}}
\\
{\scriptstyle \cdot}\!\!\!&
{\scriptstyle \cdot}\!\!\!&{\scriptstyle \ }\!\!\!&
{\scriptstyle \cdot}
\\
{\scriptstyle \cdot}\!\!\!&
{\scriptstyle \cdot}\!\!\!&{\scriptstyle \ }\!\!\!&
{\scriptstyle \cdot}
\\
{\scriptstyle \cdot}\!\!\!&
{\scriptstyle \cdot}\!\!\!&{\scriptstyle \ }\!\!\!&
{\scriptstyle \cdot}
\\
{\scriptstyle {0 \choose s_m-1}x_m^{1-s_m}}\!\!\!&
{\scriptstyle {1 \choose s_m-1}x_m^{2-s_m}}\!\!\!&{\scriptstyle \cdots}\!\!\!&
{\scriptstyle {S-T-1 \choose s_m-1}x_m^{S-T-s_m}}
\endmatrix\right].
\tag A.3
$$
Then we have
$$
\det N =\frac{\prod_{1\leq i<j\leq m}(x_j-x_i)^{s_is_j}\prod_{1\leq i<j\leq n}(z_j-z_i)^{t_it_j}}
{\prod_{i=1}^m\prod_{j=1}^n(-x_i-z_j)^{s_it_j}}.\tag A.4
$$

\endproclaim

\flushpar
Our proof relies on Proposition A.2 below which presents some interest in its own right.

\flushpar
Let $F$ be an analytic function, and let ${\bold f}$ be the $S$-vector
$$
\spreadlines{2\jot}
\align
&
{\bold f}=[
F(x_1),\frac{1}{1!}F'(x_1),\dotsc,\frac{1}{(s_1-1)!}F^{(s_1-1)}(x_1),
\\
&\ \ \ \ \ \ 
F(x_2),\frac{1}{1!}F'(x_2),\dotsc,\frac{1}{(s_2-1)!}F^{(s_2-1)}(x_2),
\\
&\ \ \ \ \ \ 
\dotsc,
\\
&\ \ \ \ \ \ 
F(x_m),\frac{1}{1!}F'(x_m),\dotsc,\frac{1}{(s_m-1)!}F^{(s_m-1)}(x_m)],\tag A.5
\endalign
$$
where $s_1\geq\cdots\geq s_m$.

For each $1\leq i<j\leq m$, consider the following two lists of vectors of type $[{\bold u},{\bold v}]$,
where ${\bold u}$ and ${\bold v}$ are vectors of lengths $s_i$ and $s_j$, respectively. 

The first list, $L_{ij}^{(1)}$, consists of the following $s_i$ groups, totaling
$s_j(s_i-s_j+1)+1+2+\dotsc+(s_j-1)=s_is_j-s_j(s_j-1)/2$ vectors (as before, ${\bold 0}_k$ denotes the
$k$-vector with all coordinates 0):
$$
\align
&\ \ \ \ \ \ \ \ \ \ \ \
\matrix
\ & \ & {\bold u} & \ & \ &         \ &              \ & \ & {\bold v}  & \ & \ 
\endmatrix
\\
\\
&
\left.
\matrix
{\bold 0}_0 & {0\choose0} & 0 & 0 & \cdots & 0 & | & 1 & 0 & 0& \cdots &0\\
{\bold 0}_0 & 0 & {1\choose1} & 0 & \cdots & 0 & | & 0 & 1 & 0& \cdots &0\\
{\bold 0}_0 & 0 & 0 & {2\choose2} & \cdots & 0 & | & 0 & 0 & 1& \cdots &0\\
\cdot & \cdot & \cdot & \cdot & \cdot & \cdot & \cdot & \cdot & \cdot & \cdot & \cdot & \cdot \\ 
\cdot & \cdot & \cdot & \cdot & \cdot & \cdot & \cdot & \cdot & \cdot & \cdot & \cdot & \cdot \\ 
\cdot & \cdot & \cdot & \cdot & \cdot & \cdot & \cdot & \cdot & \cdot & \cdot & \cdot & \cdot \\ 
\endmatrix
\right\}s_j
\\
\\
&
\left.
\matrix
{\bold 0}_1 & {1\choose0} & 0 & 0 & \cdots & 0 & | & 1 & 0 & 0& \cdots &0\\
{\bold 0}_1 & 0 & {2\choose1} & 0 & \cdots & 0 & | & 0 & 1 & 0& \cdots &0\\
{\bold 0}_1 & 0 & 0 & {3\choose2} & \cdots & 0 & | & 0 & 0 & 1& \cdots &0\\
\cdot & \cdot & \cdot & \cdot & \cdot & \cdot & \cdot & \cdot & \cdot & \cdot & \cdot & \cdot \\ 
\cdot & \cdot & \cdot & \cdot & \cdot & \cdot & \cdot & \cdot & \cdot & \cdot & \cdot & \cdot \\ 
\cdot & \cdot & \cdot & \cdot & \cdot & \cdot & \cdot & \cdot & \cdot & \cdot & \cdot & \cdot \\ 
\endmatrix
\right\}s_j
\\
&\ \ \ \ \ \ \ \ \ \ \ \ \ \ \ \ \ \ \ \ \ \ \ \ \ \ \ \ \ \ 
\cdot
\\
\\
&\ \ \ \ \ \ \ \ \ \ \ \ \ \ \ \ \ \ \ \ \ \ \ \ \ \ \ \ \ \ 
\cdot
\\
\\
&\ \ \ \ \ \ \ \ \ \ \ \ \ \ \ \ \ \ \ \ \ \ \ \ \ \ \ \ \ \ 
\cdot
\\
\\
&
\left.
\matrix
{\bold 0}_{s_i-s_j} & {s_i-s_j\choose0} & 0 & 0 & \cdots & 0 & | & 1 & 0 & 0& \cdots &0\\
{\bold 0}_{s_i-s_j} & 0 & {s_i-s_j+1\choose1} & 0 & \cdots & 0 & | & 0 & 1 & 0& \cdots &0\\
{\bold 0}_{s_i-s_j} & 0 & 0 & {s_i-s_j+2\choose2} & \cdots & 0 & | & 0 & 0 & 1& \cdots &0\\
\cdot & \cdot & \cdot & \cdot & \cdot & \cdot & \cdot & \cdot & \cdot & \cdot & \cdot & \cdot \\ 
\cdot & \cdot & \cdot & \cdot & \cdot & \cdot & \cdot & \cdot & \cdot & \cdot & \cdot & \cdot \\ 
\cdot & \cdot & \cdot & \cdot & \cdot & \cdot & \cdot & \cdot & \cdot & \cdot & \cdot & \cdot \\ 
\endmatrix
\right\}s_j
\endalign
$$
$$
\align
&
\left.
\matrix
{\bold 0}_{s_i-s_j+1} & {s_i-s_j+1\choose0} & 0 & 0 & \cdots & 0 & | & 1 & 0 & 0& \cdots &0\\
{\bold 0}_{s_i-s_j+1} & 0 & {s_i-s_j+2\choose1} & 0 & \cdots & 0 & | & 0 & 1 & 0& \cdots &0\\
{\bold 0}_{s_i-s_j+1} & 0 & 0 & {s_i-s_j+3\choose2} & \cdots & 0 & | & 0 & 0 & 1& \cdots &0\\
\cdot & \cdot & \cdot & \cdot & \cdot & \cdot & \cdot & \cdot & \cdot & \cdot & \cdot & \cdot \\ 
\cdot & \cdot & \cdot & \cdot & \cdot & \cdot & \cdot & \cdot & \cdot & \cdot & \cdot & \cdot \\ 
\cdot & \cdot & \cdot & \cdot & \cdot & \cdot & \cdot & \cdot & \cdot & \cdot & \cdot & \cdot \\ 
\endmatrix
\right\}s_j-1
\\
&\ \ \ \ \ \ \ \ \ \ \ \ \ \ \ \ \ \ \ \ \ \ \ \ \ \ \ \ \ \ 
\cdot
\\
&\ \ \ \ \ \ \ \ \ \ \ \ \ \ \ \ \ \ \ \ \ \ \ \ \ \ \ \ \ \ 
\cdot
\\
&\ \ \ \ \ \ \ \ \ \ \ \ \ \ \ \ \ \ \ \ \ \ \ \ \ \ \ \ \ \ 
\cdot
\\
&
\left.
\matrix
{\bold 0}_{s_i-2} & {s_i-2\choose0} & 0 & | & 1 & 0 & 0& \cdots &0\\
{\bold 0}_{s_i-2} & 0 & {s_i-1\choose1} & | & 0 & 1 & 0& \cdots &0\\
\endmatrix
\right\}2
\\
&
\left.
\matrix
{\bold 0}_{s_i-1} & {s_i-1\choose0} & | & 1 & 0 & 0& \cdots &0
\endmatrix
\right\}1.\tag A.6
\endalign
$$
The second list, $L^{(2)}_{ij}$, consists of the following $s_j-1$ groups 
$$
\align
&\ 
\matrix
{\bold u}& \ & \ & \ & \ & \ & {\bold v} & \ & \ &\    
\endmatrix
\\
\\
&
\left.
\matrix
{\bold 0}_{s_i} & | & {\bold 0}_{0} & -{s_i \choose 1} & 1 & 0 & 0 & \cdots & 0 & 0\\
{\bold 0}_{s_i} & | & {\bold 0}_{0} & -{s_i \choose 2} & 0 & 1 & 0 & \cdots & 0 & 0\\
\cdot & \cdot & \cdot & \cdot & \cdot & \cdot & \cdot & \cdot & \cdot & \cdot \\ 
\cdot & \cdot & \cdot & \cdot & \cdot & \cdot & \cdot & \cdot & \cdot & \cdot \\ 
\cdot & \cdot & \cdot & \cdot & \cdot & \cdot & \cdot & \cdot & \cdot & \cdot \\ 
{\bold 0}_{s_i} & | & {\bold 0}_{0} & -{s_i \choose s_j-1} & 0 & 0 & 0 & \cdots & 0 & 1
\endmatrix
\right\}s_j-1
\\
&
\left.
\matrix
{\bold 0}_{s_i} & | & {\bold 0}_{1} & -{s_i \choose 1} & 1 & 0 & 0 & \cdots & 0 & 0\\
{\bold 0}_{s_i} & | & {\bold 0}_{1} & -{s_i \choose 2} & 0 & 1 & 0 & \cdots & 0 & 0\\
\cdot & \cdot & \cdot & \cdot & \cdot & \cdot & \cdot & \cdot & \cdot & \cdot \\ 
\cdot & \cdot & \cdot & \cdot & \cdot & \cdot & \cdot & \cdot & \cdot & \cdot \\ 
\cdot & \cdot & \cdot & \cdot & \cdot & \cdot & \cdot & \cdot & \cdot & \cdot \\ 
{\bold 0}_{s_i} & | & {\bold 0}_{1} & -{s_i \choose s_j-2} & 0 & 0 & 0 & \cdots & 0 & 1
\endmatrix
\right\}s_j-2
\\
&\ \ \ \ \ \ \ \ \ \ \ \ \ \ \ \ \ \ \ \ \ \ \ \ \ \ \ \ \ \ 
\cdot\\
&\ \ \ \ \ \ \ \ \ \ \ \ \ \ \ \ \ \ \ \ \ \ \ \ \ \ \ \ \ \ 
\cdot\\
&\ \ \ \ \ \ \ \ \ \ \ \ \ \ \ \ \ \ \ \ \ \ \ \ \ \ \ \ \ \ 
\cdot
\\
&
\left.
\matrix
{\bold 0}_{s_i} & | & {\bold 0}_{s_j-3} & -{s_i \choose 1} & 1 & 0\\
{\bold 0}_{s_i} & | & {\bold 0}_{s_j-3} & -{s_i \choose 2} & 0 & 1
\endmatrix
\right\}2
\\
&
\left.
\matrix
{\bold 0}_{s_i} & | & {\bold 0}_{s_j-2} & -{s_i \choose 1} & 0 & 1
\endmatrix
\right\}1,\tag A.7
\endalign
$$
and has a total of $1+2+\dotsc+(s_j-1)=s_j(s_j-1)/2$ vectors.

Define the list $L_{ij}$ to consist of the vectors 
$$
[{\bold 0}_{s_1},\dotsc,{\bold 0}_{s_{i-1}},-{\bold u},{\bold 0}_{s_{i+1}},\cdots,
{\bold 0}_{s_{j-1}},{\bold v},{\bold 0}_{s_{j+1}},\cdots,{\bold 0}_{s_{m}}],\tag A.8
$$
where $[{\bold u},{\bold v}]$ ranges first over the list (A.6) and then over the list (A.7).
Then $L_{ij}$ consists of $s_is_j$ vectors of length $S$.

Let $L$ be the list obtained by concatenating the lists (A.8) as follows:
$$
L=[L_{12},L_{13},\dotsc,L_{1m},L_{23},L_{24},\dotsc,L_{2m},\dotsc,L_{m-1,m}].\tag A.9
$$
Then $L$ consists of
$\sum_{1\leq i<j\leq m}s_is_j$ $S$-vectors of type
$$
[0,\dotsc,0,-\alpha,0,\dotsc,0,\beta,0,\dotsc,0].\tag A.10
$$
If the two non-zero coordinates of (A.10) are in positions $k$ and $l$, $k<l$, associate with the 
vector (A.10) the operation 
$$
f_l:=\frac{\beta f_l-\alpha f_k}{x_l-x_k}\tag A.11
$$
on the coordinates of the vector ${\bold f}=[f_1,\dotsc,f_{S}]$ given by (A.5).

\proclaim{Proposition A.2} After applying the coordinate operations {\rm (A.11)} associated to the 
vectors of the list $L$ to the vector ${\bold f}$ given by {\rm (A.5)} and setting $x_m=\cdots=x_1$, 
the resulting vector is
$$
[F(x_1),\frac{1}{1!}F'(x_1),\frac{1}{2!}F''(x_1),\dotsc,\frac{1}{(S-1)!}F^{(S-1)}(x_1)].
\tag A.12
$$

\endproclaim

Our proof of this proposition is based on the next three Lemmas. They concern the special case
$m=2$, $s_1=a\geq b=s_2$ of Proposition A.2. In this case the vector ${\bold f}$ becomes
$$
[F(x_1),\frac{1}{1!}F'(x_1),\dotsc,\frac{1}{(a-1)!}F^{(a-1)}(x_1),
F(x_2),\frac{1}{1!}F'(x_2),\dotsc,\frac{1}{(b-1)!}F^{(b-1)}(x_2)].\tag A.13
$$

Denote by $R_a(f,x_0)(x)$ the remainder of the truncated Taylor expansion of $f$ around $x_0$:
$$
f(x)=f(x_0)+\frac{f'(x_0)}{1!}(x-x_0)+\cdots+\frac{f^{(a-1)}(x_0)}{(a-1)!}(x-x_0)^{a-1}+
(x-x_0)^a R_a(f,x_0)(x).\tag A.14
$$

\proclaim{Lemma A.3} Consider the $ab$ coordinate operations {\rm (A.11)} corresponding to the vectors
$[-{\bold u},{\bold v}]$, where the pairs $({\bold u},{\bold v})$ are the ones obtained from {\rm (A.6)}
and {\rm (A.7)} by replacing $s_i$ by $a$ and~$s_j$~by~$b$.

After applying these operations to the vector {\rm (A.13)}, its $a+k+1$-st coordinate becomes
$$
\spreadlines{2\jot}
\align
&
\frac{1}{k!}\frac{1}{(x_2-x_1)^k}
\left\{
R_{a-k}(F^{(k)},x_1)(x_2)-{k\choose1}(a)_1R_{a-k+1}(F^{(k-1)},x_1)(x_2)+\cdots
\right.
\\
&\ \ \ \ \ \ \ \ \ \ \ \ \ \ \ \ \ \ \ \ \ \ \ \ \ \ \ \ \ \ \ \ \ \ \ \ \ \ \ \ \ \ \ \ \ \ \ \ \ \ 
\ \ \ \ \ \ \ \ \ \ \ \ \ \ \ 
\left.
+(-1)^k {k \choose k}(a)_kR_{a}(F,x_1)(x_2)\right\},
\endalign
$$
for $k=0,1,\dotsc,b-1$.

\endproclaim

\pf We proceed by induction on $k$. Identify the vectors in the lists (A.6) and (A.7) corresponding to 
$s_i=a$, $s_j=b$ with the coordinate operations (A.11) they encode.

For $k=0$ the statement follows from the form of the coordinate operations (A.6)--(A.7) and the 
definition (A.14) of $R_a(F,x_1)(x_2)$.

As indicated in (A.7), group the $b(b-1)/2$ operations of (A.7) in groups
of sizes $b-1,b-2,\dotsc,1$. Note that after the first coordinate operation in the $i$-th group,
coordinate $a+i+1$ already reaches its final form. On the other hand, the earliest time coordinate
$a+i+1$ is multiplied by a constant and added to some coordinate that comes after it is in the
$i+1$-st group of (A.7). Since coordinate $a+1$ reaches its final form already after operations (A.6),
it follows that by the time coordinate $a+i$ is multiplied by a constant and
added to row $a+k+1$, $k\geq i$, row $a+i$ has already reached its final form, for all $i\geq1$.

Therefore, for the induction step we need to prove that
$$
\spreadlines{2\jot}
\align
&
\frac{1}{(x_2-x_1)^k}
\left\{
R_{a-k}({\scriptstyle \frac{1}{k!}}F^{(k)})
-{a \choose k}R_a(F)-{a \choose k-1}\frac{1}{1!}[R_{a-1}(F')-(a)_1R_a(F)]
\right.
\\
&\ \ \ \ \ \ \ \ \ \ \ \ \ \ \ \ 
-{a \choose k-2}\frac{1}{2!}[R_{a-2}(F'')
-2(a)_1R_{a-1}(F')
+(a)_2R_a(F)]-\cdots
\\
&\ \ \ \ \ \ \ \ \ \ \ \ \ \ \ \ 
-{a \choose 1}\frac{1}{(k-1)!}[R_{a-k+1}(F^{(k-1)})
-{k-1 \choose 1}(a)_1R_{a-k+2}(F^{(k-2)})+\cdots
\\
&\ \ \ \ \ \ \ \ \ \ \ \ \ \ \ \ \ \ \ \ \ \ \ \ \ \ \ \ \ \ \ \ \ \ \ \ \ \ \ \ \ \ \ \ \ \ \ \ 
\ \ \ \ \ \ \ \ \ \ \ \ \ \ \ \ 
\left.
+(-1)^{k-1}{k-1 \choose k-1}(a)_{k-1}R_a(F)]
\right\}
\\
&
=
\frac{1}{k!}\frac{1}{(x_2-x_1)^k}
[R_{a-k}(F^{(k)})-{k \choose 1}(a)_1R_{a-k+1}(F^{(k-1)})+\cdots
+(-1)^{k}{k \choose k}(a)_{k}R_a(F)].\tag A.15
\endalign
$$

Extracting the coefficients of $R_{a-i}(F^{(i)})$ on both sides, (A.15) is seen to be equivalent to
$$
-\sum_{j=0}^{k-i-1}{a \choose k-i-j}\frac{1}{(i+j)!}(-1)^j{i+j \choose j}(a)_j
=
\frac{1}{k!}(-1)^{k-i}{k \choose k-i}(a)_{k-i},\tag A.16
$$
for $i=0,\dotsc,k-1$.
After some manipulation (A.16) turns out to be equivalent to 
$$
\sum_{j=0}^{n+1}(-1)^j\frac{(a+1-j)_{n}}{(n+1-j)!j!}=0,\tag A.17
$$
where $n=k-i$. However, the left hand side of (A.17) can be expressed in terms of hypergeometric
series as
$$
\sum_{j=0}^{n+1}(-1)^j\frac{(a+1-j)_{n}}{(n+1-j)!j!}
=
\frac{(a+1)_n}{(n+1)!}
{}_2 F_1\!\!\left[\matrix -a,-n-1\\ -a-n\endmatrix;1\right].
$$
The resulting ${}_2 F_1$ is zero by Gauss' summation (4.9). \epf

\proclaim{Lemma A.4} For $k=0,1,\dotsc,a-1$ one has
$$
\spreadlines{2\jot}
\align
&
\frac{d^k}{dx_2^k} R_a(F,x_1)(x_2)=
\frac{1}{(x_2-x_1)^k}
\left\{
R_{a-k}(F^{(k)},x_1)(x_2)
\right.
\\
&\ \ \ \ \ \ \ \ \ \ \ \ \ \ \ \ \ \ \ \ \ \ \ \ \ \ \ \ \ \ \ \ \ \ \ \ \ \ \ \ \ \ \ \ \ \ \ \ 
-{k\choose1}(a)_1R_{a-k+1}(F^{(k-1)},x_1)(x_2)+\cdots
\\
&\ \ \ \ \ \ \ \ \ \ \ \ \ \ \ \ \ \ \ \ \ \ \ \ \ \ \ \ \ \ \ \ \ \ \ \ \ \ \ \ \ \ \ \ \ \ \ \ 
\left.
+(-1)^k {k \choose k}(a)_kR_{a}(F,x_1)(x_2)\right\}.\tag A.18
\endalign
$$

\endproclaim

\pf By definition (A.14) we have
$$
R_a(F,x_1)(x_2)=F(x_2)(x_2-x_1)^{-a}-\sum_{j=0}^{a-1}\frac{F^{(j)}(x_1)}{j!}(x_2-x_1)^{-a+j}.
$$
This implies that
$$
\frac{d^k}{dx_2^k}R_a(F,x_1)(x_2)
=
\sum_{i=0}^k {k \choose i} (-1)^i (a)_i \frac{F^{(k-i)}(x_2)}{(x_2-x_1)^{a+i}}
-
(-1)^{k}\sum_{j=0}^{a-1}\frac{F^{(j)}(x_1)}{j!}\frac{(a-j)_k}{(x_2-x_1)^{a+k-j}}.\tag A.19
$$

Definition (A.14) also implies
$$
\spreadlines{1\jot}
\align
&
R_{a-i}(F^{(i)},x_1)(x_2)=
\frac{1}{(x_2-x_1)^{a-i}}
\left\{
F^{(i)}(x_2)-F^{(i)}(x_1)-\frac{F^{(i+1)}(x_1)}{1!}(x_2-x_1)-\cdots
\right.
\\
&\ \ \ \ \ \ \ \ \ \ \ \ \ \ \ \ \ \ \ \ \ \ \ \ \ \ \ \ \ \ \ \ \ \ \ \ \ \ \ \ \ \ \ \ \ \ \ \ \ \ 
\ \ \ \ \ \ \ \ \ \ \ \ \ \ \ \ \ \ \ \ \ \ \ \ 
\left.
-\frac{F^{(a-1)}(x_1)}{(a-i-1)!}(x_2-x_1)^{a-i-1}
\right\},
\endalign
$$
for $i=0,\dotsc,k$. Therefore, writing for brevity $x=x_2-x_1$, $F_i=F^{(i)}(x_1)$ and 
$G_i=F^{(i)}(x_2)$, the right hand side of (A.18) becomes
$$
\spreadlines{1\jot}
\align
&
\frac{1}{x^k}
\left\{
\frac{1}{x^{a-k}}[G_k-\frac{F_k}{0!}-\frac{F_{k+1}}{1!}x-\cdots-\frac{F_{a-1}}{(a-k-1)!}x^{a-k-1}]
\right.
\\
&\ \ \ \ \ \ \ \ \ \ 
-{k \choose 1} \frac{(a)_1}{x^{a-k+1}} 
[G_{k-1}-\frac{F_{k-1}}{0!}-\frac{F_{k}}{1!}x-\cdots-\frac{F_{a-1}}{(a-k)!}x^{a-k}]
\\
&\ \ \ \ \ \ \ \ \ \ 
+\cdots
\\
&\ \ \ \ \ \ \ \ \ \ 
\left.
+(-1)^k {k \choose k} \frac{(a)_k}{x^{a}}
[G_{0}-\frac{F_{0}}{0!}-\frac{F_{1}}{1!}x-\cdots-\frac{F_{a-1}}{(a-1)!}x^{a-1}]
\right\}.\tag A.20
\endalign
$$

To prove (A.18) we need to show that the above expression equals (A.19). The coefficient of $G_i$ in
(A.20) is
$$
(-1)^{k-i}{k \choose k-i} \frac{(a)_{k-i}}{x^{a-i}}\frac{1}{x^k},
$$
which clearly agrees with the coefficient of $G_i$ in (A.19). Agreement of the coefficients of $F_i$
on the other hand turns out to amount to
$$
k!\sum_{j=0}^i \frac{(-1)^j}{j!(i-j)!} \frac{(a)_{k-j}}{(k-j)!}
=
\frac{(a-i)_k}{i!},\tag A.21
$$
for $i=0,\dotsc,k$.

The left hand side of (A.21) becomes after some manipulation
$$
k!\sum_{j=0}^i \frac{(-1)^j}{j!(i-j)!} \frac{(a)_{k-j}}{(k-j)!}
=
\frac{(a)_k}{i!}\sum_{j=0}^i \frac{(-k)_j(-i)_j}{j!(-a-k+1)_j}
=
\frac{(a)_k}{i!}\,
{}_2 F_1\!\!\left[\matrix -k,-i\\ -k-(a-1)\endmatrix;1\right].
$$
Gauss' summation (4.9) implies that when $a$ is a large enough negative integer the above expression
agrees with the right hand side of (A.21). Then (A.21) follows since both its sides are polynomials 
in $a$. \epf

\proclaim{Lemma A.5} For all $k\geq0$ we have
$$
\left.\frac{d^k}{dx^k} R_a(F,x_0)(x)\right|_{x=x_0}=\frac{k!}{(a+k)!}F^{(a+k)}(x_0).
$$

\endproclaim

\pf Since $F$ is analytic we have by (A.14) that
$$
R_a(F,x_0)(x)=\frac{F^{(a)}(x_0)}{a!}+\frac{F^{(a+1)}(x_0)}{(a+1)!}(x-x_0)+\cdots.
$$
This implies
$$
\spreadlines{1\jot}
\align
&
\frac{d^k}{dx^k} R_a(F,x_0)(x)
\\
&\ \ 
=
\frac{k!}{0!}(x-x_0)^0\frac{1}{(a+k)!}F^{(a+k)}(x_0)+
\frac{(k+1)!}{1!}(x-x_0)^1\frac{1}{(a+k+1)!}F^{(a+k+1)}(x_0)+\cdots.
\endalign
$$
Take $x=x_0$ to obtain the statement of the Lemma. \epf

{\it Proof of Proposition A.2.} We proceed by induction on $m$. For $m=1$ the statement is clear.
Perform on the vector ${\bold f}$ the coordinate operations (A.11) corresponding to the initial
segment $[L_{12},\dotsc,L_{1m}]$ of the list $L$---a total of $s_1s_2+\cdots+s_1s_m$ coordinate
operations. By Lemmas A.3 and A.4, the last $S-s_1$ coordinates become
$$
\spreadlines{2\jot}
\align
&
[G(x_2),\frac{1}{1!}G'(x_2),\dotsc,\frac{1}{(s_2-1)!}G^{(s_2-1)}(x_2),
\\
&\ \ \ \ \ \ 
G(x_3),\frac{1}{1!}G'(x_3),\dotsc,\frac{1}{(s_3-1)!}G^{(s_3-1)}(x_3),
\\
&\ \ \ \ \ \ \ \ \ \ \ \ 
\dotsc,
\\
&\ \ \ \ \ \ \ \ \ \ \ \ \ \ \ \ \ \ 
G(x_m),\frac{1}{1!}G'(x_m),\dotsc,\frac{1}{(s_m-1)!}G^{(s_m-1)}(x_m)],
\endalign
$$
where $G(x)=R_{s_1}(F,x_1)(x)$.

By the induction hypothesis, applying the remaining $\sum_{2\leq i<j\leq m}s_is_j$ coordinate operations
and letting $x_m=\cdots=x_2$, the above vector becomes
$$
[G(x_2),\frac{1}{1!}G'(x_2),\dotsc,\frac{1}{(S-s_1)!}G^{(S-s_1)}(x_2)].\tag A.22
$$
However, since $G(x)=R_{s_1}(F,x_1)(x)$, we obtain by Lemma A.5 that
$$
G^{(k)}(x_1)=\left.\frac{d^k}{dx^k}R_{s_1}(F,x_1)(x)\right|_{x=x_1}=\frac{k!}{(s_1+k)!}F^{(s_1+k)}(x_1).
$$
This shows that when appending the $x_2=x_1$ specialization of (A.22) to the first $s_1$ coordinates 
of~${\bold f}$, the resulting vector is precisely the one given by (A.12). \epf

Let $F(x)=1/(-x-z)$ and $G_l(x)=x^l$, $l\geq0$. One readily checks that
$$
\frac{1}{k!}\frac{d^k}{dx^k}\left[\frac{1}{l!}F^{(l)}(x)\right]=
\frac{{\displaystyle {k+l \choose k} }}{(-x-z)^{k+l+1}}
$$
and
$$
\frac{1}{k!}G_l^{(k)}(x)={l \choose k}x^{l-k},
$$
for all $k,l\geq0$.

It follows from these equalities that if we let $F_j(x)=1/(-x-z_j)$, $j=0,\dotsc,n$, and define the
$S$-vector ${\bold h}(x)$ by
$$
\spreadlines{2\jot}
\align
&
{\bold h}(x)=[
F_1(x),\frac{1}{1!}F_1'(x),\dotsc,\frac{1}{(t_1-1)!}F_1^{(t_1-1)}(x),
\\
&\ \ \ \ \ \ \ \ \ \ \ \ \ 
F_2(x),\frac{1}{1!}F_2'(x),\dotsc,\frac{1}{(t_2-1)!}F_2^{(t_2-1)}(x),
\\
&\ \ \ \ \ \ \ \ \ \ \ \ \ \ \ 
\dotsc,
\\
&\ \ \ \ \ \ \ \ \ \ \ \ \ \ \ \ \ \ 
F_n(x),\frac{1}{1!}F_n'(x),\dotsc,\frac{1}{(t_n-1)!}F_n^{(t_n-1)}(x),
G_0(x),G_1(x),\dotsc,G_{S-T-1}(x)
],\tag A.23
\endalign
$$
then the matrix $N$ in the statement of Theorem A.1 can be written as
$$
\spreadmatrixlines{1\jot}
N=
\left[\matrix
{\bold h}(x_1)\\
\frac{1}{1!}{\bold h}'(x_1)\\
\cdots\\
\frac{1}{(s_1-1)!}{\bold h}^{(s_1-1)}(x_1)
\\
\\
{\bold h}(x_2)\\
\frac{1}{1!}{\bold h}'(x_2)\\
\cdots\\
\frac{1}{(s_2-1)!}{\bold h}^{(s_2-1)}(x_2)
\\
\cdot\\
\cdot\\
\cdot
\\
{\bold h}(x_m)\\
\frac{1}{1!}{\bold h}'(x_m)\\
\cdots\\
\frac{1}{(s_m-1)!}{\bold h}^{(s_m-1)}(x_m)
\endmatrix\right].\tag A.24
$$

To present our next result, associate to each vector of type (A.10) in which the non-zero coordinates are
in positions $k<l$ the row operation
$$
R_l:=\frac{\beta R_l-\alpha R_k}{x_l-x_k}\tag A.25
$$
in matrix $N$. Then the list $L$ defined by (A.6)--(A.9) describes $\sum_{1\leq i<j\leq m}s_is_j$
row operations on~$N$.

In addition, consider the list 
$$
L'=[L'_{12},L'_{13},\dotsc,L'_{1n},L'_{23},L'_{24},\dotsc,L'_{2n},\dotsc,L'_{n-1,n}],
$$
where $L'_{ij}$ is obtained from the list $L_{ij}$ defined by (A.6)--(A.8) by replacing the $m$-tuple
$(s_1,\dotsc,$ $s_m$ by the $n$-tuple $(t_1,\dotsc,t_n)$.

Interpret each of the $\sum_{1\leq i<j\leq n}t_it_j$ vectors of $L'$ as a 
{\it column} operation on matrix $N$ using the analog of (A.25) in which the $x_i$'s are replaced by $z_i$'s.

\proclaim{Proposition A.6}{\rm (a).} After applying the $\sum_{1\leq i<j\leq m}s_is_j$ row operations 
{\rm (A.25)} of the list $L$ to matrix $N$ and setting $x_m=\cdots=x_1=x$, the resulting matrix is
$$
\spreadmatrixlines{2\jot}
N_1=
\left[\matrix
{\bold h}(x)\\
\frac{1}{1!}{\bold h}'(x)\\
\cdots\\
\frac{1}{(S-1)!}{\bold h}^{(S-1)}(x)
\endmatrix\right].
$$

{\rm (b).} After applying the  $\sum_{1\leq i<j\leq n}t_it_j$ column operations encoded by
$L'$ to matrix $N_1$ and setting $z_n=\cdots=z_1=z$, the resulting matrix $N_2$ is of the form
$$
\spreadmatrixlines{2\jot}
\left[\matrix
F(x,z) & \frac{1}{1!}\frac{\partial F(x,z)}{\partial z} & \cdots & 
\frac{1}{(T-1)!} \frac{\partial^{T-1}F(x,z)}{\partial z^{T-1}} & 1 & * & \cdots & *\\
\frac{1}{1!}\frac{\partial F(x,z)}{\partial x} & 
\frac{1}{1!1!}\frac{\partial^2 F(x,z)}{\partial x\partial z} & \cdots & 
\frac{1}{1!(T-1)!}\frac{\partial^T F(x,z)}{\partial x\partial z^{T-1}} & 
0 & 1 & \cdots & *\\
\cdot & \cdot & & \cdot & \cdot & \cdot & & \cdot \\
\cdot & \cdot & & \cdot & \cdot & \cdot & & \cdot \\
\cdot & \cdot & & \cdot & \cdot & \cdot & & \cdot \\
\frac{1}{(S-1)!}\frac{\partial^{S-1}F(x,z)}{\partial x^{S-1}} &
\frac{1}{(S-1)!1!}\frac{\partial^{S}F(x,z)}{\partial x^{S-1} \partial z} & \cdots & 
\frac{1}{(S-1)!(T-1)!}\frac{\partial^{S+T-2}F(x,z)}{\partial x^{S-1}\partial z^{T-1}}& 
0 & 0 & \cdots & 0
\endmatrix\right],\tag A.26
$$
where $F(x,z)=1/(-x-z)$.

\endproclaim

\pf (a). This follows using the form (A.24) of matrix $N$ and applying Proposition A.2 to each of its
column vectors.

(b).
By (A.23) the portion of ${\bold h}^{(k)}(x)$ consisting of its first $S-T$ coordinates
has the form of the vector ${\bold f}$ in the statement of Proposition~A.2, for all $k\geq 0$. Therefore,
applying Proposition~A.2 to each row of $N_1$ we obtain that $N_2$ has the stated form in the portion
contained in its first $T$ columns.

Along the last $S-T$ columns the matrix $N_2$ is the same as $N_1$. Since $G_l(x)=x^l$ in (A.23), part
(a) implies that the portion of $N_1$ contained in its last $S-T$ columns has the form indicated in (A.26).
\epf

\proclaim{Lemma A.7} One has the matrix identity
$$
\left[\left({l+i+j \choose i}x^{i+j+1}\right)_{0\leq i,j<n}\right]
=
\left[\left({l+i \choose l+j}x^{i-j}\right)_{0\leq i,j<n}\right]
\left[\left({j \choose i}x^{i+j+1}\right)_{0\leq i,j<n}\right].
$$

\endproclaim

\pf The stated equality is equivalent to
$$
\sum_{k=0}^{n-1}{l+i \choose l+k}{j \choose k}={l+i+j \choose i},
$$
for all $0\leq i,j<n$. This follows by extracting the coefficient of $x^i$ on both sides of the 
identity $(x+1)^{l+i}(x+1)^j=(x+1)^{l+i+j}$. \epf

\smallpagebreak
{\it Proof of Theorem A.1.} From (8.11) and (8.7) we deduce that
$$
\det N =c\,\frac{\prod_{1\leq i<j\leq m}(x_j-x_i)^{s_is_j}\prod_{1\leq i<j\leq n}(z_j-z_i)^{t_it_j}}
{\prod_{i=1}^m\prod_{j=1}^n(-x_i-z_j)^{s_it_j}},
$$
where $c\in\Q$. Write this as
$$
\frac{\det N}{\prod_{1\leq i<j\leq m}(x_j-x_i)^{s_is_j}\prod_{1\leq i<j\leq n}(z_j-z_i)^{t_it_j}}
=
\frac{c}{\prod_{i=1}^m\prod_{j=1}^n(-x_i-z_j)^{s_it_j}}.\tag A.27
$$
By Part (b) of Proposition A.6 we obtain, denoting $f(x)=-1/x$ and using $f^{(n)}(-1)=n!$, that
$$
\spreadlines{2\jot}
\align
&
\left.
\frac{\det N}{\prod_{1\leq i<j\leq m}(x_j-x_i)^{s_is_j}\prod_{1\leq i<j\leq n}(z_j-z_i)^{t_it_j}}
\right|_{x_1=\cdots=x_m=-1 \atop z_1=\cdots=z_n=0}
\\
&\ \ \ \ \ \ \ \ \ \ \ \ \ \ \ \ \ \ \ \ \ \ \ \ \ \ \ \ \ \
=
\det\!\left[\!\left(\frac{1}{k!l!}f^{(k+l)}(-1)\right)_{S-T\leq k \leq S-1 \atop 0\leq l\leq T-1}\!
\right]
\\
&\ \ \ \ \ \ \ \ \ \ \ \ \ \ \ \ \ \ \ \ \ \ \ \ \ \ \ \ \ \
=
\det \left[\left({S-T+k+l \choose l} \right)_{0\leq k,l \leq T-1}
\right].\tag A.28
\endalign
$$
However, by Lemma A.7, the last matrix is the product of a lower triangular and an upper triangular 
matrix, both having all diagonal entries equal to 1---so its determinant is 1. Then (A.28) implies that
$c=1$ in (A.27), thus completing the proof of the Theorem.~\epf

\smallpagebreak
{\smc Remark A.8.} Theorem A.1 generalizes both the Cauchy determinant, which is obtained when
$m=n$ and all $s_i$'s and $t_j$'s equal 1, and the Vandermonde determinant, obtained when $n=0$, 
$s_1=\cdots=s_m=1$. 

The case $m=rn$, $s_1=\cdots=s_{rn}=1$, $t_1=\cdots=t_n=r$ is stated as an exercise 
in~\cite{\Muir, Ex.42,\,p.360}. 


{\bf Acknowledgments.} I would like to thank the anonymous referee for pointing out Sheffield's results 
\cite{\SheffRS} on the existence and uniqueness of invariant Gibbs measures and for helping to properly place
our result in the context of previous results in the literature. The intuitive rationale for
electrostatics to emerge in the dimer model based on the fact that the discrete Gaussian free field model,
which is known to have other close connections with the dimer model, leads to electrostatics when considering
questions analogous to the correlation of holes, is also due to the referee.

\mysec{References}
{\openup 1\jot \frenchspacing\raggedbottom
\roster

\myref{\Bout}
  C. Boutillier, Pattern densities in fluid dimer models, arXiv:math.PR/0603324.

\myref{\ri}
  M. Ciucu, Rotational invariance of quadromer correlations on the hexagonal lattice, 
{\it Adv. in Math.} {\bf 191} (2005), 46--47.

\myref{\sc}
  M. Ciucu, A random tiling model for two dimensional electrostatics, {\it Mem. Amer. Math. Soc.} {\bf 178} 
(2005), no. 839, 1--104.

\myref{\ppone} 
  M. Ciucu, Plane partitions I: A generalization of MacMahon's
formula, {\it Mem. Amer. Math. Soc.} {\bf 178} (2005),
no. 839, 106--144. 

\myref{\ef}
  M. Ciucu, The emergence of the electrostatic field as a Feynman sum in random tilings with holes, preprint,
www.math.gatech.edu/$\sim$ciucu/ef.pdf

\myref{\CEP}
  H. Cohn, N. Elkies, and J. Propp, Local statistics for random domino tilings of the 
Aztec diamond, {\it Duke Math. J.} {\bf 85} (1996), 117-166.

\myref{\CKP}
  H. Cohn, R. Kenyon, and J. Propp, A variational principle for domino tilings, {\it J. Amer. Math. Soc.}  
{\bf 14} (2001), 297--346

\myref{\CLP}
  H. Cohn, M. Larsen, J. Propp, The shape of a typical boxed plane partition, 
{\it New York J. of Math.} {\bf 4} (1998), 137--165.

\myref{\Fone}
  R. P. Feynman, ``The Feynman Lectures on Physics,'' vol. I, Addison-Wesley, Reading, 
Massachusetts, 1963.

\myref{\FS}
  M. E. Fisher and J. Stephenson, Statistical mechanics of dimers on a plane 
lattice. II. Dimer correlations and monomers, {\it Phys. Rev. (2)} {\bf 132} (1963),
1411--1431.


\myref{\Har} 
  R. E. Hartwig, Monomer pair correlations, {\it J. Mathematical Phys.} {\bf 7}
(1966), 286--299.

\myref{\HKMS}
  D. A. Huse, W. Krauth, R. Moessner and S. L. Sondhi, Coulomb and liquid dimer models in three 
dimensions, arxiv.org/abs/cond-mat/0305318, May 2003.

\myref{\Jor}
  C. Jordan, ``Calculus of finite differences,'' Chelsea, New York, 1960.

\myref{\Kone}
  R. Kenyon, Local statistics of lattice dimers, {\it Ann. Inst. H. Poincar\'e Probab.
  Statist.} {\bf 33} (1997), 591--618. 

\myref{\Ktwo}
  R. Kenyon, Long-range properties of spanning trees. Probabilistic
  techniques in equilibrium and nonequilibrium statistical physics, J. Math. Phys. {\bf 41} (2000), 
 1338--1363.

\myref{\Kthree}
  R. Kenyon, Dominos and the Gaussian free field, {\it Ann. Probab.} {\bf 29} (2001), 1128--1137.


\myref{\KOS}
  R. Kenyon, A. Okounkov and S. Sheffield, Dimers and amoebae, {\it Ann. of Math.(2)}  163, no. 3 (2006), 
1019--1056.


\myref{\Muir}
  T. Muir, ``A treatise on the theory of determinants,'' revised and enlarged by 
W.~H.~Metzler,
Longmans, Green and Co., New York, 1933.


\myref{\Ol}
  F. W. J. Olver, Asymptotics and special functions, Academic Press, New York, 1974.

\myref{\Perc}
  J. K. Percus, One more technique for the dimer problem, {\it J. Math. Phys.} {\bf 10} (1969), 
1881--1888.


\myref{\SheffRS}
  S. Sheffield, Random Surfaces, {\it Ast\'erisque}, 2005, No. 304. 

\myref{\SheffGFF}
  S. Sheffield, Gaussian free fields for mathematicians, arXiv:math.PR/0312099.

\myref{\Sl}
  L. J. Slater, ``Generalized hypergeometric series,'' Cambridge University Press, Cambridge, 1966.

\myref{\Sta}
  R. P. Stanley, ``Enumerative Combinatorics,'' vol. I, Cambridge University Press, Cambridge, 1997.

\myref{\Statwo}
  R. P. Stanley, ``Enumerative Combinatorics,'' vol. II, Cambridge, University Press, Cambridge, 1999.

\endroster\par}

\enddocument